\documentclass[a4paper,fleqn,usenatbib]{mnras}

\usepackage{newtxtext,newtxmath,amsmath}

\usepackage[T1]{fontenc}
\usepackage[utf8]{inputenc}
\usepackage{ae,aecompl}

\usepackage{graphicx}	
\usepackage{amsmath}	
\usepackage{amssymb}	
\usepackage{hyperref}
\hypersetup{
    colorlinks=true,
    linkcolor=blue,
    filecolor=magenta,      
    urlcolor=cyan,
}

\newcommand{\codenamefirstns}{\texttt{adaptiveBox}}
\newcommand{\codenamesecondns}{\texttt{butterfly}}
\newcommand{\codenamefirst}{\texttt{adaptiveBox }}

\newcommand{\facc}{$f_{\rm acc}$ }

\title[Stellar halos in TNG100]{A missing outskirts problem? Comparisons between stellar halos in the Dragonfly Nearby Galaxies Survey and the TNG100 simulation}

\author[Merritt et al.]{
Allison Merritt,$^{1}$\thanks{E-mail: merritt@mpia.de}
Annalisa Pillepich,$^{1}$
Pieter van Dokkum,$^{2}$ 
Dylan Nelson, $^{3}$
\newauthor
Lars Hernquist, $^{4}$ 
Federico Marinacci, $^{5}$
and Mark Vogelsberger $^{6}$
\\
$^{1}$Max-Planck-Institut f{\"u}r Astronomie, K{\"o}nigstuhl 17, D-69117 Heidelberg, Germany\\
$^{2}$ Astronomy Department, Yale University, New Haven, CT 06511, USA \\
$^{3}$ Max-Planck-Institut f{\"u}r Astrophysik, Karl-Schwarzschild-Str 1, D-85741 Garching, Germany \\
$^{4}$ Harvard-Smithsonian Center for Astrophysics, 60 Garden Street, Cambridge, MA 02138, USA \\
$^{5}$ Department of Physics and Astronomy, University of Bologna, Via Gobetti 93/2, I-40129, Bologna, Italy \\
$^{6}$ Department of Physics, Kavli Institute for Astrophysics and Space Research, MIT, Cambridge, MA 02139, USA \\
}

\date{Accepted XXX. Received YYY; in original form ZZZ}

\pubyear{2019}

\begin{document}
\label{firstpage}
\pagerange{\pageref{firstpage}--\pageref{lastpage}}
\maketitle

\begin{abstract}
Low surface brightness galactic stellar halos provide a challenging but promising path towards unraveling the past assembly histories of individual galaxies. Here, we present detailed comparisons between the stellar halos of Milky Way-mass disk galaxies observed as part of the Dragonfly Nearby Galaxies Survey (DNGS) and stellar mass-matched galaxies in the TNG100 run of the IllustrisTNG project. We produce stellar mass maps as well as mock $g$ and $r$-band images for randomly oriented simulated galaxies, convolving the latter with the Dragonfly PSF and taking care to match the background noise, surface brightness limits and spatial resolution of DNGS. We measure azimuthally averaged stellar mass density and surface brightness profiles,
and find that the DNGS galaxies generally have less stellar mass (or light) at large radii (>20 kpc) compared to their mass-matched TNG100 counterparts, and that simulated galaxies with similar surface density profiles tend to have low accreted mass fractions for their stellar mass. We explore potential solutions to this apparent "missing outskirts problem" by implementing several ad-hoc adjustments within TNG100 at the stellar particle level. Although we are unable to identify any single adjustment that fully reconciles the differences between the observed and simulated galaxy outskirts, we find that artificially delaying the disruption of satellite galaxies and reducing the spatial extent of in-situ stellar populations result in improved matches between the outer profile shapes and stellar halo masses, respectively. Further insight can be achieved with higher resolution simulations that are able to better resolve satellite accretion, and with larger samples of observed galaxies.

\end{abstract}




\section{Introduction}\label{sec:intro}

Stellar halos are an inevitable outcome of the $\Lambda$CDM cosmological paradigm, and provide a promising path towards unraveling the past assembly histories of their central galaxies. As a result of structure formation in this theory, dark matter halos build up their mass over time through a series of mergers with other halos as well as smooth accretion \citep{whiterees1978,moore1999}; and, as the dark matter halos are disrupted, so are their stellar components. One byproduct of a galaxy's merger history, therefore, is a diffuse and often highly substructured envelope of stars surrounding the main body of the galaxy --- the stellar halo --- that holds clues to the number, masses, and timing of past accretion events. As a consequence of dynamical friction, most accreted mass will end up residing in the bulge or disk regions of the galaxy \citep{pillepich2015} and become nearly impossible to distinguish observationally; however, the material in the outskirts is preserved over relatively long dynamical timescales \citep{bullockjohnston2005}. 

Unfortunately, reliable observations of low surface brightness stellar halos are notoriously difficult to obtain and have resulted in relatively small sample sizes thus far. Star counts can in some cases provide an extremely deep and panoramic view of the stellar density, color, and inferred age and metallicity of galactic stellar halos \citep[see, for example,][]{ibata2014,okamoto2015,crnojevic2016}, but as the apparent brightness of individual stars decreases with the square of the distance, these studies become prohibitively difficult beyond $\sim 5-7$ Mpc from the ground \citep{danieli2018}. The same technique applied to space-based instruments has been proven successful out to $\sim 20$ Mpc and has revealed a striking degree of diversity in both the masses and stellar populations of the stellar halos of massive disk galaxies \citep[e.g.][]{monachesi2013,harmsen2017}, but these are expensive observations and often suffer from sparse area coverage. 

Integrated light observations provide a more time-efficient way forward, but suffer from degeneracies in age and metallicity, complicating any inferences of stellar populations. More importantly, these observations are typically heavily contaminated by scattered light from stars or other compact sources within or near the field of view \citep[see e.g.][]{slater2009}: performing any quantitative analyses requires a careful characterization of the point spread function (PSF) and separation of scattered light from galaxy light. The difficulty of carrying out this exercise over angular size scales comparable to stellar halos has typically limited the number of individual galaxies observed down to surface brightness limits deemed deep enough to robustly explore stellar halos (approximately $30$ mag arcsec $^{-2}$; \citealt{johnston2008}), although there are some recent notable exceptions \citep[e.g.][]{huang2018,rich2019}.

Stacking thousands of galaxies has proved to be a useful way to explore a number of trends between stellar halo and galaxy properties \citep[e.g.][]{dsouza2014,wang2019}, but information on the galaxy-to-galaxy variation is lost \citep[see e.g. ][for a measure and discussion of scatter in the Illustris simulation]{pillepich2014}. Furthermore, \cite{dejong2008} showed that in stacks of thousands of SDSS disk galaxies, up to 80 percent of the light in apparently anomalous ``red halos'' can be attributed to an incomplete accounting of the PSF. Finally, the presence of Galactic cirrus contaminates stellar halo detection and any following analysis \citep[see e.g.][]{roman2019}, and integrated light surveys are forced to simply avoid regions in the sky with relatively high cirrus.

Simulating stellar halos is not any easier than observing them --- the outskirts of stellar halos are multiple orders of magnitude less dense than the central regions of galaxies and require high spatial and mass resolution, which presents a difficult computational challenge. Analytic models \citep{purcell2007} and dark matter-only N-body simulations coupled with semi-analytic prescriptions \citep{bullockjohnston2005} or stellar particle tagging \citep{cooper2010,amorisco2017a} have provided useful insight into the expected amount and distribution of accreted material for galaxies across a wide range of stellar masses; however, \cite{bailin2014} demonstrated that the simplifying assumptions necessary when modeling stellar halos using dark matter-only methods can lead to a systematic underestimation of both concentration and amount of substructure. Hydrodynamic zoom-in simulations such as Eris \citep{pillepich2015}, AURIGA \citep{monachesi2019}, APOSTLE \citep{oman2017}, FIRE \citep{sanderson2017} or ARTEMIS \citep{font2020} are able to more self-consistently model stellar halos, but lack the statistics to fully explore galaxy-to-galaxy variation. 

Careful comparisons between observations and simulations of stellar halos are nevertheless key to understanding the merger histories of individual galaxies. Their observationally measured stellar masses, colors, shapes, slopes, and substructure can be analyzed side by side with predictions from simulations. Where they agree, we can use the simulations to provide meaningful physical context to the observations; and where they disagree, we have the opportunity to learn something useful about the assumptions and limitations that factor into both types of datasets.

In \cite{merritt2016a}, we presented measurements of the stellar halos of 8 Milky Way-mass spiral galaxies observed with the Dragonfly Telephoto Array \citep[][]{abraham2014} as part of the Dragonfly Nearby Galaxy Survey (DNGS). In spite of our expectation of finding clear signatures of stellar halos (either in diffuse light or coherent substructures) around every galaxy in our sample\footnote{Detecting stellar halos around spiral galaxies was, in fact, one of the original reasons for designing Dragonfly.}, we instead saw a remarkable diversity in the outskirts of these galaxies. Intriguingly, a comparison between our estimates of the outer stellar halo mass fractions and the accreted fractions reported by existing simulations suggested a possible tension between observations and simulations --- specifically that simulations could be overpredicting the average amount of mass in stellar halos, and underpredicting the galaxy-to-galaxy scatter in this measurement. 

However, carrying out ``apples-to-apples'' comparisons between observations and simulations is not a trivial pursuit, and ultimately limited our ability to understand whether (or to what extent) any real tension existed. The challenge is partly due to widely varying definitions of the stellar halo, and partly due to the fact that simulations track accreted (ex-situ) material regardless of where in the galaxy it ultimately ends up, while observations can only hope to disentangle ex-situ from in-situ (i.e., stars that formed within the galaxy from cold, condensed gas) structure in the outskirts of galaxies --- and even in this regime the separation between these components is not straightforward \citep[e.g.][]{purcell2010,dorman2013}.

With the latest advances in cosmological hydrodynamic simulations \citep[e.g.][]{vogelsberger2014b,dubois2014,schaye2015,dave2016,remus2017,henden2019}, however, this is a good moment to examine the differences between observed and simulated stellar halos in greater detail  \citep[e.g.][]{cook2016,elias2018,dsouzabell2018}.  In this paper, we compare and contrast the DNGS stellar halos with those in the TNG100 simulation of the IllustrisTNG project \citep{pillepich2018b,naiman2018,springel2018,nelson2018,marinacci2018}.

The scope of this analysis is multifold. First, we focus on carrying out ``apples-to-apples'' comparisons as closely as possible.
Section \ref{sec:data} describes the two datasets as well as our sample selection and stellar mass-matching procedures within TNG100. In Section \ref{sec:part2im}, we summarize our methodology for converting stellar particle data from the simulation into 2D images that can be analyzed analogously to the observations; and we provide a library of example images in Section \ref{sec:measured:images}. Section \ref{sec:measured:profiles} presents the TNG100 stellar mass density profiles and mock observed surface brightness profiles, and provides a comparison with DNGS.
Second, we explore different metrics for quantifying stellar halos in Section \ref{sec:metrics}. We delve into a number of different definitions for the masses of stellar halos, and explicitly distinguish these from accreted stellar masses: the former is an observable quantity, while the latter is the total stellar mass acquired through mergers or accretion, and is not directly observable.
Next, we track individual stellar particles in our TNG100 galaxies across the full simulation in order to better understand the buildup of their stellar halos. In Section \ref{sec:ptcls}, we explore what we can learn about observed stellar halos via comparisons to simulated galaxies matched in both stellar mass and profile shape.
Finally, in Section \ref{sec:missing} we discuss the myriad of caveats involved in comparing cosmological hydrodynamic simulations with observations, and examine the extent to which we are able to put constructive and physical constraints on simulations using observations of stellar halos.
We summarize our findings and discuss directions for future work in Section \ref{sec:conclusions}. Throughout our analysis we assumed a $\Lambda$CDM cosmology, and used the cosmological parameters as detailed in \cite{planckcollab2016} where applicable. All quantities are reported in physical (rather than comoving) units unless specified otherwise, and we make a conscious effort to limit our analysis to properties that are either directly observable or possible to derive from optical imaging.

\section{Data}\label{sec:data}
\begin{table*}
	\centering
	\caption{The galaxies included in the DNGS sample included in \citet{merritt2016a}. Absolute magnitudes and distances were taken from \citet{tully2009}, and inclination angles from HyperLeda \citep{makarov2014}. Both of the stellar masses quoted are integrated quantities. $M_{\rm stell,obs}$ is the ``observed'' stellar mass, measured by integrating the surface density profile down to $10^{4}M_{\odot}$ kpc$^{-2}$ (see Section \ref{sec:data:simulations}), and $M_{\rm stell}(< 30 \, {\rm kpc})$ is the stellar mass enclosed within a semi-major axis of 30 kpc. $R_{\rm half}$ is the radius that contains half of the stellar mass of the galaxy (assuming $M_{\rm stell,obs}$). Details of the surface density profiles are given in \citet{merritt2016a} and in Section \ref{sec:measured:profiles} of this paper.}
	\label{tab:dngs}
	\begin{tabular}{ccccccc} 
		\hline
		\hline
		Galaxy & $M_{B}$ [mag] & Distance [Mpc] & Inclination [deg] & $M_{\rm stell,obs}$ [$10^{10} M_{\odot}$] & $M_{\rm stell}(< 30 \, {\rm kpc})$ [$10^{10} M_{\odot}$] & $R_{\rm half}$ [kpc] \\
		\hline
		NGC 1042 & -20.27 & 17.3 & $58.1$ & $1.51 \pm 0.48$ & $1.51 \pm 0.48$ & $5.31$ \\
		NGC 1084 & -20.41 & 17.3 & $49.9$ & $4.31 \pm 1.41$ & $4.21 \pm 1.38$ & $3.4$ \\
		NGC 2903 & -20.3 & 8.5 & $67.1$ & $4.89 \pm 1.56$ & $4.87 \pm 1.55$ & $3.9$ \\
		NGC 3351 & -20.36 & 10.0 & $54.6$ & $5.79 \pm 1.94$ & $5.79 \pm 1.94$ & $2.08$ \\
		NGC 3368 & -20.03 & 7.24 & $51.1$ & $8.84 \pm 2.85$ & $8.81 \pm 2.84$ & $2.56$ \\
		NGC 4220 & -19.31 & 17.1 & $90.0$ & $6.06 \pm 1.94$ & $6.03 \pm 1.93$ & $3.2$ \\
		NGC 4258 & -20.2 & 7.61 & $68.3$ & $7.55 \pm 2.40$ & $7.44 \pm 2.36$ & $4.88$ \\
		M101 & -20.2 & 7.0 & $16.0$ & $5.84 \pm 1.85$ & $5.82 \pm 1.84$ & $6.54$ \\
		\hline
	\end{tabular}
\end{table*}

\subsection{Observations: The Dragonfly Nearby Galaxies Survey}\label{sec:data:observations}
We used observations of the stellar halos of eight spiral disk galaxies presented as part of the Dragonfly Nearby Galaxies Survey (DNGS) in \cite{merritt2016a}. We observed our sample from $2013-2016$ with the Dragonfly Telephoto Array (``Dragonfly'' for short), a robotic, nearly-autonomous refracting array of telephoto lenses optimized to detect extended, low surface brightness optical emission \citep{abraham2014}\footnote{Dragonfly is presently a 48-lens array, but over this time period it grew from 8 to 24 lenses.} and equipped with Sloan $g$- and $r$-band filters. The telescope has a native spatial resolution of 2.85 arcsec pixel$^{-1}$ (this is resampled to 2.5 arcsec pixel$^{-1}$ during the data reduction process), and the PSF has a FWHM of 7 arcsec (derived empirically from DNGS data; see \citealt{merritt2016a} and \citealt{zhang2018} for details).

Dragonfly is ideally suited for surveying diffuse stellar halos --- sub-wavelength structure anti-reflective coatings on each optical element suppress scattered light by an order of magnitude relative to standard instruments, resulting in a steeply-declining and stable wide angle PSF; and the wide field of view ($2\times 3$ degrees) comfortably encompasses each galaxy as well as its stellar halo and satellite system. A combination of dithering and small, deliberate pointing offsets between individual cameras ensures multiple independent lines of sight to each target galaxy, removing any remaining concerns over confusion from optical artifacts.\footnote{In the Southern hemisphere, the Huntsman Telescope is similarly built from Canon telephoto lenses;  \citealt{spitler2019} provide a detailed characterization of flat fields and demonstrate that the corresponding 5$\sigma$ surface brightness floor (due to flat fielding errors) is as low as 33 mag arcsec$^{-2}$ across the field of view.}

The instrument and observing strategy are paired with a low surface brightness-optimized reduction pipeline \citep[described in][]{danieli2019}. One key step is that we monitor individual images for any changes in the wide-angle PSF which are thought to be atmospheric in origin (perhaps caused by high-atmosphere aerosols; \citealt{devore2013}), unlike the inner regions of the PSF which obtain their structure mostly from the optical components of the instrument or telescope \citep{slater2009}. These atmospheric variations can happen on timescales of minutes, and affect images at surface brightnesses too low to be detected by visual inspection. In the presence of even very thin atmospheric cirrus, light is thrown from the central regions of the PSF to the outskirts, which conveniently manifests as a shift in the photometric zeropoint \citep[see][for a more detailed description]{zhang2018}. We therefore exclude (among other criteria) any image that deviates from the nominal value by more than $\sim 0.1$ mag, and caution that, in general, deep images produced from either long exposures ($\geq 10$ minutes) or by stacking a large number of shorter exposures are not immune to this effect.

One other critical reduction step worth noting is our approach to sky subtraction. A particular pitfall that should be avoided is the over-subtraction of the sky, caused by including pixels with light from low surface brightness galaxy outskirts in its calculation. To ensure accurate sky values we perform sky subtraction in two phases: in the first, we heavily mask out all sources before fitting and subtracting a $3^{\rm rd}$ degree polynomial from the images; and in the second, we create a more aggressive mask using a (preliminary) stack of all available images for a given target and repeat the sky modeling and subtraction \citep[see][]{merritt2016a,zhang2018}. As a result of our choice to use a $3^{\rm rd}$ degree polynomial we are confident in our results over angular scales of $\sim 45$ arcmin (or physical scales of $\sim 130$ kpc and $260$ kpc at distances of $10$ Mpc and $20$ Mpc, respectively.)

\subsubsection{The DNGS Sample}
The galaxies were selected based \text{only} on their absolute magnitude ($M_{B} < -19.3$) and proximity; that is, we made sure to include only the nearest (and therefore apparently largest) galaxies above this luminosity threshold, as Dragonfly performs best when dealing with extended objects. We avoided regions of significant Galactic cirrus with a cut on FIR flux ($F_{100 \mu m} \leq 1.5$ mJy/Sr along the line of sight, obtained from IRAS maps). We also explicitly excluded galaxies in the Local Group (defined loosely as having distances within 2 Mpc) as well as known members of the Virgo cluster. No other galaxy properties were considered or given priority, and as a result, the only common property among the sample is their spiral disk morphology (see Figure 1 of \citealt{merritt2016a} for images of each galaxy). The DNGS galaxies cover a wide range in inclination angle ($16-90$ degrees), color, physical size, and disk-to-bulge ratios, and lie at distances between $7-18$ Mpc. However, in \cite{merritt2016a} our goal was to characterize the stellar halos of Milky Way-like galaxies, and we therefore selected galaxies from DNGS with visually identifiable spiral structure. 
We note that, in an effort to take advantage of Dragonfly's large field of view, we also searched for additional spiral galaxies above our magnitude limit (effectively just relaxing the proximity requirement) in fields that we had already observed; this lead to the inclusion of NGC 4220 in our sample.

We observed each galaxy (field) for $15-20$ hours with Dragonfly, and were able to measure azimuthally averaged surface brightness profiles down to limits of $30-32$ mag arcsec$^{-2}$.\footnote{On smaller angular scales relevant for individual substructure detection, the limiting surface brightness ranges from 28.6-29.2 mag arcsec$^{-2}$ and 29-30 mag arcsec$^{-2}$ (measured in 12 and 60 arcsecond apertures, respectively).} In \cite{merritt2016a} we also presented stellar mass surface density profiles for our sample, which we obtained by combining our surface brightness profiles with mass-to-light ratio profiles estimated from optical $g-r$ color profiles. 

We measured total stellar masses by integrating these surface density profiles down to a common density threshold of $10^{4} M_{\odot}$ kpc$^{-2}$ and found that, despite being broadly similar to the Milky Way in stellar mass ($1.5-9\times 10^{10}M_{\odot}$), the spiral galaxies in DNGS display a remarkable degree of diversity in their outskirts, with some featuring large streams and others lacking any detectable signatures of past merger events. This variation in the low surface brightness outskirts suggests a similarly large variation in the assembly histories of the galaxies (and see also \citealt{pillepich2014} and \citealt{monachesi2016}, who demonstrated that this diversity is also reflected in the metallicity and color profiles of Milky Way-mass galaxies). In this work, we used the observed profiles measured in \cite{merritt2016a} directly, with no changes or scalings compared to the published paper.\footnote{The imaging data from this survey can be accessed at \url{https://www.dragonflytelescope.org/data-access.html}.} Table \ref{tab:dngs} highlights some of the basic properties of our DNGS sample.

\subsection{Simulations: TNG100 of IllustrisTNG}\label{sec:data:simulations}
We paired our observations with simulated galaxies from The
Next Generation Illustris (hereafter IllustrisTNG, or simply TNG) project, a
suite of state-of-the-art cosmological magnetohydrodynamic simulations \citep{springel2018,nelson2018,pillepich2018b,naiman2018,marinacci2018}
that serves as the successor to Illustris \citep{vogelsberger2014a,vogelsberger2014b,genel2014}. The TNG physical model for galaxy formation \citep{weinberger2017,pillepich2018a}, which builds upon the original Illustris model \citep{vogelsberger2013,torrey2014}, features several updates to the growth of and feedback from supermassive black holes, as well as stellar evolution, gas enrichment, and the injection of galactic winds.
Most notably for studies of stellar halos, low mass ($M_{\rm stell} \leq 10^{10}M_{\odot}$) galaxies in TNG are smaller in size relative to Illustris, bringing them into better qualitative agreement with observations \citep{pillepich2018a,genel2018}. Moreover, the TNG100 galaxy stellar mass functions are in closer agreement with observations; together, these changes affect the number and typical stellar masses of shredded satellite galaxies that build up the stellar halos of more massive galaxies. We note that although TNG100 has been shown to successfully reproduce a remarkable number of observed properties of galaxies, there are also a handful of known areas of tension \citep[see Section 5.1 of][as well as Section \ref{sec:discussion:toymodels} here for a more in depth discussion]{nelson2019}.

IllustrisTNG is comprised of the TNG50, TNG100, and TNG300 cosmological boxes, spanning 51.7, 110.7, and 302.6 Mpc on a side, respectively. We chose to work with TNG100, as it represents a compromise between statistical sample sizes of Milky Way-mass galaxies and the resolution needed to quantify the relatively sparse stellar halo regions in a physically meaningful way. TNG100 contains $\sim 1.2\times 10^{10}$ resolution elements in total, and has a baryonic mass resolution of $1.4\times 10^{6}M_{\odot}$ with a gravitational softening length of $740$ pc for stellar particles. For Milky Way-like galaxies, we typically\footnote{The values quoted here are the $25^{\rm th}$ and $75^{\rm th}$ percentiles. We also calculated the $1^{\rm st}$ and $99^{\rm th}$ percentiles; these are $\sim 1400 - 26,000$ and $\sim 50 - 7000$, respectively. These values are a strong function of galaxy stellar mass.} find between $\sim 3000-9000$ and $\sim 250-1300$ stellar particles beyond 2 and 5 half-mass radii at $z=0$, respectively.

Throughout this work, we considered any stellar particle that formed within a subhalo (i.e., not on the main progenitor branch of the central galaxy) to be an ex-situ particle. From a formation times perspective, this means that any particle that formed before becoming gravitationally bound to the central galaxy is considered ex-situ.

\subsubsection{The TNG100 Sample}

\begin{figure*}
    \centering
    \includegraphics[width=15cm]{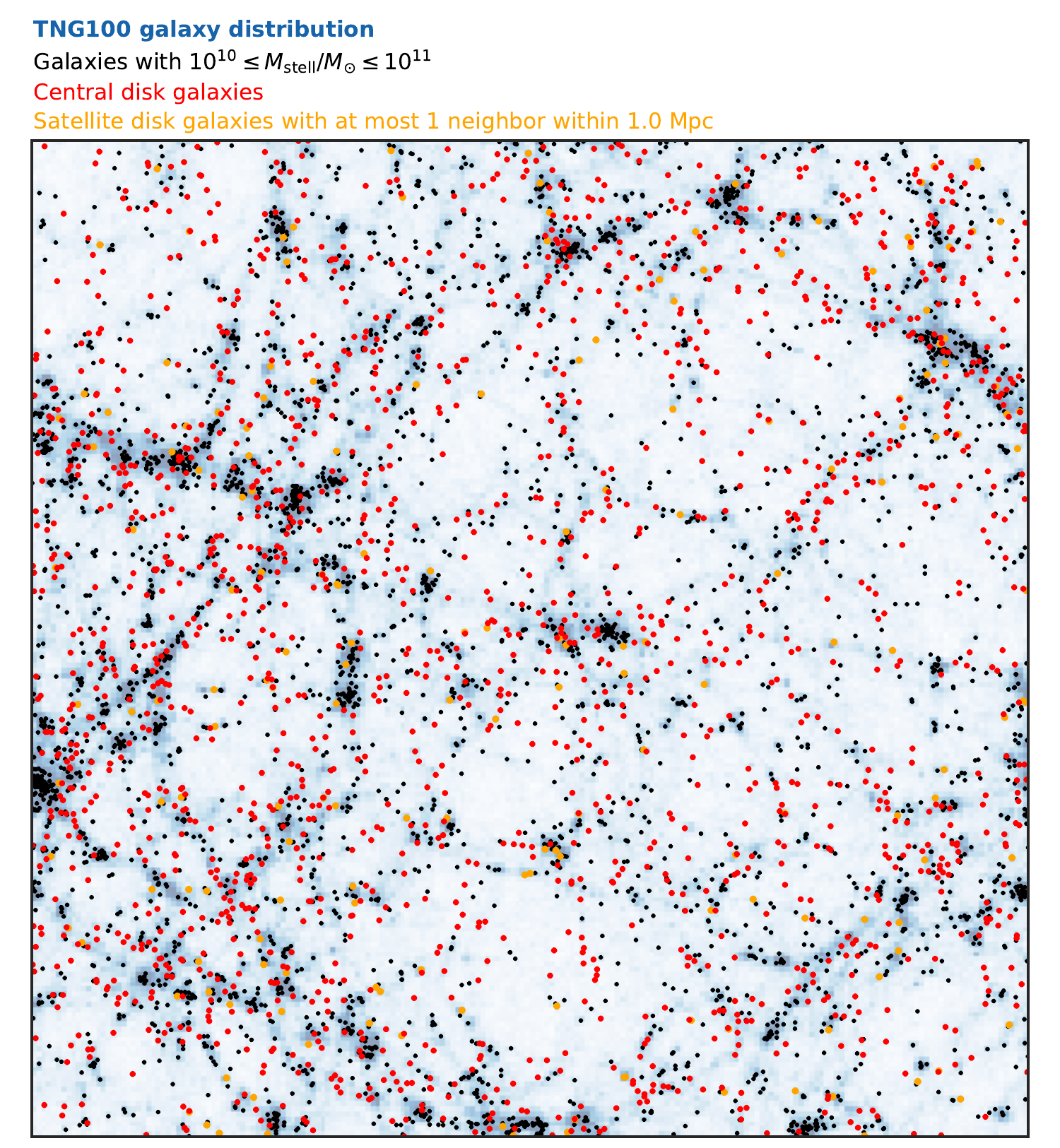}
    \caption{\textbf{An overview of our TNG100 parent galaxy sample selection.} Red dots show the positions of central disk galaxies in TNG100 with $10^{10} \leq M_{\rm stell}/M_{\odot} \leq 10^{11}$. Orange dots point to satellite disk galaxies in the same mass range; to avoid Virgo-like clusters, we require that satellite galaxies have no more than one massive ($\geq 10^{10}M_{\odot}$) neighbor within 1 Mpc. Galaxies rejected on the basis of these environmental and morphological criteria are shown in black, for reference. In the background, the spatial distribution of all galaxies in TNG100 is shown in blue.}
    \label{fig:environment}
\end{figure*}

We constructed a parent sample of "Milky Way-like" galaxies from TNG100 by identifying disk galaxies with (broadly) similar stellar masses that live in relatively low density environments at $z=0$, analogous to the DNGS sample. 

To do this, we first selected all galaxies within TNG100 with stellar masses in the range $10^{10} < M_{\rm stell}/M_{\odot} < 10^{11}$. In this context, a galaxy's stellar mass is defined as the total mass in stellar particles gravitationally bound to its host subhalo (and not including stellar particles bound to any subhalos --- satellites --- that might exist within it). Moving forward, we will refer to this stellar mass as the ``true'' stellar mass of the galaxy, $M_{\rm stell, true}$.  

We then estimated the local environment for these galaxies, using the number of massive ($M_{\rm stell, true} \geq 10^{10}M_{\odot}$) galaxies within 1 Mpc as our metric. Only central galaxies or satellite galaxies with at most 1 nearby massive neighbor are included in our sample, resulting in 3467 central galaxies with an additional 505 satellite galaxies. In practice, this means that we avoid high density cluster regions (similar to the Virgo Cluster) while including Local Group analogs. The latter choice is particularly important, as DNGS contains multiple pairs of galaxies that are members of the same loose group \citep{makarov2011}.\footnote{Not to mention the fact that to date, the two stellar halos studied in the most detail are those of the Milky Way and M31, which are near neighbors with separation less than 1 Mpc.} Including massive satellite galaxies in the sample also facilitates future studies of galaxy stellar halo variance at fixed environment.

The final prerequisite for a TNG100 galaxy being ``like'' the Milky Way is that it must be disk-dominated. Following \cite{Genel2015}, we opted to select disk galaxies kinematically, focusing on their distributions of circularity parameters. \cite{Marinacci2014} defined the circularity parameter $\epsilon$ for a stellar particle as the ratio of its specific angular momentum $J_{z}$ to the maximum possible specific angular momentum given its specific binding energy $J(E)$. Disk stellar particles can then be identified as those with high circularity (typically $\epsilon \geq$ 0.7); and, further, disk-dominated galaxies are those with high fractions of disk stars. Random motions of bulge stars can contaminate these measurements, however, and we estimated this contribution by doubling the fraction of stars with $\epsilon \leq -0.7$ under the assumption that the distribution of bulge stellar particle circularities is symmetric around zero. Formally, we used the following criteria for galaxies to be considered disks:
\begin{equation}
    f_{\rm disk} - f_{\rm bulge} \geq 0.4
\end{equation}
where $f_{\rm disk}$ and $f_{\rm bulge}$ are both mass fractions measured within 10 half-mass radii. We note that this dynamic disk selection is not perfect --- although we avoid contamination from more spheroidal galaxies, we are likely missing some disky galaxies with especially massive bulges or with significant mass in tidal features (e.g., from recent massive mergers).

Figure \ref{fig:environment} shows the spatial positions of our final parent sample projected within the TNG100 box. In the end, we have 1656 central and 188 satellite disk galaxies with stellar masses between $10^{10}-10^{11}M_{\odot}$ (red and orange dots, respectively), with total accreted stellar mass fractions -- defined as the ratio of the stellar mass in ex-situ stellar particles to the true stellar mass -- ranging from 0.5\% to 60\% \citep[see also][]{rodriguezgomez2016}. We also show, for reference, the positions of galaxies in the same mass range which were excluded on the basis of environment (i.e., satellite galaxies residing in high density environments) or morphology (i.e., central or satellite galaxies that are too spheroidal).
We note that, since our TNG parent sample was selected on the basis of stellar mass rather than $B$-band luminosity, these galaxies do not populate the same stellar mass-luminosity distribution as the DNGS sample. We will return to the possible implications of this in Section \ref{sec:missing:obs}.

In addition to this parent sample, we also defined smaller samples of stellar mass-matched TNG100 galaxies for each DNGS galaxy. This time we applied a more observer-friendly definition of stellar mass --- specifically the integral of the stellar mass surface density profile. We denote this ``observed'' stellar mass  as $M_{\rm stell, obs}$. When calculating TNG100 stellar masses in this way, we only integrated the density profiles down to a floor of $10^{4}M_{\odot}$ kpc$^{-2}$, as this corresponds to the typical depth of the DNGS observations (see Section \ref{sec:measured:profiles:mass} for more details). The physical radius corresponding to this surface density threshold varies between $\sim 20-70$ kpc for the DNGS sample, although we note that this is an extrapolation (from a last observed point of $\sim 10^{4.5}M_{\odot}$ kpc$^{-2}$) in two of the eight galaxies. TNG100 galaxies exhibit an even larger variation in physical extent under this definition, ranging from $\sim 40 - 140$ kpc. We can infer from these numbers that TNG100 galaxies are more extended than DNGS galaxies, and we will return to this idea in Sections \ref{sec:metrics} and \ref{sec:discussion:matching}. 

As a consequence of the shape of the stellar mass function, if we were to simply select \textit{all} TNG100 galaxies that fall within the observational errors (see Table \ref{tab:dngs}), each mass-matched sample would remain strongly weighted towards low mass galaxies by number. To alleviate this issue, we instead used Rejection Sampling \citep{press1992} to obtain a normal distribution of TNG100 galaxy stellar masses for each DNGS galaxy. We calculated the probability $P(M_{\rm stell,obs})$ of drawing each TNG100 stellar mass from a Gaussian distribution with a mean and standard deviation set by $M_{\rm stell,DNGS}$ and $\sigma_{M_{\rm stell,DNGS}}$, respectively, and accepted the first 50 TNG100 galaxies with $P(M_{\rm stell,obs})$ less than a random number drawn from the same distribution.  

We emphasize that, in addition to the definition of the mass-matched samples, all comparisons between observations and simulations were done using quantities measured from TNG100 galaxies (for example, surface density profiles or stellar halo mass fractions) in a manner identical to the DNGS sample. When exploring parameter space that contains unobservable properties, however, we will occasionally use the ``true'' stellar mass instead. In the following sections we will explicitly distinguish between the parent and mass-matched samples, and between ``true'' and ``observed'' stellar masses where applicable.

\section{From stellar particles to images}\label{sec:part2im}
\begin{figure*}
    \centering
    \includegraphics[width=\linewidth]{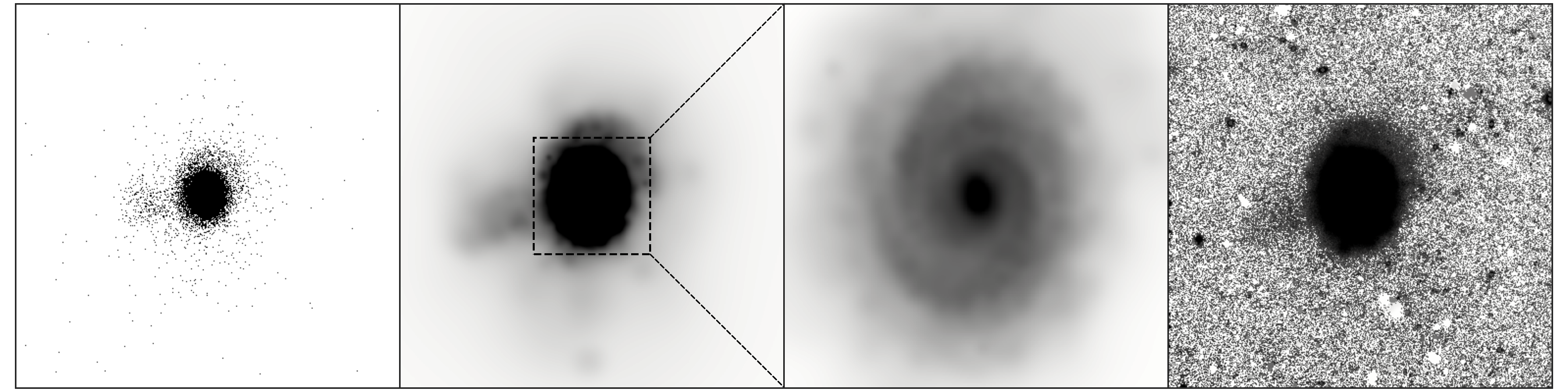}
    \caption{\textbf{From stellar particles to images. Left Panel:} A 2D histogram of the stellar particles in Subhalo 483900 at $z=0$ (chosen randomly, as an illustrative example). This galaxy has a stellar mass of $\sim 6 \times 10^{10} M_{\odot}$. \textbf{Second Panel:} The stellar mass distribution for the same galaxy; however, this time adaptive smoothing has been applied via \codenamefirst (Section \ref{smoothing:physical}). The kernel size is the 3D distance to the $k^{\rm th}$ nearest neighbor (here we use the fiducial value of $k=3$), and the 3D grid cells are 0.5 kpc. The final image pixel size is 2.5 arcsec (via \codenamesecondns, to match the Dragonfly detectors; Section \ref{smoothing:observational}), assuming a distance of 10 Mpc. Two low density streams and a diffuse envelope can be clearly seen here. \textbf{Third Panel:} A zoom in on the central $\sim 60 \times 60$ kpc; in addition to highlighting structure in the faint outskirts of galaxies, \codenamefirst also preserves bulge and spiral structure in the brighter central regions. \textbf{Right Panel:} The $g$-band mock observation for this galaxy, after being passed through \codenamesecondns. The larger stream is still prominent, although the fainter feature to the south of the galaxy is difficult to detect by eye. We note that the residuals from the star subtraction process are masked out during the profile fitting stage.}
    \label{fig:smoothingexample}
\end{figure*}

The first step towards comparing TNG100 galaxies to observations is creating 2D images from the stellar particle data --- specifically, stellar mass maps as well as $g$ and $r$ band images. This is a necessary step before moving on to measurements of surface density profiles or calculations of stellar halo masses, because the simulation provides a point-like sampling of stellar mass (or light) in 3D space, whereas observations sample stellar light integrated over pixel scales in a 2D projected space (however, as shown in Appendix \ref{app:images}, our results do not change even when no smoothing is applied).

We split this process into a ``physical smoothing'' step (Section \ref{smoothing:physical}) and an ``observational smoothing'' step (Section \ref{smoothing:observational}). The former allows us to go from stellar particle data to 2D ``idealized'' maps of stellar mass or light, while the latter folds in the effects of the point spread function (PSF) and realistic surface brightness limits.

\subsection{Physical smoothing}\label{smoothing:physical}

We developed \codenamefirst to produce 2D mass and light distributions of each TNG100 galaxy in our parent sample. First, we constructed (initially empty) 3D pixel grids, with the total size determined by the choice of spatial resolution (in this case, we adopt a default size of $0.5$ kpc per pixel\footnote{While $0.5$ kpc$/$pixel may seem large, it is comparable to the gravitational softening length of TNG100 and easily small enough to study the properties of the diffuse galaxy outskirts. See Appendix \ref{app:images} for more details.}) and distance to the galaxy (we placed all of the TNG100 galaxies at $10$ Mpc, which is typical for the DNGS sample). The grids span up to $200$ kpc on a side, ensuring our ability to measure profiles out as far as the deepest observations of stellar halos in the Local Volume \citep[e.g.][]{ferguson2002,gilbert2012,deason2013,okamoto2015,cohen2016,merritt2016a,harmsen2017,medina2018}. We converted stellar particle coordinates from comoving to physical units, and shifted each galaxy such that the stellar particle with the deepest potential was centered in the frame. For consistency with the DNGS sample, we used random orientations for the TNG100 galaxy images, rather than enforcing an edge-on or face-on geometry.

The three fundamental strategies for filling in the 3D pixel grid with stellar mass or flux are: $(1)$ make a simple 3D histogram, meaning that the only ``smoothing'' applied to the stellar particles is their placement into the nearest pixel-sized bin; $(2)$ smooth stellar particles with a fixed-width kernel; and $(3)$ smooth stellar particles \textit{adaptively}, i.e. with a variable kernel width \citep[see e.g.][]{springel2005}. We chose the latter strategy, as the galaxies have a large dynamic range in stellar mass surface densities (varying by up to $\sim 5$ orders of magnitude). This choice is particularly important for studies of stellar halos, where we expect a diffuse component as well as narrower, higher density streams or other structures. By scaling the width of the smoothing kernel with the local stellar particle density, we were able to resolve structure --- including bulges, spiral structure, and traces of minor mergers --- in the higher density regions of galaxies while preserving the diffuse stellar halo on larger scales.

We modeled stellar particles as 3D Gaussians, and defined the value of each 3D pixel in the grid to be the sum of the contributions from every stellar particle\footnote{Due to our choice to apply uniform dimensions to every image, some small fraction of mass or light from the outermost stellar particles associated with a given galaxy are not included in the image.} belonging to the galaxy. Following \citet{torrey2015} and \citet{rodriguezgomez2019}, we set the smoothing kernel size to be the distance to its $k^{th}$ nearest stellar neighbor $d_{k}$ (fiducially, $k=3$, but see Appendix \ref{app:images} for a summary of the effects of changing the value of $k$). Then, for a pixel located at coordinates $(x_{0}, y_{0}, z_{0})$, the mass (or flux) from a stellar particle at coordinates $(x, y, z)$ was calculated as:

\begin{multline}
g(x,y,z) = A \, {\rm exp}\left[ -\frac{1}{2} \left( \left(\frac{x-x_{0}}{\sigma_{x}}\right)^{2} +  \left(\frac{y-y_{0}}{\sigma_{y}}\right)^{2} + 
\left(\frac{z-z_{0}}{\sigma_{z}}\right)^{2} \right) \right]
\end{multline}

where $\sigma_{x} = \sigma_{y} = \sigma_{z} = d_{k}$, and the normalization $A$ is set by requiring that the volume of the Gaussian is equal to the mass or flux of the stellar particle. However, we note that for computational efficiency, we only smooth the mass or flux of each stellar particle out to a maximum radius of $5 d_{k}$. Finally, after establishing the 3D distribution of mass or flux, we collapsed the pixel grids along the $z$ axis to end up with 2D projections of each galaxy.

Figure \ref{fig:smoothingexample} highlights the effects of adaptive smoothing for Subhalo 483900, one of the TNG100 galaxies in our parent sample. The left panel shows the ``raw'' simulation data in the form of a 2D histogram of its stellar particles (i.e., with minimal smoothing), while the following two panels show the adaptively-smoothed image. The difference is visually striking, particularly in the outer regions of the galaxy, as adjacent pixels in the original image can (and frequently do) jump between $0$ and $\sim10^{6} M_{\odot}$ (the typical mass of a single stellar particle), whereas pixels in the smoothed stellar mass map are significantly less noisy.

\subsection{Observational effects}\label{smoothing:observational}
Whenever possible, observers attempt to convert surface brightness profiles to stellar mass surface density profiles --- especially when comparing to simulations, where a stellar particle's stellar mass is more ``fundamental'' than its light in a given band. This requires deep, high quality photometry in two bands ($g$ and $r$ in Dragonfly's case), as we can utilize the relation between optical color and stellar mass-to-light ratios. 

In reality, however, ``apples-to-apples`` comparisons between observations and simulations using either stellar mass \textit{or} optical light profiles are riddled with complications. In one case, we treat a derived measurement of an unobservable property (stellar mass) in the same way as a direct simulation output; in the other, a directly observable property (stellar light) is analyzed alongside a derived simulation output. Neither choice is ideal, and therefore for completeness we carry out both comparisons in the following sections (although we note that the majority of our analysis is based on the stellar mass density profiles; see Section \ref{sec:measured:profiles}).
We note that we used the ``raw'' SDSS stellar photometry outputs from TNG100, which were computed using \cite{bruzualcharlot2003} stellar population synthesis models and do not include the effects of dust obscuration. \cite{nelson2018} found that the integrated $g-r$ colors of TNG100 galaxies become $\sim 0.2$ magnitudes redder when dust modeling is included; however, dust column densities scale with neutral gas density and are therefore lower in the outskirts of galaxies.

In this second phase, we wrote \codenamesecondns\footnote{The name of the code is a hat tip to the number of times, in the Dragonfly Telephoto Array's early years, that the first author was entertained by requests to elaborate on the exciting new ``Butterfly Telescope''.} to incorporate typical observational effects, namely the PSF and surface brightness limits. Starting from the 2D images produced by \codenamefirstns, we rebinned the pixel sizes to match the spatial resolution of Dragonfly observations (i.e., $2.5$ arcsec per pixel) at 10 Mpc. We converted the TNG100 fluxes to Dragonfly counts using photometric zeropoints determined from Dragonfly data, and scaled the images appropriately before convolving the data with the relevant Dragonfly PSF (derived empirically from $g$ or $r$ band DNGS data). The Dragonfly PSF has a FWHM of $\sim 7$ arcsec; see  \cite{merritt2016a} or \cite{zhang2018} for more details.

The last step was to place each galaxy into a relatively empty region of a (star-subtracted) DNGS field. This allowed us to simultaneously adopt both background noise and surface brightness limits representative of the survey (in particular, DNGS backgrounds are sometimes confusion-limited due to Dragonfly's large pixels, and this method naturally incorporates this source of systematic error). For simplicity, we used the same background field for every TNG100 galaxy -- namely, a patch of sky from the M101 field, which has surface brightness limits of 29.5 mag arcsec$^{-2}$ and $29.8$ mag arcsec$^{-2}$ on scales of 10 arcseconds in the $g$ and $r$ band images, respectively, and a large scale peak-to-peak variation (over the full $3\times 2.8$ degree field) of $\sim 0.2$ percent \citep{vandokkum2014,merritt2014,merritt2016a}. We show the $g$-band mock observation of Subhalo 483900 in the righthand panel of Figure \ref{fig:smoothingexample}.

\begin{figure*}
    \centering
    \includegraphics[width=\linewidth]{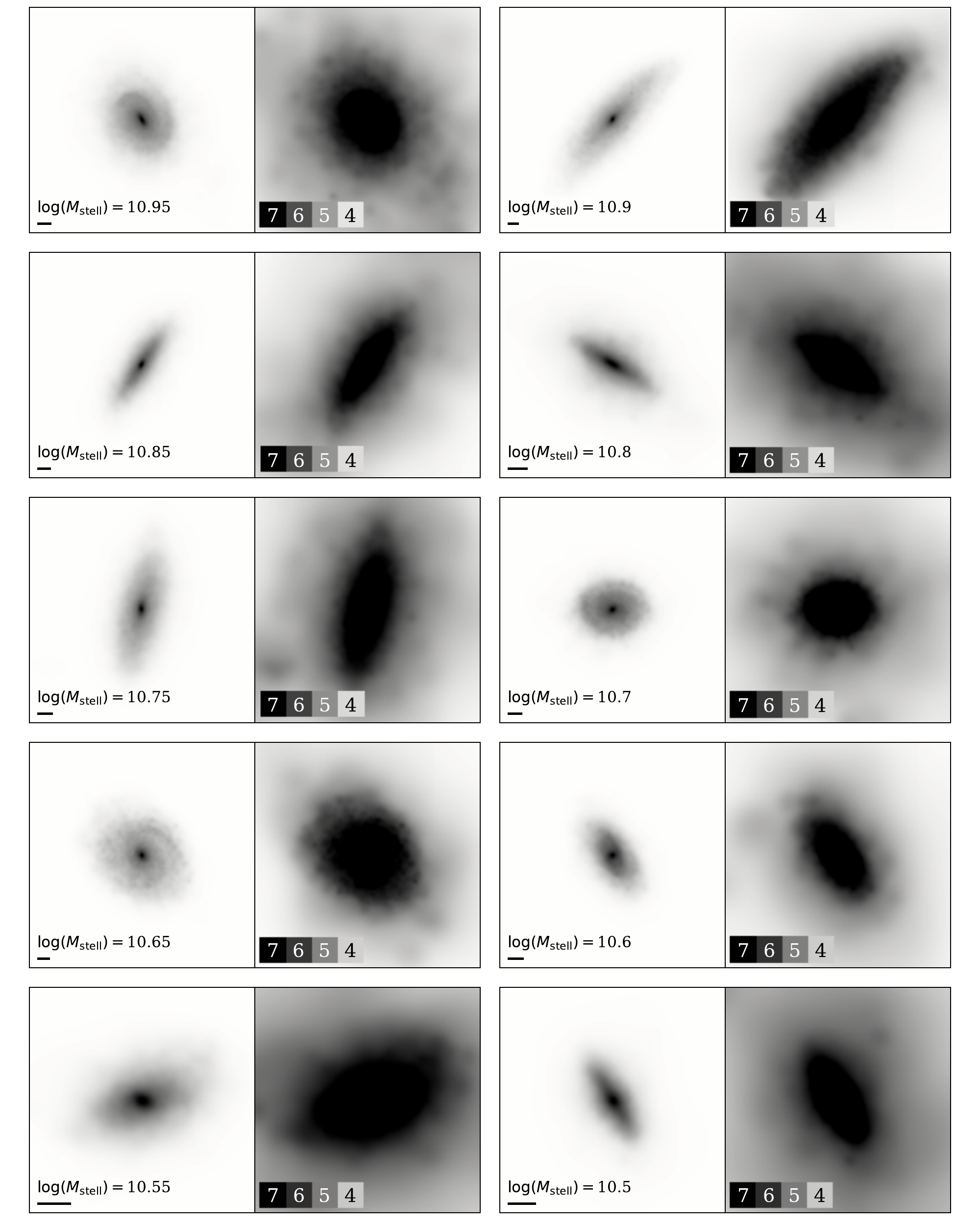}
    \caption{\textbf{A visual overview of our sample of Milky Way-like disks in TNG100.} Each pair of stellar mass maps shows an example galaxy selected randomly from a small mass bin (indicated on the left panel). For each galaxy, the \textbf{left} panel shows a more ``classic'' view, scaled to highlight the higher density regions, and the \textbf{right} panel is stretched to bring out the lower density outskirts. To facilitate comparisons between galaxies of different stellar masses and physical sizes, all images are $20$ half-mass radii on a side, and the black lines span 10 kpc for reference. In each righthand panel, a colorbar guides the eye to recognize surface densities of $10^{4}$, $10^{5}$, $10^{6}$ and $10^{7} M_{\odot}$ kpc$^{-2}$.  }
    \label{fig:library1}
\end{figure*}

\begin{figure*}
    \centering
    \includegraphics[width=\linewidth]{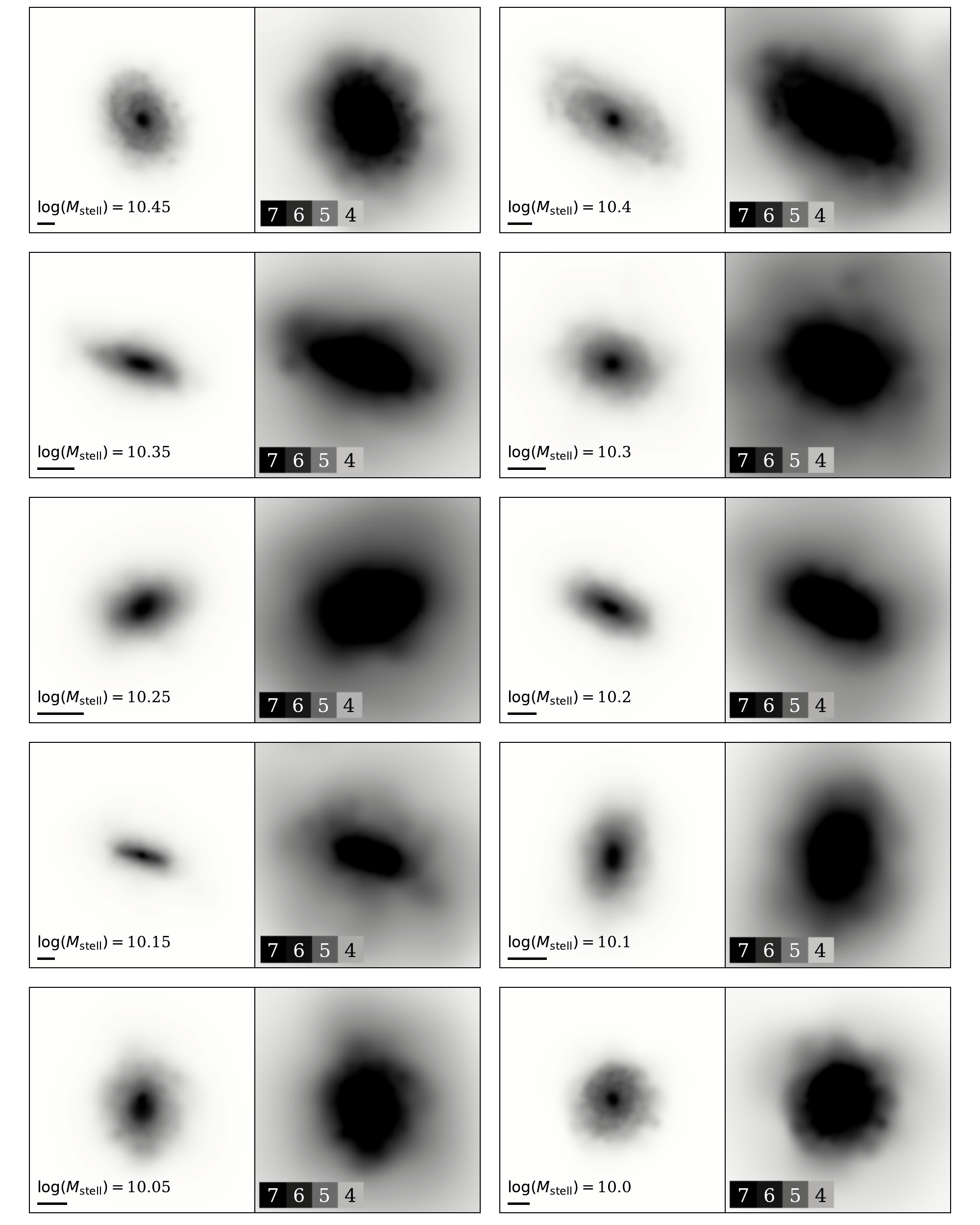}
    \caption{A continuation of Figure \ref{fig:library1}.}
    \label{fig:library2}
\end{figure*}


\section{Stellar halos: a visual inspection}\label{sec:measured:images}

Figures \ref{fig:library1} and \ref{fig:library2} present a ``library'' of example simulated galaxy stellar mass maps from our parent sample of Milky Way disks. Each figure shows 10 galaxies (yielding 20 in total), and we move through the sample in order of decreasing stellar mass in steps of 0.05 dex (choosing one galaxy randomly from each bin). In order to provide a fair representation of each galaxy, we present two different views --- both are stretched logarithmically; however, the first is scaled to highlight the high surface density structure and the central morphology of the galaxies (left panels), while the second brings the lower surface density regions into focus. All images span $20$ half-mass radii on a side, and we indicate the mass range and physical extent of 10 kpc on each panel. A colorbar indicates surface densities of $10^{4}$, $10^{5}$, $10^{6}$ and $10^{7} M_{\odot}$ kpc$^{-2}$ for reference.

The galaxies cover a wide range of physical sizes \citep[driven mostly by the size-mass relation, e.g.][although some variation is present even at $\sim$ fixed stellar mass]{genel2018} and bulge-to-disk ratios. As noted previously, we did not enforce any particular orientation when producing the mass maps, so the galaxies appear at all inclination angles. 

The morphology and structure in the lower density stellar halos display a remarkable diversity --- depending on the galaxy, we can identify stellar streams and shells, as well as a smoother component. The amount of visible substructure scales with stellar mass, consistent with what we would expect from merger rates. Interestingly, however, it is clear that \textit{all} of the galaxies in these figures have stellar mass out in the farthest reaches of the fields of view plotted here (i.e., galacto-centric distances of 10 half-mass radii), indicating that the simulations produce spatially extended stellar halos which enable meaningful comparisons with observations. Further, in the majority of cases, surface densities remain above our canonical ``floor'' of $10^{4}M_{\odot}$ kpc$^{-2}$ even out at $10R_{\rm half}$.

We note that in the outermost regions, the stellar particle nature of the data is still discernible. For the purposes of this work, however, this degree of smoothing is sufficient to make stable measurements (see discussion in Section \ref{sec:measured:profiles:mass} and Appendix \ref{app:images}).

\section{Stellar halos in profile}\label{sec:measured:profiles}
A key strategy for quantifying stellar halos observationally --- whether via star counts or integrated light --- is to measure surface brightness profiles. Major or minor axis profiles (or, more generally, wedge profiles) allow us to zoom in on specific regions of the disks or halos of galaxies and quantify substructure along a particular direction, providing a means to place constraints on the number and timing of accretion events. Particularly in the regime of integrated light imaging, where the size of the field of view is generally not a limiting factor, azimuthally averaged radial profiles (which trace the average value within an elliptical annulus as a function of its semi-major axis) have proven useful for probing the outermost regions of stellar halos, thanks to their (typically) lower surface brightness reaches. Despite the unavoidable loss of some shape and structure information, detailed visual comparisons between azimuthally averaged profiles and galaxy and stellar halo morphology have confirmed consistent features \citep[e.g.][]{merritt2016a}.

\begin{figure*}
    \centering
    \includegraphics[width=\linewidth]{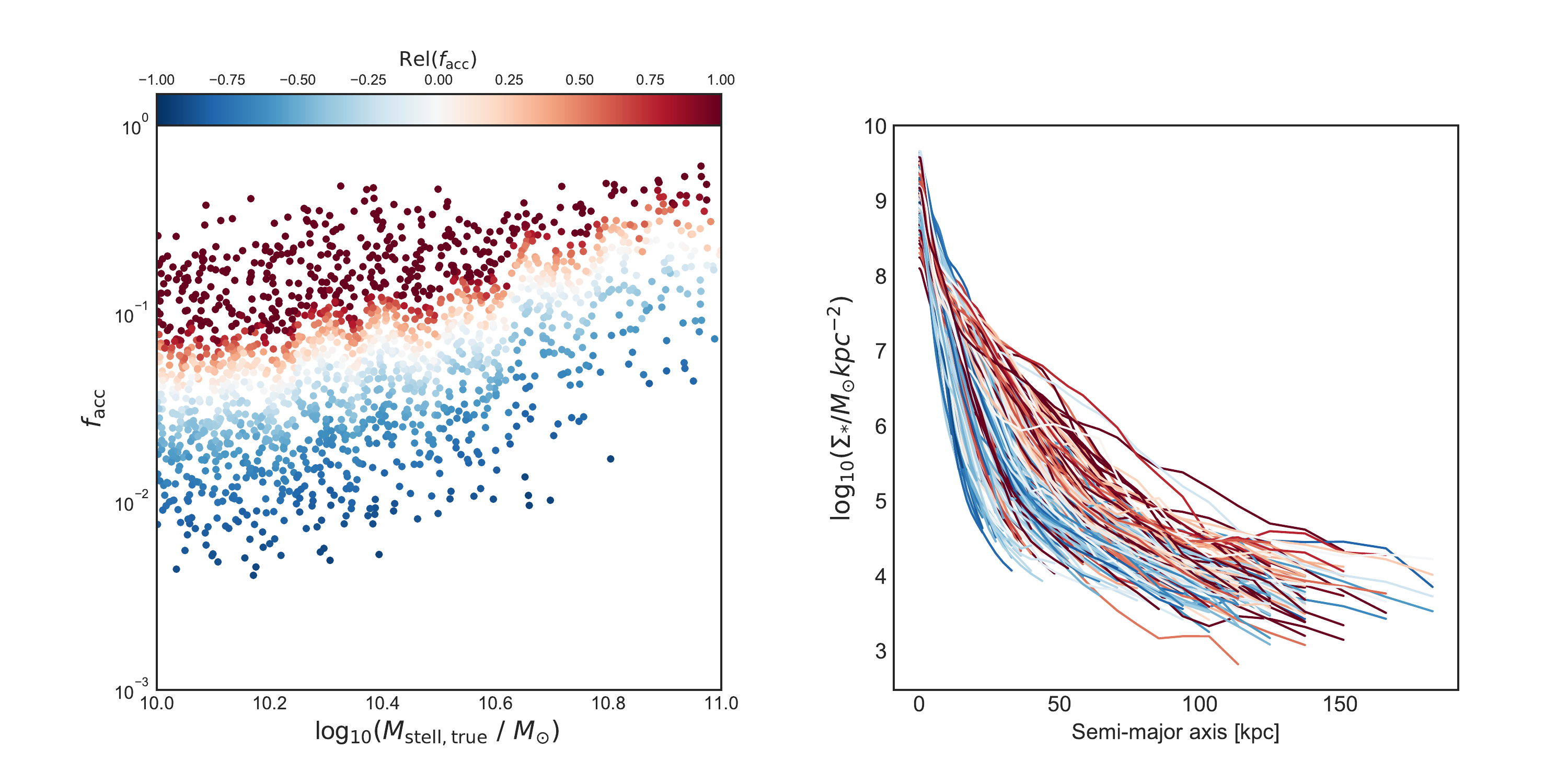}
    \caption{\textbf{Profile shape vs. accreted stellar mass fractions in TNG100. Left Panel:} The distribution of TNG100 galaxies in accretion fraction
    ($f_{\rm acc}$) and stellar mass. The color coding illustrates the relative distance of
    each galaxy in \facc from the median \facc at fixed $M_{\rm stell,true}$,
    as defined by equation \ref{eqn:relX}; that is, red/blue points denote galaxies that have a high/low \facc for their stellar mass.
    \textbf{Right Panel:} Azimuthally averaged stellar mass surface density profiles for each TNG100 galaxy in our parent sample, color-coded in the same way. A correlation between the (relative) accretion fraction and profile shapes of the TNG100 galaxies is visible, in spite of the random orientations of our TNG100 galaxies.}
    \label{fig:profiles:facc:ms}
\end{figure*}

\begin{figure*}
    \centering
    \includegraphics[width=\linewidth]{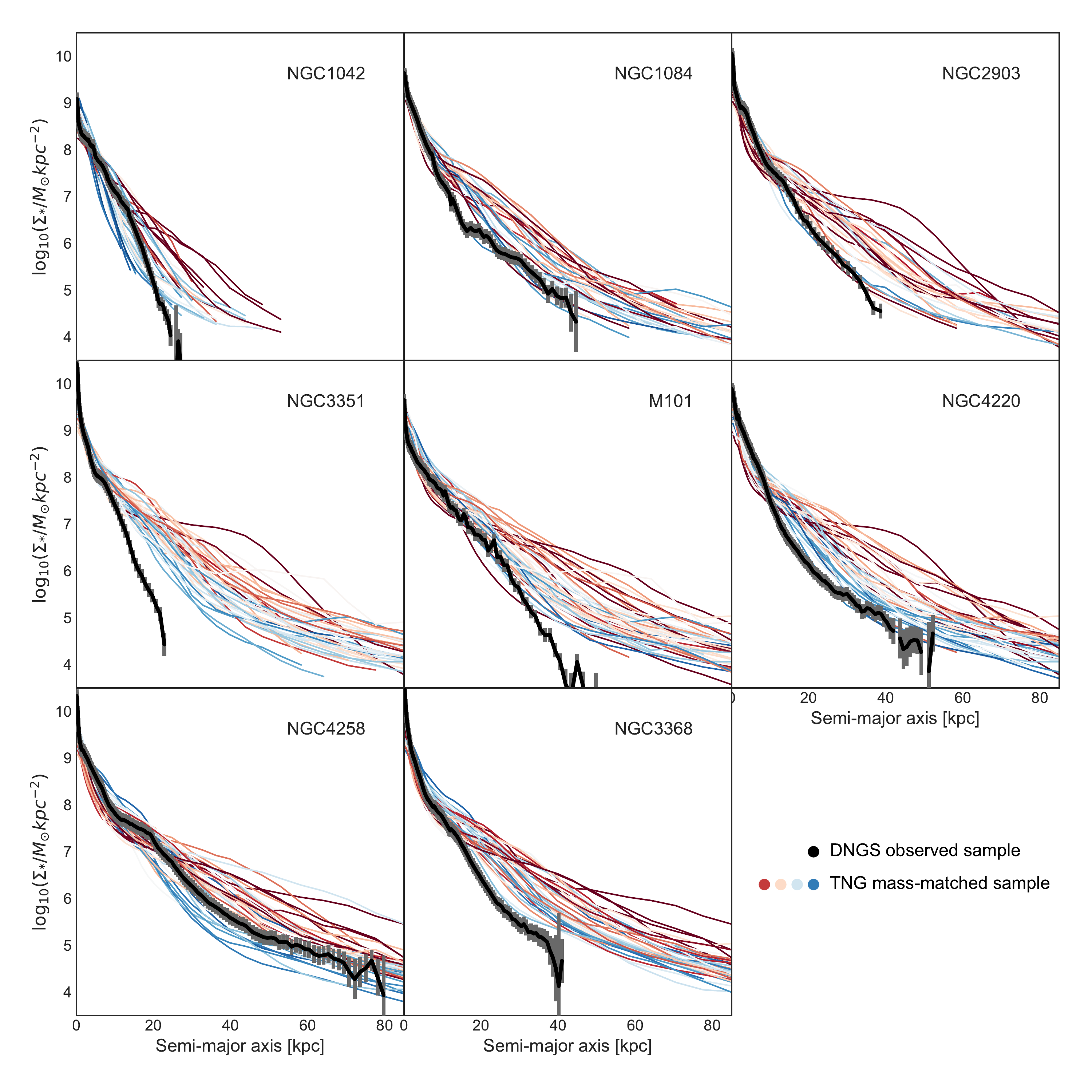}
    \caption{\textbf{DNGS vs TNG100 stellar mass surface density profiles.} Each panel here shows a comparison between the surface density profile of a DNGS galaxy and the surface density profiles of stellar mass-matched TNG100 galaxies (see Section \ref{sec:data:simulations} for details of the mass-matching process). We only show the profiles out as far as the distance to the $50^{\rm th}$ outermost stellar particle to give a rough visual sense of the robustness of the surface density measurements. The color coding is the same as in Figure \ref{fig:profiles:facc:ms}; it describes the relative distance of each TNG100 galaxy in \facc from the median \facc at fixed $M_{\rm stell,true}$. The observed DNGS galaxies are consistent with the profiles of TNG100 galaxies with unusually low \facc, and in many cases are steeper than \textit{all} TNG100 galaxy profiles at its particular stellar mass. Galaxies are ordered by increasing stellar mass.}
    \label{fig:profiles:facc:ms:dngsmatched}
\end{figure*}

\begin{figure*}
    \centering
    \includegraphics[width=\linewidth]{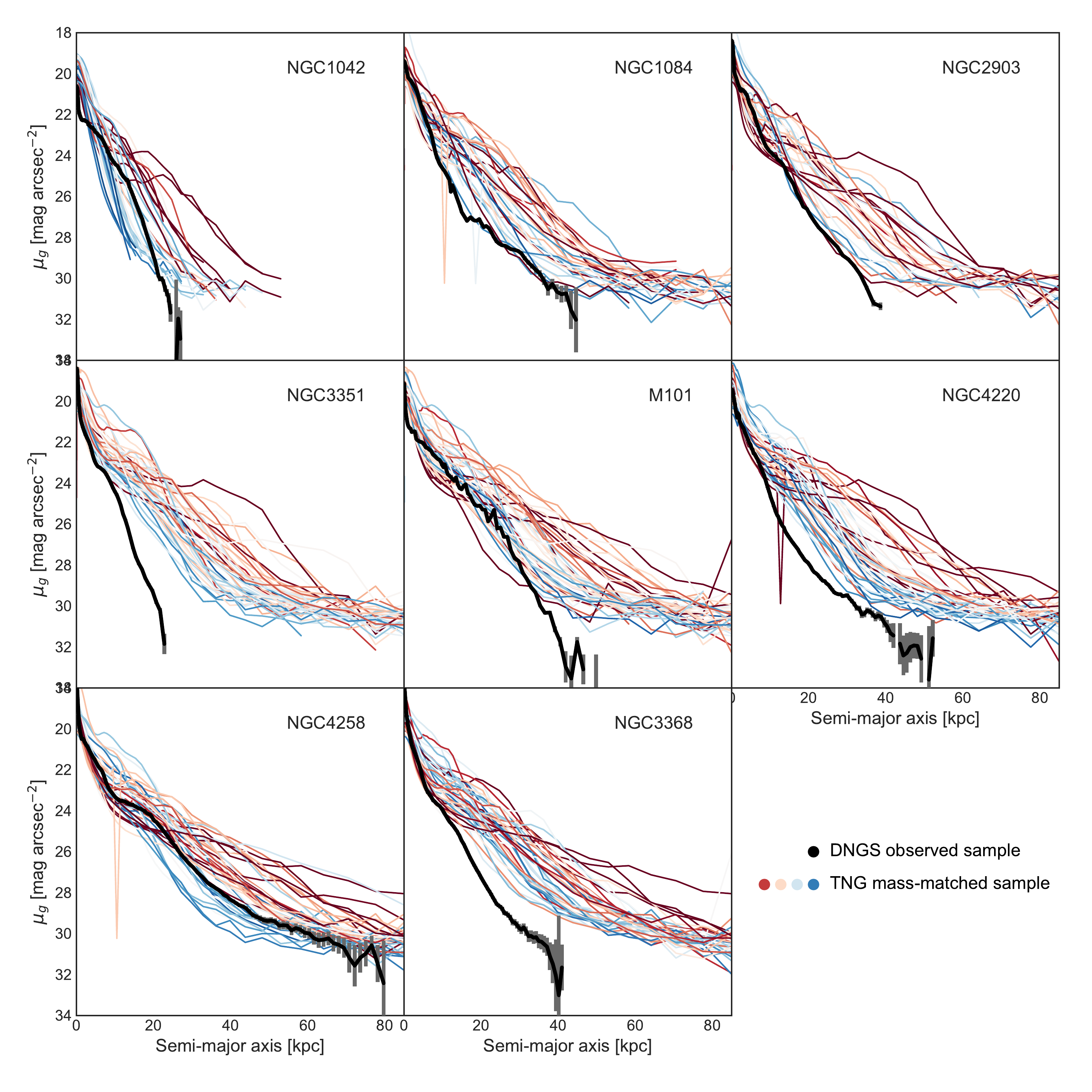}
    \caption{\textbf{DNGS vs TNG100 $g$-band surface brightness profiles.} Same as Figure \ref{fig:profiles:facc:ms:dngsmatched}, but here we show the $g$-band surface brightness profiles. Although the mock-observed surface brightness profiles are noisier than the mass-matched TNG100 stellar mass surface density profiles, we can see that the general trends remain the same. That is, the observed DNGS galaxies are generally only consistent with the profiles of TNG100 galaxies with the lowest outer profile densities and lowest \facc for their stellar mass.}
    \label{fig:profiles:facc:sbg:dngsmatched}
\end{figure*}

\subsection{Stellar mass surface density}\label{sec:measured:profiles:mass}
To measure surface density profiles, we ran the IRAF task \texttt{ellipse} \citep{jedrzejewski1987,busko1996} on each of our TNG100 images. Following the procedure in \cite{merritt2016a}, we fit elliptical isophotes, allowing both the position angle and ellipticity to vary as a function of radius (defined in the context of density profiles to be the semi-major axis). Due to the small size of the DNGS sample, in \cite{merritt2016a} we were able to ensure the quality of the fit by eye and make adjustments where necessary (for example, making sure that the presence of spiral arms or other morphological structures did not adversely affect the fits). However, the TNG100 sample is significantly larger, and we therefore used an automated metric to assess how accurately our \texttt{ellipse} results represent the input galaxy images. We used the IRAF task \texttt{bmodel} to generate a 2D model of the galaxy using the isophotal data, and calculated the relative stellar mass error as $(M_{\rm image} - M_{\rm model}) / M_{\rm image}$. We calculated this value considering (a) all pixels in the image and (b) only pixels further than 20 kpc from the center of the image. Galaxies with total relative stellar mass error $\geq 0.01$ or with outer relative stellar mass error $\geq 0.05$ were subsequently excluded from further analysis (specifically, they are not part of any of the mass-matched samples mentioned in Section \ref{sec:data:simulations}; this amounted to $\sim 20$ percent of the parent sample). Physically, failed \texttt{ellipse} fits generally correspond to galaxies that are currently undergoing a disruptive major merger (as the algorithm steps out radially from the galaxy center, it requires a steadily declining surface density distribution in order to converge on a stable solution). We note that, as a result, we may be systematically ignoring any (very) recent or ongoing major mergers within the TNG100 sample.

Before we can make meaningful comparisons with observations, we need to quantify the strong trends between outer galaxy profile shape, stellar mass and \facc \citep[see e.g.][]{cooper2013,dsouza2014,pillepich2018b}. In the left panel of Figure \ref{fig:profiles:facc:ms}, we introduce a useful metric for understanding profile shape: the \textit{relative} accretion fraction at fixed stellar mass. In general, we define relative values as:
\begin{equation}\label{eqn:relX}
{\rm Rel}(X) \equiv \frac{X - {\rm median}(X | Y)}{{\rm median}(X | Y)}
\end{equation}
where in this case, $X$ and $Y$ are \facc and $M_{\rm stell,true}$, respectively. Here, since we are exploring the relation between stellar mass and an unobservable property (the accreted stellar mass fraction), we used the true, simulated stellar masses of the TNG100 galaxies. We measured the accreted stellar mass fractions \facc as the ratio of the total ex situ stellar mass to the total stellar mass. For each galaxy, we calculated the rolling median value of \facc over the subset of galaxies with stellar masses within 5\% of its $M_{\rm stell,true}$. 

The left panel of Figure \ref{fig:profiles:facc:ms} shows the distribution of \facc and stellar mass for the TNG100 galaxies color-coded by Rel(\facc), while the right panel applies the same color scheme to the azimuthally averaged TNG100 surface density profiles. Consistent with what we saw in the images in Figures \ref{fig:library1} and \ref{fig:library2}, the TNG100 surface density profiles reach surface densities of $10^{4} M_{\odot}$ kpc$^{-2}$ at  galacto-centric distances of 50-150 kpc. As a crude indicator of uncertainties in the outer profiles (which are likely underestimated by \texttt{ellipse}, see Section \ref{sec:metrics}), we only show the profiles out as far as the galacto-centric distance to the $50^{\rm th}$ outermost stellar particle. For low mass galaxies, this distance is typically smaller than both the size of the image and the radius at which the profile drops below $10^{4}M_{\odot} {\rm kpc}^{-2}$.

The outer profiles in particular show a large spread in (log) surface density at fixed radius, with $1\sigma$ scatter of $0.42$ dex at 10 kpc and $0.79$ dex at 50 kpc. TNG100 galaxies with high accreted mass fractions (or a greater number of significant mergers) have, on average, shallower density profiles with higher normalization in the outskirts relative to galaxies with low accreted mass fractions. If we limit the sample to galaxies with \facc $\leq 0.1$ ($\geq 0.4$), the scatter at 50 kpc reduces to $0.62$ dex ($0.23$ dex), suggesting that the outer profiles are dominated by growth via accretion rather than in-situ star formation. This is consistent with results from several previous studies \citep[e.g.][]{deason2013,pillepich2014,monachesi2019}, and something that we demonstrate and explore more explicitly in later sections. We can also infer from these numbers that the scatter in the profiles is dominated by galaxies with low accretion fractions. 

Moreover, the color scheme applied in Figure \ref{fig:profiles:facc:ms} demonstrates that the shape of the profiles also correlates with the \textit{relative} accretion fraction Rel(\facc). The presence of this trend in the azimuthally averaged profiles is encouraging, considering that our simulated galaxy images are randomly oriented and it has been demonstrated that a galaxy's inclination angle can affect its overall surface density profile shape as well estimations of the contribution of its stellar halo  \citep[most recently,][]{elias2018}.

We also considered the possibility that some of the spread might be due to systematic differences between the central and satellite galaxy populations within our parent sample. We were unable to discern any environmental effects in the surface density profiles; however, we caution that this finding should be considered preliminary as our sample contains far more central galaxies than satellites, and our TNG selection criteria included an isolation requirement which likely minimizes any environmentally-driven differences between the two populations.

\subsection{Comparison between DNGS and TNG100}\label{sec:measured:profiles:compare}
In Figure \ref{fig:profiles:facc:ms:dngsmatched} we show the stellar mass surface density profiles of the mass-matched TNG100 galaxies, with one panel for each of the DNGS galaxies. The color scheme for the simulated galaxies is taken directly from Figure \ref{fig:profiles:facc:ms} (that is, red/blue lines indicate galaxies with high/low \facc for their stellar mass). We can now see that even at fixed stellar mass, galaxies with higher accreted mass fractions generally have more mass in the outer regions of their profiles.

The observed stellar mass surface density profiles from \cite{merritt2016a} are overplotted in black. Strikingly, the DNGS galaxies appear to be most closely matched by simulated galaxies with lower than average \facc for their stellar mass, with the  exception of NGC 1042 and NGC 4258, which follow the average \facc for their mass-matched samples. Galaxies such as NGC 1084 and NGC 2903 are consistent with one or more simulated galaxies, with outer profiles falling at or below the median TNG100 surface density at all radii, and in some extreme cases (NGC3351, NGC3368, NGC4220) the surface density beyond $20$ kpc is lower than all 50 mass-matched TNG100 galaxies. 

We stress that these differences between the DNGS profiles and their mass-matched samples do \textit{not} necessarily mean that TNG100 lacks galaxies with surface density profiles that resemble the DNGS galaxies. We selected TNG100 analogs based on matches in stellar mass ($M_{\rm stell,obs}$), but we could have matched directly on surface density profile shape instead, and might have found analogs with (e.g.) different stellar masses (see Section \ref{sec:data:simulations} for details and Section \ref{sec:discussion:matching} for further discussion).

In Figure \ref{fig:profiles:facc:sbg:dngsmatched}, we show the same comparison between the two datasets, but now with the $g$-band surface brightness rather than stellar mass, with the TNG100 profiles measured from mock light images produced with \codenamesecondns. \footnote{We used the same \texttt{ellipse} apertures as in the stellar mass profile measurements.} As expected, the profiles are noisier than the stellar density profiles (as described in Section \ref{smoothing:observational}, we convolved our light images with the Dragonfly PSF and included a realistic sky background). However, the same trend seen in Figure \ref{fig:profiles:facc:ms:dngsmatched} remains, demonstrating that any profile differences are not caused by the conversion of light to stellar mass used by \cite{merritt2016a}. 

These apparent structural differences between the simulated and observed profiles at fixed $M_{\rm stell,obs}$, as well as the finding that DNGS profiles are best described by TNG100 galaxies with relatively low \facc for their stellar mass, constitute the central result of this paper and will be quantified and explored in the subsequent sections. This result is consistent with the original analysis in \cite{merritt2016a}, where the data were compared to older models by \cite{cooper2010}, \cite{pillepich2015}, and \cite{cooper2013}. 

Recently, \cite{elias2018} found a similar result when comparing DNGS galaxies to the (original) Illustris simulation, although we note that their sample was defined based on total halo mass $M_{200}$ rather than stellar mass and included the full range of possible morphologies within that mass bin. Using the 5\% lowest accretion fraction, disk-dominated subset of their sample, they showed that the observed inclination angle of a galaxy has a negligible effect on the \textit{slopes} of measured surface brightness profiles, but can influence the absolute values at the level of approximately $1.5$ (2) mag arcsec$^{-2}$ beyond $\sim 25$ (70) kpc. This is particularly relevant for our study, as some DNGS galaxies (NGC1042 and M101 in particular) have very low inclination angles. From Figure \ref{fig:profiles:facc:sbg:dngsmatched}, however, we can infer that a drop in the surface brightness profiles by this amount (effectively converting edge-on to face-on profiles) would not change the result that DNGS galaxies seem to align best with low \facc TNG100 galaxies.\footnote{This would actually be an overly conservative correction, as some of our randomly-oriented TNG100 galaxies are already face-on.} We discuss the effects of inclination angle more comprehensively in Section \ref{sec:missing}.

\section{Quantifying stellar halos}\label{sec:metrics}

\begin{figure*}
    \centering
    \includegraphics[width=\linewidth]{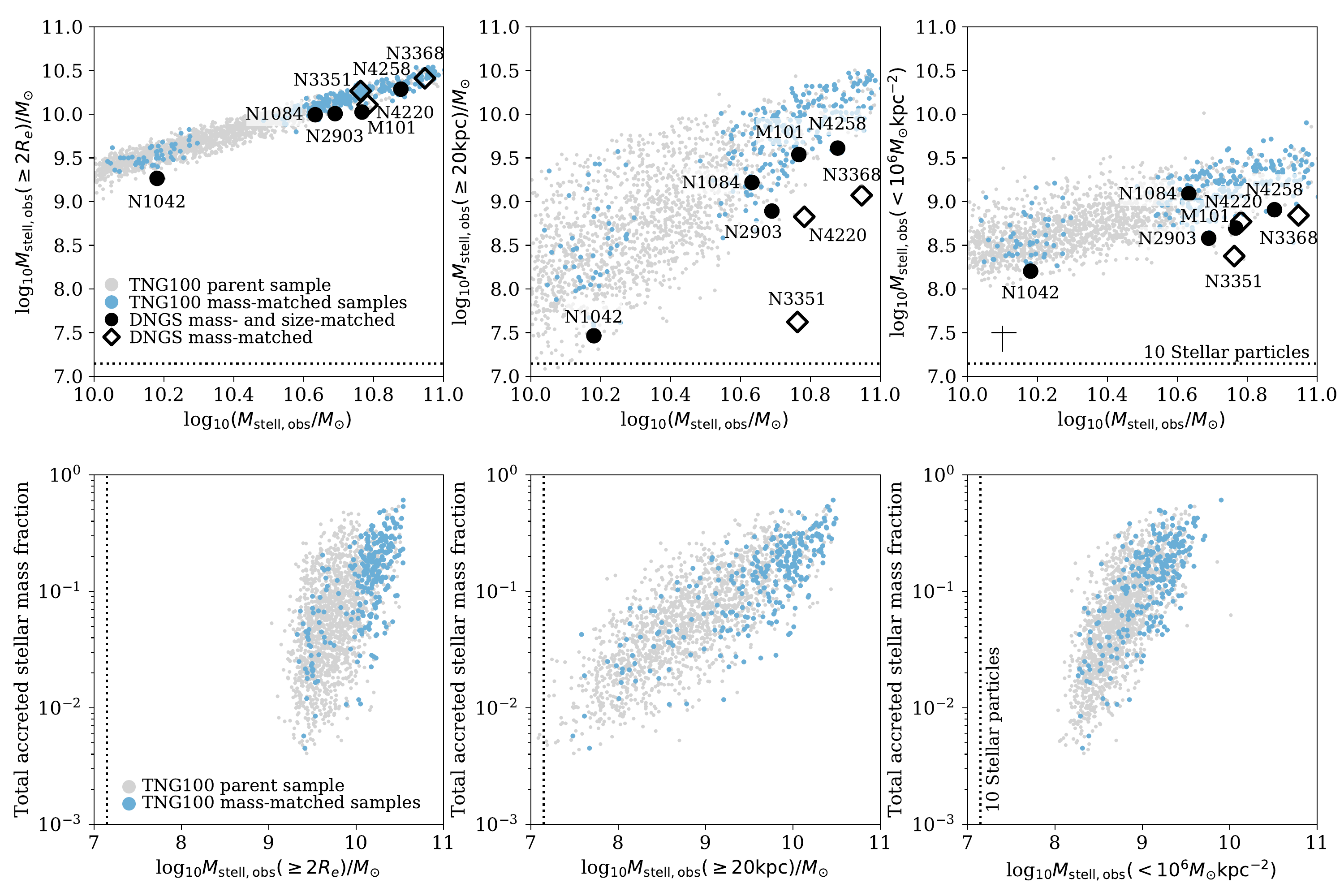}
    \caption{\textbf{Different estimates of stellar halo mass. Left panel:} We compare the amount of stellar mass beyond 2 half-mass radii as a function of galaxy stellar mass for TNG100 galaxies (blue points) and DNGS galaxies (black points). The regularity of the simulated galaxies is remarkable. For the observations, closed circle symbols indicate DNGS galaxies that we were able to match to TNG100 in both stellar mass \textit{and} size, while open diamond symbols point to DNGS galaxies that could not be matched in size (specifically, the TNG100 galaxies at fixed mass were larger). \textbf{Middle panel:} The same, except we now consider the stellar mass beyond a fixed radius of 20 kpc. \textbf{Right panel:} Again, the same idea as the left panel; but here we adopt a stellar surface density threshold ($\leq 10^{6}M_{\odot}$ kpc$^{-2}$) as the working definition for the stellar halo.  For additional context, we indicate the mass scales associated with 10 stellar particles with dotted lines. In all cases the stellar mass is integrated along the surface density profiles down to a threshold of $10^{4}M_{\odot}$ kpc$^{-2}$, and in the righthand panel, we indicate the typical error in the ``observed'' galaxy and stellar halo mass measurements.}
    \label{fig:masses:compare}
\end{figure*}

In the previous section we examined the surface brightness and stellar mass  surface density profiles of the DNGS and TNG100 galaxies and found that at fixed stellar mass ($M_{\rm stell,obs}$), shallower (steeper) TNG100 profiles correspond to higher (lower) relative accretion fractions. We also saw that the observed DNGS galaxies are consistent with TNG100 galaxies with relatively \textit{low} accretion fractions at fixed stellar mass. In this section we will make an effort to quantify this statement more robustly, and to outline a method to map as directly as possible from observable properties of the outskirts of galaxies to a (total) accreted mass fraction.

Moving to more quantitative metrics serves as a key step towards understanding a galaxy's assembly history and maximizing the information content from both observations and simulations; unfortunately, however, there is no easy answer to the question ``what is the best metric to use when quantifying stellar halos?'' The challenge of defining a stellar halo and attaining unbiased estimates of the accreted mass in a galaxy is outlined in detail by \cite{sanderson2017}, who used the FIRE-2 cosmological zoom-in suite of simulations to demonstrate that (single-parameter) metrics commonly used by observers/theorists tend to underestimate/overestimate the true accreted stellar mass. Even comparisons between different sets of observational data are not straightforward, although conversions between chosen metrics are sometimes possible \citep[for example, see][for a comparison between DNGS and GHOSTS galaxies]{harmsen2017}.

In an effort to match the DNGS stellar halo masses as reported by \cite{merritt2016a} as closely as possible, we first attempted to measure the excess stellar mass beyond 5 half-mass radii relative to a pure disk$+$bulge model for our TNG100 galaxies. However, in practice this is a significantly more complicated endeavor for the simulations than the observations. First, a key aspect of the method described in \cite{merritt2016a} is that the disk$+$bulge model was only fit within the inner regions of the galaxy to ensure that the disk and bulge were the dominant components and to prevent any existing stellar halo from affecting the results. In that study, the working definition of ``inner region'' was the maximum radial extent out to which spiral arms could be identified \citep[see also][]{sanderson2017}. In our TNG100 sample, however, not all galaxies have discernible spiral arms, which renders this approach ineffective (particularly at the low mass end). Given the magnitude of the effect that changing the fitting region of the profile has on the outcome (driven by any substructure signatures), we did not feel that this was an appropriate or fair comparison to make. Second, out at galacto-centric distances of $5$ half-mass radii we begin to run into resolution effects in TNG100. Specifically, differences in profile shapes due to differences in assembly history are reduced relative to measurements made further in.

\subsection{Stellar halo masses}
\label{sec:metrics:masses}
Emphasizing the difficulty in choosing a single parameter definition for the ``stellar halo'' of a galaxy, the top row of Figure \ref{fig:masses:compare} shows the fraction of stellar mass beyond 2 half-mass radii (left panel), beyond 20 kpc (middle panel), and below $10^{6} M_{\odot}$ kpc$^{-2}$ (right panel; following \citealt{cooper2013} and \citealt{sanderson2017}) as a function of ``observed'' galaxy stellar mass for the TNG100 parent sample (grey points) and mass-matched samples (blue points). In the righthand panel, we illustrate the typical error incurred by measuring the stellar masses of the galaxies and their stellar halos by integrating the \texttt{ellipse} surface density profiles (the error bars span the $5^{\rm th}-95^{\rm th}$ percentiles). DNGS measurements are overlaid in black points, and we distinguish between galaxies that have similar sizes (2D half-mass radii, as measured along the semi-major axis of the surface density profiles) compared to their mass-matched TNG100 sample and those that do not with filled and empty symbols, respectively. 

We can see that each panel seems to tell a slightly different story. DNGS and TNG100 galaxies have a fairly comparable amount of mass beyond 2 half-mass radii (NGC 4258 even has a \textit{higher} amount of mass outside this point relative to its mass-matched sample). However, several DNGS galaxies contain less stellar mass outside of 20 kpc relative to TNG100 galaxies, and the situation is similar for a stellar density threshold of $10^{6} M_{\odot}$ kpc$^{-2}$. Each panel also comes with its own set of caveats. Plotting the fraction of mass outside of 2 half-mass radii against total stellar mass essentially shows that TNG100 galaxy outskirts are relatively uniform --- we can imagine this relation as a single curve with no scatter if every galaxy had the same S\'ersic index and followed the mass-size relation. On the other hand, definitions such as ``beyond $20$ kpc'' or ``below $10^{6} M_{\odot}$ kpc$^{-2}$'' require choices about exactly which values to use (e.g., why use 20 kpc and not 30 kpc?).  

In the bottom row of Figure \ref{fig:masses:compare}, we demonstrate the effect that the choice of stellar halo metric has on estimations of the total accreted mass fraction, using the three definitions shown in the top row. These results cast some doubt on the idea that any estimate of stellar halo mass can be a successful proxy for the total accreted stellar mass of a galaxy. We confirm that in general, TNG100 galaxies with more mass in the outskirts (or at lower surface densities) have had more active merger histories overall; however, there is a significant amount of scatter here. The scatter in total \facc at fixed ``stellar halo'' mass is driven to varying extents by the total stellar masses of galaxies and by the 
spatial distribution of accreted mass, as the exact fraction of accreted stellar particles that wind up within or beyond a given threshold in a galaxy will depend on the details of its assembly history. Crucially, the utility of each of these three  definitions relies on the assumption that the stellar mass beyond the appropriate threshold is dominated by accreted material rather than an in-situ disk, and that the accreted stellar mass interior to a given threshold is proportional to the accreted stellar mass beyond it. Neither of these will be completely true in detail, and are issues that are only amplified for cases where the sizes of TNG100 galaxies are systematically larger (by a factor of $2-4$) than those of DNGS galaxies (primarily affecting more massive galaxies; see open symbols in Figure \ref{fig:masses:compare}).

We note that \cite{dsouza2014} measured the fraction of light in the stellar halo (parametrized as by the light in an outer S\'ersic component) as a function of galaxy stellar mass for stacks of low-concentration galaxies in the SDSS, and quoted values ranging from $\sim 0.03-0.2$ over the stellar mass range $10^{10}-10^{11}M_{\odot}$. Although a direct comparison to our results is beyond the scope of this work, we can approximately translate their numbers to our own framework by assuming a stellar mass-to-light ratio of 1.0 and re-scaling by the stellar masses of galaxies. This yields a range in stellar halo masses of $8.5 \lesssim {\rm log}_{10}M_{\rm stell,D'Souza}/M_{\odot} \lesssim 10.3$, which is most consistent with our stellar halo masses measured beyond 20 kpc as presented in Figure \ref{fig:masses:compare}.

Figure \ref{fig:masses:compare} also reveals that the tightest correlation between stellar halo mass and total accreted mass fraction occurs when we measure the mass outside of 20 kpc. Motivated by this, as well as by the simplicity of the definition, we adopt the mass fraction outside of 20 kpc as our working definition for the stellar halo mass fraction. Similarly, in the following sections, the terms ``galaxy outskirts'' and ``stellar halo'' will refer explicitly to material beyond 20 kpc. This choice makes our results overall less susceptible to biases resulting from low numbers of stellar particles; the trade-off is that we expect more significant contribution from (in-situ) disk stellar particles at these distances.

\subsection{Stellar halo slopes}
\label{sec:metrics:slopes}

\begin{figure}
    \centering
    \includegraphics[width=\linewidth]{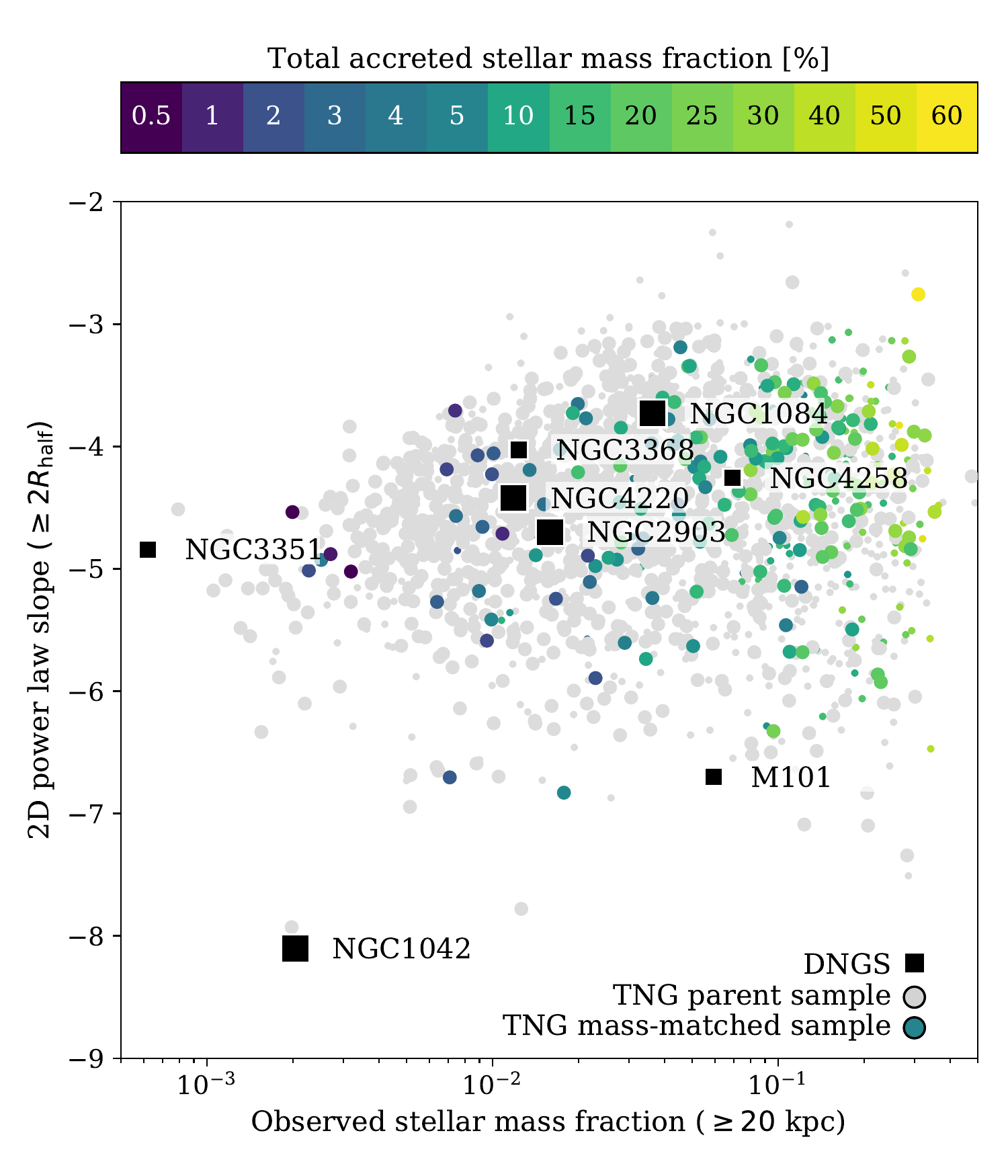}
    \caption{\textbf{A two parameter estimate of accreted mass fractions.} The distribution of stellar mass fractions and 2D power law slopes for the mass-matched sample of TNG100 galaxies, color-coded by the fraction of stellar mass beyond 2 half-mass radii. The full TNG100 parent sample is shown in grey points. Smaller symbols indicate galaxies for which a single power law was not an acceptable assumption, included here only for completeness (approximately $\sim 37$\% of the sample require a more complex fit; see Section \ref{sec:metrics:slopes} for details). In general, shallower slopes correspond to higher accreted fractions, albeit with significant scatter. At fixed power law slope, galaxies with higher stellar mass fractions in the outskirts have higher total accreted fractions.}
    \label{fig:massfrac:plaw:facc}
\end{figure}

As a second measure, we also explored the possibility of a two-parameter metric for the outskirts of galaxies, as this more closely approximates the information in the full density or surface brightness profiles. The outer mass fractions serve as the ``normalization'' of the profile, and we chose to measure the power law slope of the profiles in the outskirts as well. 

We made the simplifying assumption that the outer density profiles can be characterized by the functional form $y = Ax^{\alpha}$, and fit for the 2D power law slope ($\alpha$) between 2 half-mass radii and the point at which the surface density dropped below $10^{4}M_{\odot}$ kpc$^{-2}$ (see Section \ref{sec:measured:profiles:mass}). 
We determined the maximum likelihood and associated uncertainties for the power law slopes using \texttt{emcee} \citep{foremanmackey2013}, an implementation of the affine-invariant Monte Carlo Markov Chain (MCMC) ensemble sampler \citep{goodman2010}. Rather than use the directly measured errors from \texttt{ellipse}, we modeled the ``true'' errors as $s_{i}^{2} = \sigma_{i}^{2} + f^{2}m_{i}^{2}$, where $\sigma_{i}$ and $m_{i}$ are the measured error and model value at a given position --- in other words, assuming that the true errors have been systematically underestimated by some fraction $f$. The motivation behind this was to account for shot noise present in the low surface density measurements at large radii; however, it also proved to be a useful diagnostic for identifying poor power law fits (i.e., if $f \sim 1$).

Figure \ref{fig:massfrac:plaw:facc} compares these two independently-measured metrics for the stellar halo: the stellar mass fraction outside of 20kpc and the 2D slope of the outer surface density profiles measured beyond 2 half-mass radii.\footnote{Although the exact parametrization differs, both this work and \cite{merritt2016a} treat the \textit{shape} of the density profile as the fundamental source of information on galaxy assembly histories.} Mass-matched TNG100 galaxy samples are shown in circular points, color-coded by their total accreted stellar mass fractions. For reference, we include the full TNG100 parent sample in the background in grey points. DNGS galaxies are overlaid in black squares; for both observations and simulations, smaller symbols represent galaxies with large values of $f$. Over the stellar mass range $10^{10}-10^{11}M_{\odot}$, galaxies in the TNG100 parent sample (grey points) exhibit 2D slopes from -3 to nearly -7, with a median of $\sim -4.5$. Broadly speaking, we can see that at fixed power law slope, galaxies with higher outer stellar mass fractions have higher total \facc, consistent with what we saw more qualitatively from the surface density profiles.  Approximately $37$\% of our galaxies were not well fit by our simple assumption of a single power law; a broken power law may be
more appropriate in these cases.

Figures \ref{fig:masses:compare} and \ref{fig:massfrac:plaw:facc} also raise the question of \textit{why} these observations seem to be ``missing'' stellar mass beyond $20$ kpc (or, phrased differently, why TNG100 galaxies appear to be more massive and extended than DNGS galaxies at fixed stellar mass). No galaxy in DNGS has a stellar halo mass fraction above 10 percent (when measured outside of 20 kpc), whereas the TNG100 galaxies in Figure \ref{fig:massfrac:plaw:facc} can have up to 40 percent of their stellar mass beyond this radius. 

One possible explanation is that DNGS galaxies may have low total accreted mass fractions relative to their mass-matched counterparts in TNG100. It is worth pausing to point out, however, that our capacity to ask (and answer) this question depends heavily on the validity of three assumptions:
\begin{enumerate}
    \item The parameters we chose to observationally estimate the accreted mass fraction (stellar halo mass fractions and power law slopes) are able to provide an appropriate description of the outer profiles of galaxies -- and, further, that these profiles contain sufficient information to characterize the assembly histories of galaxies.
    \item TNG100 reproduces observed properties of galaxies accurately enough that we should expect agreement in stellar halo properties.
    \item DNGS represents an unbiased sample of Milky Way-mass spiral galaxies.
\end{enumerate}

We will delve more deeply into these assumptions in Section \ref{sec:missing}. First, however, we will proceed in Section \ref{sec:ptcls} as though each of these are true, and attempt to carve out a more complete physical understanding of galaxies with low accreted mass fractions.

\section{Insights from stellar particles}\label{sec:ptcls}
\begin{figure}
    \centering
    \includegraphics[width=\linewidth]{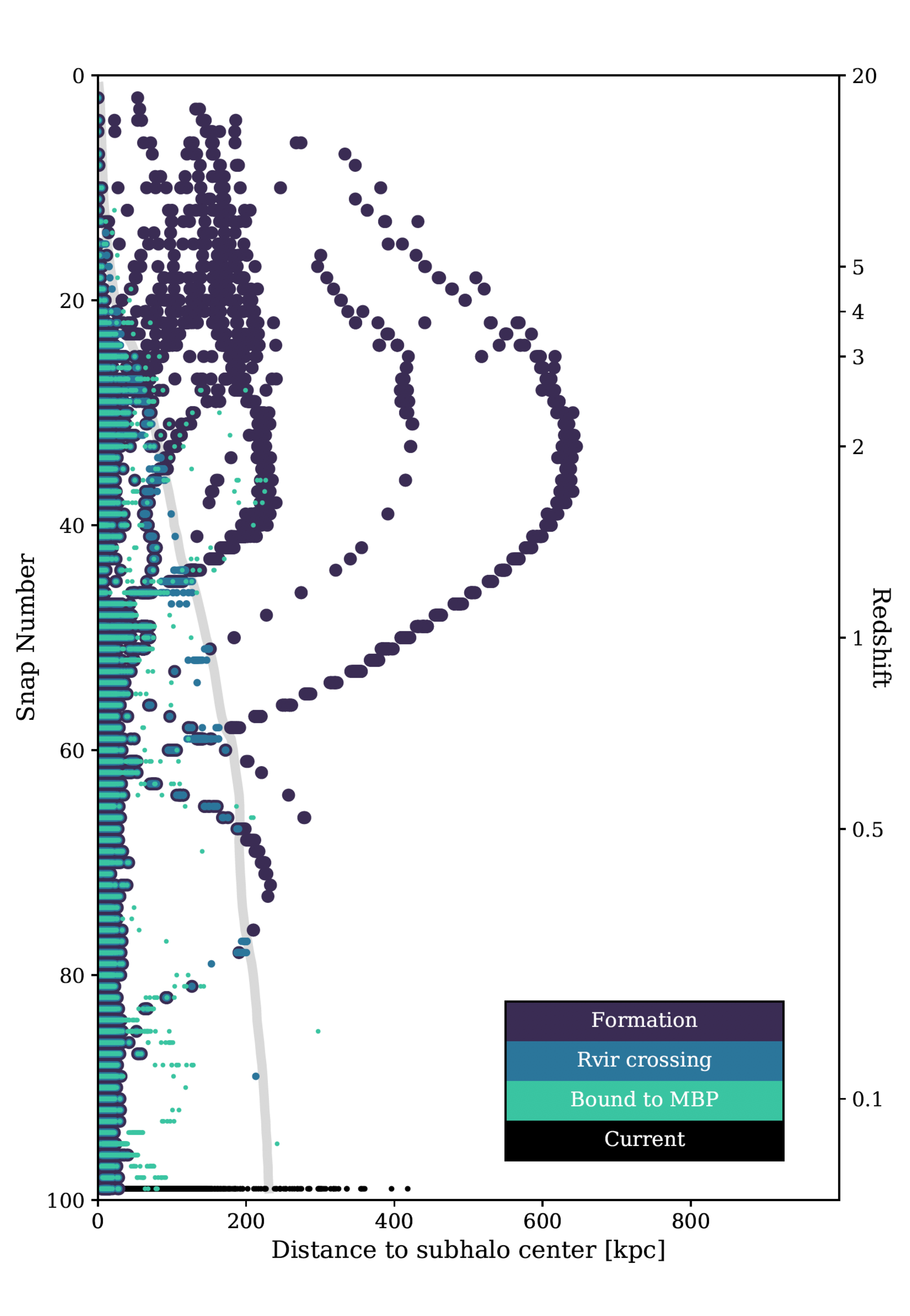}
    \caption{\textbf{Visualizing the particle tracker.} A representation of the (stellar) formation history for Subhalo 483900 of TNG100: different colored points indicate the Snapshot/redshift and galacto-centric distance when stellar particles formed (purple), crossed the main progenitor branch virial radius for the first time (blue), and were stripped from their progenitor to become gravitationally bound to the $z=0$ host (green). The grey line shows the host virial radius at each snapshot, and black points chart out the present-day positions of stellar particles. Note that stellar particles formed in situ can be identified as those having overlapping purple/blue/green points (i.e., stellar particles formed inside the virial radius and were immediately gravitationally bound to the main progenitor branch host).}
    \label{fig:distsnapnum}
\end{figure}

We have seen from Figures \ref{fig:profiles:facc:ms:dngsmatched}, \ref{fig:profiles:facc:sbg:dngsmatched} and \ref{fig:massfrac:plaw:facc} that DNGS galaxies appear to have low accretion fractions relative to their mass-matched TNG100 galaxies. The next logical question, then, is: ``What are the assembly histories of (TNG100) galaxies with low accretion fractions?'' If, for example, this subset of galaxies shares any additional aspects of their assembly histories, we can potentially gain physical insight into our observations. In this section we explore several different properties of our mass-matched TNG100 galaxies, combining information from the density profiles and individual stellar particles to characterize their past assembly as robustly as possible.

\subsection{Building up galaxy outskirts, stellar particle by stellar particle}\label{sec:ptlcs:methods}

Answering questions along the lines of \textit{``What is the typical stellar mass of galaxies that deposit stellar particles in the outskirts of galaxies?''} or \textit{``What is the contribution from stellar particles that formed from accreted material after crossing the host virial radius?''} requires a complete description of every stellar particle residing in the outskirts of galaxies at $z=0$.

We tracked stellar particles across the simulation outputs from Snapshot 0 ($z=20$) through Snapshot 99 ($z=0$) for every central galaxy in our parent sample. We identified and recorded the times/redshifts at which stellar particles formed ($z_{\rm form}$), crossed the virial radius of their $z=0$ host for the first time ($z_{\rm cross}$), and were removed from the galaxy they formed in to become gravitationally bound to their $z=0$ host ($z_{\rm strip}$; see also \citealt{pop2018} for a similar procedure). For each of these points in time, we also kept track of the (true) stellar masses of the galaxy that the stellar particle was gravitationally bound to; this allowed us to define a progenitor galaxy stellar mass ($M_{\rm stell,prog}$) for each individual stellar particle. For in-situ stellar particles, this quantity is always simply the stellar mass of the host galaxy at that point in time; for ex-situ stellar particles, it is the stellar mass of the galaxy the stellar particle was bound to just before being stripped and thus becoming bound to its final ($z=0$) host. We note that $M_{\rm stell,prog}$ is \textit{not} necessarily the maximum stellar mass reached by the progenitor galaxy, and furthermore, a set of stellar particles associated with the disruption of a single galaxy will exhibit a spread in $M_{\rm stell,prog}$ due to the finite timescale of the merger, since the progenitor galaxy experiences mass loss as it merges with the central host galaxy but may also continue to form stars for some time.

Figure \ref{fig:distsnapnum} provides a visualization of our stellar particle tracker --- different colored points indicate $z_{\rm form}$ (purple), $z_{\rm cross}$ (blue) and $z_{\rm strip}$ (green) along with the corresponding galacto-centric distances for stellar particles in example Subhalo 483900 (first featured in Figure \ref{fig:smoothingexample}). We indicate the virial radius of the subhalo at each snapshot with a grey line. In situ stellar particles can be readily identified as those with all three points overlapping (in other words, they formed inside the virial radius and were therefore immediately gravitationally bound to the main progenitor branch host). This particular galaxy also experienced a number of accretion events, taking place predominantly between $0.5 \leq z \leq 3$. 

\begin{figure}
    \centering
    \includegraphics[width=\linewidth]{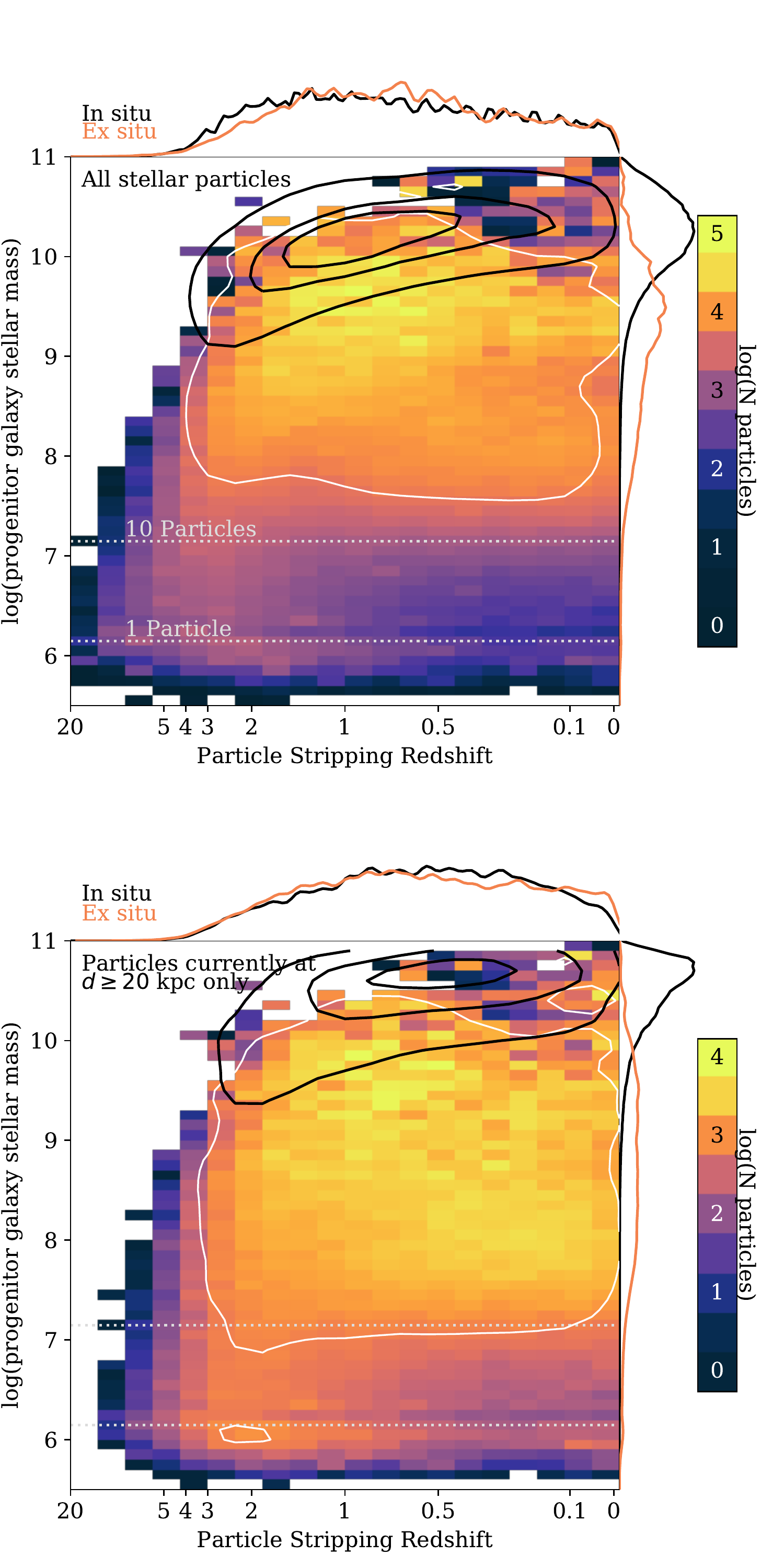}
    \caption{\textbf{The distributions of ex-situ particle progenitor stellar masses and stripping redshifts. Top panel:} The distribution of progenitor galaxy stellar masses and $z_{\rm strip}$ for every stellar particle in all 1656 central Milky Way disk galaxies in the TNG100 parent sample (that is, $10^{10} \leq M_{\rm stell,true}/M_{\odot} \leq 10^{11}$). Black contours enclose 25\%, 50\% and 90\% of the in situ stellar mass; these simply trace the build-up of the main progenitor host galaxy over time (in this case, $z_{\rm strip}$ is, more accurately, the formation redshift). Ex situ stellar particles are shown in 2D histograms with a logarithmic color scale; the white contours enclose 90\% of the ex situ stellar mass. To the right, the colorbar shows the number of particles in each bin. \textbf{Bottom panel:} The same, except only for stellar particles that are found beyond 20 kpc at $z=0$.}
    \label{fig:distall}
\end{figure}

\begin{figure*}
    \centering
    \includegraphics[width=\linewidth]{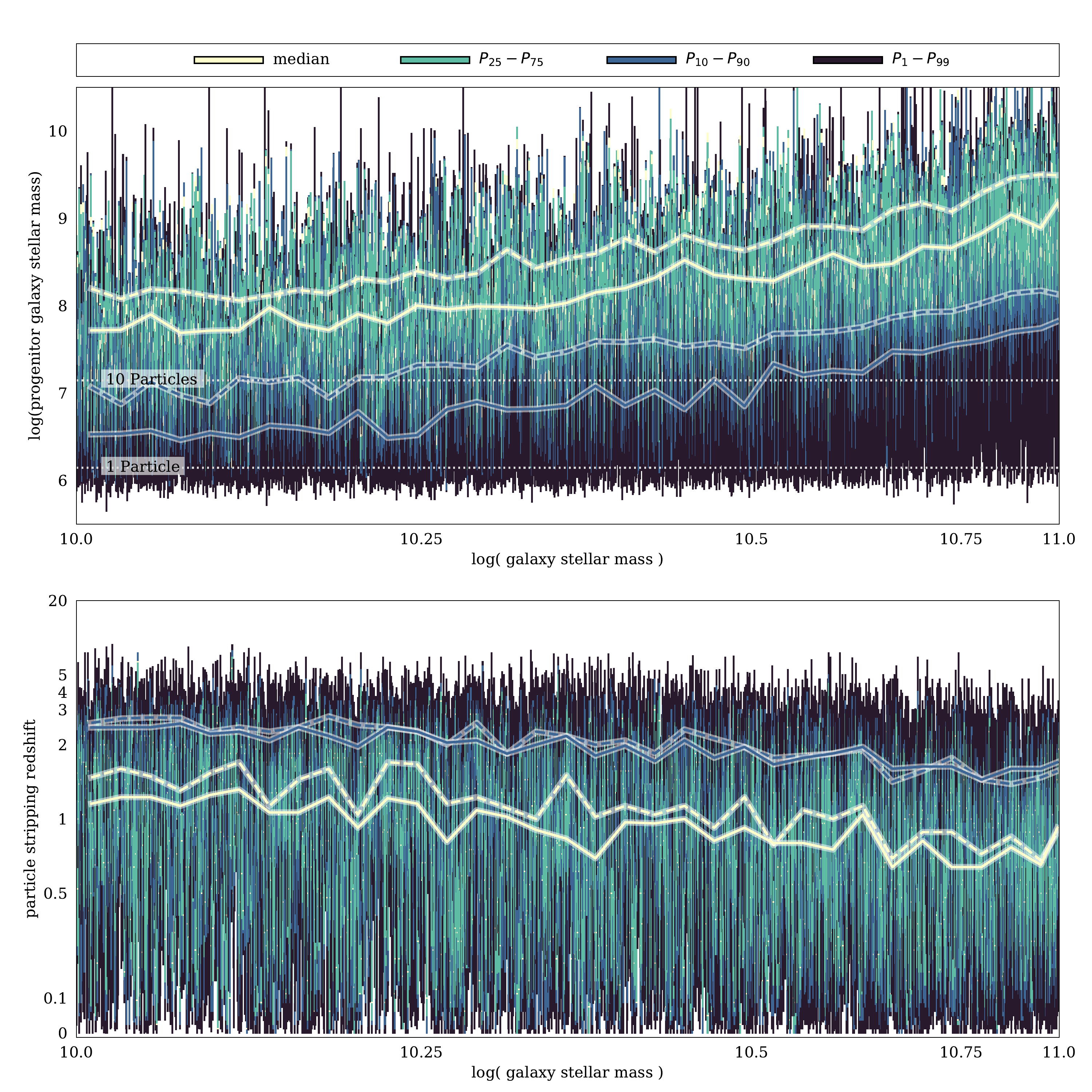}
    \caption{\textbf{Percentiles of ex-situ particle progenitor galaxy masses and stripping times. Top:} Highlighted yellow and blue lines in the foreground trace the running median and minimum progenitor galaxy stellar mass (such that 50\% and 90\% of ex situ stellar particles were contributed by galaxies at or above this threshold, respectively) for the \textit{outskirts} of our sample of central disk galaxies in TNG100. In the background, each column represents the distribution of progenitor galaxy stellar masses for ex situ stellar particles located beyond 20 kpc for a \textit{single} galaxy: we indicate the median values of $M_{\rm stell, prog}$ with yellow regions, as well as the $25^{\rm th}-75^{\rm th}$, $10^{\rm th}-90^{\rm th}$, and $1^{\rm st}-99^{\rm th}$ percentiles in green, blue, and black regions, respectively. As expected, the outskirts of more massive galaxies have a more significant contribution from massive satellites; however, the substantial galaxy-to-galaxy scatter is apparent. For reference, the dashed yellow and blue lines show the running median and minimum progenitor masses for \textit{all} ex-situ stellar particles (i.e., not limited to the outskirts). We note that the trend between galaxy stellar mass and median progenitor mass is similar between entire galaxies and galaxy outskirts, although the median progenitor masses are lower in the outskirts. \textbf{Bottom:} The same, except we now investigate the distributions of ex-situ stellar particle stripping redshifts for individual galaxies. We note a weak trend between galaxy mass and median $z_{\rm strip}$, such that more massive galaxies acquired their ex-situ stellar components slightly later than low mass galaxies. Interestingly, we can also see that there is very little difference between the median/minimum $z_{\rm strip}$ for entire galaxies and galaxy outskirts.}
    \label{fig:mprogzstrippercentilesoutskirts}
\end{figure*}

\subsection{Where do the outskirts of simulated galaxies come from?}\label{sec:ptlcs:overview}

\subsubsection{Merger masses and timescales}\label{sec:discussion:nmerge}
It has been established for some time now that the spatial distribution of the debris of a disrupted satellite is affected by the mass ratio and timing of the merger event, as well as the orbital properties or concentration of the satellite \citep[e.g.][to name a few]{johnston2008,rodriguezgomez2016,amorisco2017a}. This knowledge has been applied to a variety of efforts to characterize the assembly histories of nearby galaxies --- for example, \cite{deason2013} and \cite{dsouzabell2018} argued that the Milky Way has had a quieter merger history weighted more strongly towards early times relative to M31's more extended merger history. 

Figure \ref{fig:distall} displays the full distributions of the stellar particle stripping redshifts ($z_{\rm strip}$) and progenitor galaxy stellar masses ($M_{\rm stell,prog}$) for every stellar particle in our parent sample of central disk Milky Way -like galaxies. We emphasize that for in-situ stellar particles, $z_{\rm strip}$ is more comfortably thought of as $z_{\rm form}$. In the top panel, we can see that the full distribution of ex-situ stellar particles peaks at progenitor masses around $10^{9.5} M_{\odot}$ over approximately $2 > z > 0.5$. The bottom panel focuses on stellar particles in the outskirts (i.e., located at or beyond 20 kpc at $z=0$) alone, and suggests that the majority of the stellar halo is constructed from stellar particles that span a similar range in $z_{\rm strip}$ relative to the full galaxy, albeit contributed by slightly lower progenitor stellar masses ($\sim 10^{8}-10^{9.5}M_{\odot}$). 

The wide distributions in both panels of Figure \ref{fig:distall} are largely driven by trends between a galaxy's (true) stellar mass and its assembly history. 
Note, for example, that the distribution of progenitor galaxy stellar masses for in-situ particles peaks at \textit{higher} stellar masses in the outskirts. This is because, when we look at the parent sample as a whole, the information in the outskirts is dominated by high mass galaxies who have both more particles in total and, specifically, more particles in the outskirts relative to low mass galaxies.
To more thoroughly investigate the role that galaxy stellar mass plays in determining the particular distribution of progenitor masses in its stellar halo, we compute the $25^{\rm th}-75^{\rm th}$, $10^{\rm th}-90^{\rm th}$, and $1^{\rm st}-99^{\rm th}$ percentiles of $M_{\rm stell,prog}$ for each galaxy. The results are visualized in the top panel of Figure \ref{fig:mprogzstrippercentilesoutskirts}, which shows these percentiles side by side in order of increasing host galaxy stellar mass. To guide the eye, we also trace the running median and minimum progenitor galaxy stellar masses (such that 50\% and 90\% of stellar particles were brought in by progenitor galaxies above this threshold) in yellow and blue solid lines, respectively. The median $M_{\rm stell,prog}$ outside of 20 kpc increases from $10^{8}M_{\odot}$ to $10^{9}$, and the minimum $M_{\rm stell,prog}$ increase from $10^{6.5}M_{\odot}$ to $10^{7.5}M_{\odot}$.

If we consider ex-situ stellar particles across the entire galaxy (dashed lines), we see a similar trend with progenitor masses $\sim 0.5$ dex higher than in the outskirts. This same offset in progenitor stellar masses between the central regions and outskirts of galaxies in TNG100 and TNG300 was reported by \cite{pillepich2018b}, although we note that in that study the comparison was between stellar populations at $< 30$ kpc and $>100$ kpc. \cite{pillepich2018b} also quote systematically higher values of $M_{\rm stell,prog}$ than we show in Figure \ref{fig:mprogzstrippercentilesoutskirts} ($\sim 10^{8}-10^{9}M_{\odot}$ as opposed to $\sim 10^{7.5}-10^{8}M_{\odot}$ for galaxy stellar masses $\gtrsim 10^{10.75}$); however, this is due to an important difference in definition: the authors define the progenitor stellar mass as the \textit{maximum} stellar mass reached by the satellite, but the majority of stellar particles will be stripped either before or after this maximum mass is achieved.

In the lower panel of Figure \ref{fig:mprogzstrippercentilesoutskirts}, we visualize the distributions of $z_{\rm strip}$ for the particles in the outskirts of each galaxy in the TNG100 parent sample. A weak trend exists between the median value of $z_{\rm strip}$ and stellar masses of galaxies, but values remain close to $z_{\rm strip} \sim 1$ across the entire mass range. Additionally, unlike progenitor stellar masses, the typical stripping times in the outskirts of galaxies closely match those averaged over galaxies as a whole (this can be seen by comparing the solid and dashed lines). 

Figure \ref{fig:mprogzstrippercentilesoutskirts} also highlights the diversity in the assembly of the outskirts of galaxies.
Despite the fact that we are able to track an overall rise in progenitor masses with host galaxy mass and identify a characteristic stripping redshift, there is a remarkable degree of scatter even between galaxies of nearly identical mass. This can be most clearly seen by following the \textit{individual} values of the median $M_{\rm stell,prog}$ or $z_{\rm strip}$ --- two consecutive galaxies can vary by up to an order of magnitude in median progenitor stellar mass, or jump between median $z_{\rm strip} \sim 0.5-3$.
Some of this scatter may simply be due to the scatter in the stellar mass to halo mass relation. However, at fixed dark matter halo mass, the peak to peak scatter in stellar mass within our parent sample is at most $\sim 0.8$ dex for halo masses of $\sim 10^{12}M_{\odot}$; the peak-to-peak scatter in stellar halo mass, on the other hand, is approximately 2 orders of magnitude at these mass scales. This suggests that the heterogeneity in the assembly histories of galaxies is the more important driver of variation in characteristic progenitor galaxy masses and particle stripping times.

\begin{figure}
    \centering
    \includegraphics[width=0.99\linewidth]{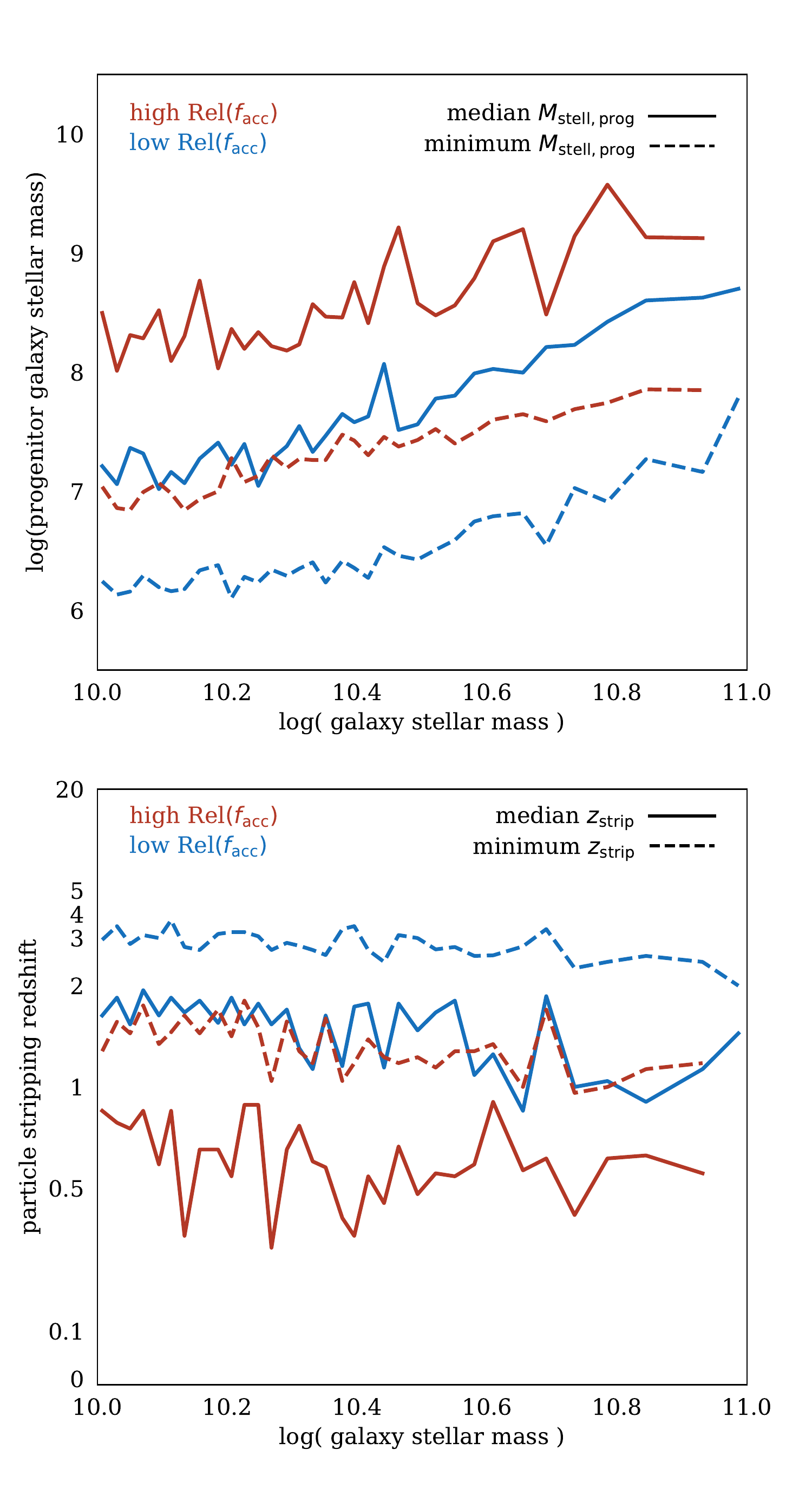}
    \caption{\textbf{Effects of progenitor galaxy masses and stripping times on accretion fractions. Top:} Continuing from Figure \ref{fig:mprogzstrippercentilesoutskirts}, we show the running median (solid lines) and minimum (dashed lines) values of progenitor galaxy stellar masses for ex-situ stellar particles in the outskirts of TNG100 galaxies, but here we split our parent sample into those with high (red) and low (blue) relative accretion fractions. Outside of 20 kpc, ex-situ stellar particles in galaxies with high Rel(\facc) are contributed by satellites that are an order of magnitude more massive than for galaxies with low Rel(\facc).  \textbf{Bottom:} The same, but now with the running median values of $z_{\rm strip}$ split by relative accretion fraction. There are no discernible trends with galaxy stellar mass; however, the outskirts of galaxies with high Rel(\facc) were built up later than galaxies with low Rel(\facc).}
    \label{fig:mprogpercentilesoutskirtsfacc}
\end{figure}

We can gain further insight into this variation by separately tracing the median $M_{\rm stell,prog}$ and $z_{\rm strip}$ for galaxies with high and low fractions of accreted material for their stellar mass (defined here as Rel(\facc) $> 0.5$ and Rel(\facc) $< -0.5$, respectively). Figure \ref{fig:mprogpercentilesoutskirtsfacc} shows that the stellar halos of galaxies with high Rel(\facc) were built by progenitor galaxies that were on average a factor of 10 more massive than the progenitors of galaxies with low Rel(\facc), and that galaxies with high/low Rel(\facc) systematically have later/earlier assembly histories (i.e., median $z_{\rm strip} \sim 2$ as opposed to $\sim 0.75$).

\subsubsection{Contributions from in situ stellar particles}\label{sec:discussion:insitu}

Stellar halos are sometimes assumed to be composed entirely of accreted material. However, observations have shown that mergers and accretion events are not the only processes that place stars at large radii: disk stars can migrate to larger radii \citep{radburnsmith2012,ruizlara2017}, or get kicked out of the plane of the disk due to a perturbation caused by a close satellite passage or a minor accretion event \citep{sheffield2012,dorman2013,pricewhelan2015}.

Simulations have also produced in-situ stellar halo stars \citep{roskar2008,zolotov2009,purcell2010,tissera2013}, and although they generally agree that the in-situ components of stellar halos are limited to the inner regions, the transition radius to the accretion-dominated stellar halo is less well determined.
Recently, \cite{font2020} discussed the notion that different implementations of star formation and feedback processes between hydrodynamic simulations in particular can impact the relative contributions from in-situ and ex-situ stars.
\cite{rodriguezgomez2016} showed that this radius is a function of galaxy stellar mass, and \cite{pillepich2015} demonstrated that it can change over the course of an individual galaxy's assembly history. Furthermore, using the Auriga simulations \cite{monachesi2019} showed that, even out at 100 kpc, the in-situ component can account for up to 20-30\% of the stellar mass for approximately one third of their galaxy sample. 

The top panel of Figure \ref{fig:fex} shows the ex-situ fractions in our TNG100 parent sample as a function of radius, and split into 5 smaller stellar mass bins with width of 0.2 dex. The fraction of accreted material rises with increasing distance from the center of the galaxies, and the typical transition radius to the ex-situ dominated stellar halo increases with galaxy stellar mass (the latter is likely driven at least in part by the low numbers of stellar particles at these radii, particularly for the lowest mass galaxies). For the TNG100 galaxies, a significant in-situ population persists beyond 20 kpc. Even out at 50-100 kpc, our galaxies have median in-situ fractions of up to 20 percent, with some dependence on the stellar mass bin. This is broadly consistent with the Auriga and ARTEMIS stellar halos \citep{monachesi2019, font2020}, but stands in contrast to the Eris stellar halo \citep{pillepich2015}, which has a negligible contribution from in-situ stars beyond $\sim 40$ kpc.

\subsubsection{Post-infall star formation}\label{sec:discussion:postinfall}
Since we tracked the exact times at which each stellar particle was stripped from its satellite to become bound to the central, we are able to distinguish between two types of ex-situ stellar particles: those that formed before their progenitor crossed the virial radius of the host halo, and those that formed after this point. We refer to the latter group as the post-infall population of ex-situ stellar particles, following \cite{pillepich2015} and \cite{rodriguezgomez2016}.

It has been shown that satellite galaxies do not typically quench immediately after crossing the virial radius of their central galaxies, but rather continue forming stars for up to several Gyr \citep[e.g.][]{tollerud2011,wetzel2012,wheeler2014}, and this is qualitatively reproduced by the original Illustris simulation \citep{sales2015}. We therefore asked the question of where these post-infall stars fall in TNG100 galaxies at $z=0$ and whether they provide a significant contribution to the stellar halo.

\begin{figure}
    \centering
    \includegraphics[width=\linewidth]{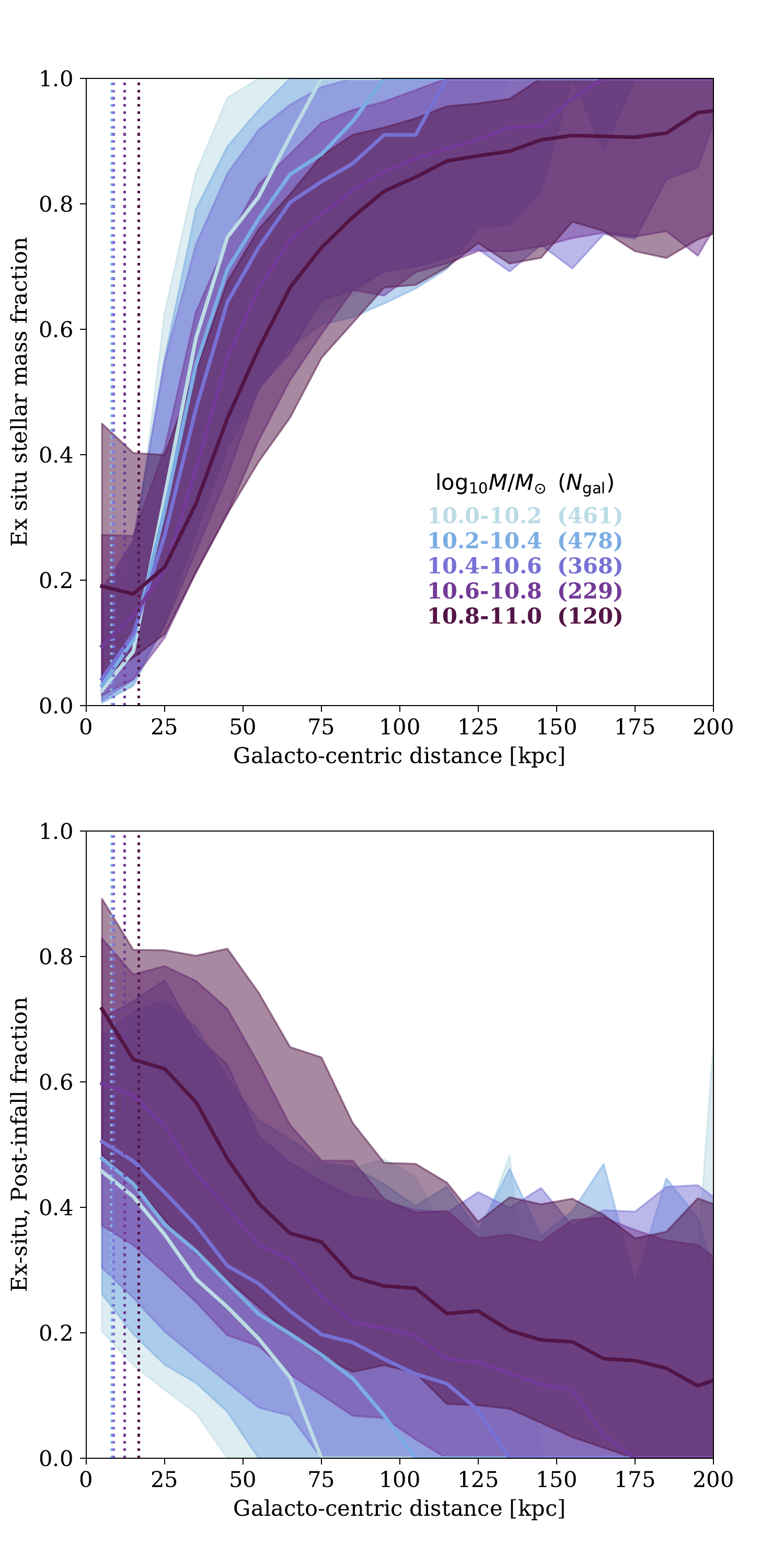}
    \caption{\textbf{Ex-situ stellar populations as a function of galacto-centric distance. Top:} The median ex situ stellar particle mass fraction as a function of galacto-centric distance for our parent sample of TNG100 galaxies, split into smaller bins of stellar mass. The galaxy-to-galaxy variation is shown via the shaded bands, which encompass the 90\% scatter in each bin. The dotted lines show the median value of $2R_{\rm half}$ for each mass bin. \textbf{Bottom:} The median fraction of ex-situ stars that formed \textit{after} crossing the host galaxy's virial radius as a function of galacto-centric distance for our parent sample of TNG100 galaxies, once again split into smaller bins of stellar mass. In the innermost regions a significant fraction of ex-situ stars formed after their progenitor galaxy crossed the virial radius of the host; however, in the outer stellar halo the majority of ex-situ stars formed pre-infall. }
    \label{fig:fex}
\end{figure}

The bottom panel of Figure \ref{fig:fex} shows the fraction of ex-situ stars that formed after crossing the host virial radius (i.e., the post-infall ex-situ population) as a function of galacto-centric radius, once again split into smaller galaxy stellar mass bins to remove any trends with stellar mass.  Inside of $\sim 20$ kpc post-infall star formation can account for up to 80 percent of the ex-situ population, but out at $\sim 50$ kpc this contribution drops below 50 percent for all but the most massive galaxies. In general, the fraction of post-infall stellar particles is lower for lower mass galaxies at fixed radius; this is driven by the fact that lower mass galaxies have smaller physical extents as well as a timing constraint --- stellar particles that have more recently crossed the virial radius (and are therefore located at large radii) have had less time to form out of gas post-infall, and as a result are more likely to have formed pre-infall.

\begin{figure*}
    \centering
    \includegraphics[width=\linewidth]{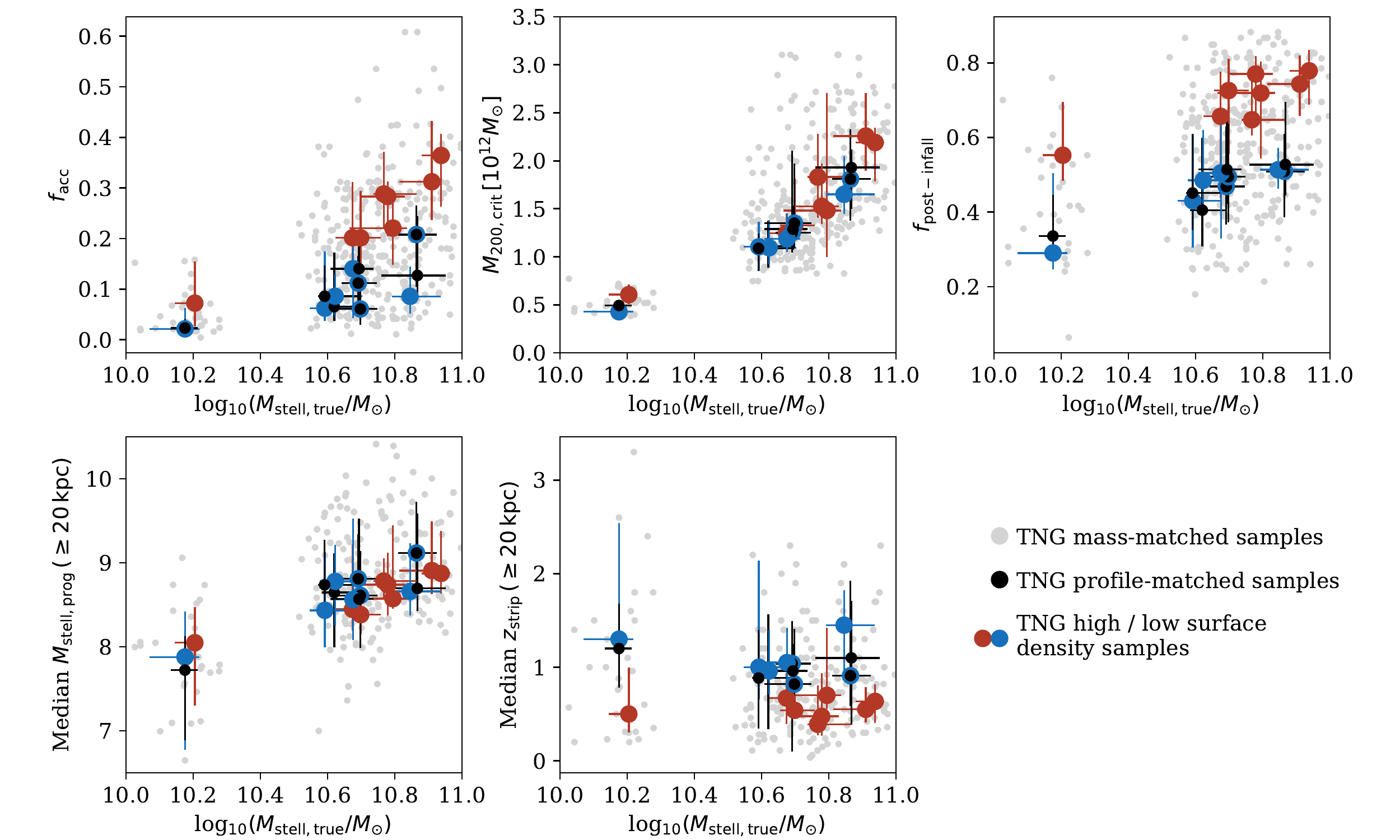}
    \caption{ \textbf{Properties of the profile-matched samples.} In every panel, small grey dots show the distribution of various details of assembly history as a function of galaxy stellar mass for each mass-matched TNG100 sample. We show the median values of the high and low surface density samples (see text for details) in red and blue points. Finally, the medians of the profile-matched samples are shown in black. The stellar masses on the x-axis are the median values within each sample, and error bars enclose the $1\sigma$ variation. We can see that for every DNGS galaxy, the accretion fractions \textbf{(top left panel)} of the profile-matched TNG100 galaxies are very similar to the accretion fractions of the low surface density samples. Similarly, the profile-matched and low surface density samples tend to have somewhat lower halo masses \textbf{(top middle panel)} and lower fractions of post-infall ex-situ stellar particles \textbf{(top right panel)}. Looking towards the outskirts of these galaxies, the median progenitor masses \textbf{(lower left panel)} of the profile-matched samples do not appear to prefer either the low or high surface density samples, whereas the median accretion redshifts of the profile-matched samples once again follow the low surface density samples \textbf{(lower right panel)}. }
    \label{fig:profilematched:bigfig}
\end{figure*}

\subsection{Pairing stellar particles and profiles: constraining the unobservables}\label{sec:ptlcs:profmatch}

In an effort to constrain the assembly histories of our individual DNGS galaxies as much as possible, we now return to the idea that the density profiles of galaxies contain valuable clues. We define a new sample of TNG100 galaxies which are matched to observations based on the shapes of their profiles in addition to their ``observed'' stellar mass. This time, rather than parametrizing profiles as a power law, we divide the stellar density profiles of each DNGS galaxy by the density profiles of their respective TNG100 mass-matched samples, and measure the median values of this ratio beyond 20 kpc. Defining $\Upsilon \equiv {\rm log}_{10}({\rm median \,\, ratio})$, we can construct three new sub-samples:
\begin{itemize}
    \item \textit{Profile-matched sample}: These are the TNG100 galaxies that most closely match the outer structure of each DNGS galaxy. Specifically, we take the subset of the mass-matched sample with the 25 percent lowest absolute values of $\Upsilon$.
    \item \textit{High surface density sample}: For each DNGS galaxy, this is the subset of the mass-matched sample with the 25 percent highest values of $\Upsilon$. 
    \item \textit{Low surface density sample}: For each DNGS galaxy, this is the subset of the mass-matched sample with the 25 percent lowest values of $\Upsilon$.
\end{itemize}

In the upper left panel of Figure \ref{fig:profilematched:bigfig}, we plot the total accreted stellar mass fraction against the true galaxy stellar masses of each of our TNG100 mass-matched samples in grey points. Red and blue points indicate the median \facc for the high and low surface density samples; consistent with what we saw in earlier sections, these separate cleanly into higher and lower values of \facc. Finally, black points point to the median values of each profile-matched sample. Error bars highlight the $1\sigma$ variation in each of the three sub-samples (and the stellar mass is the median within each sample). Consistent with Figures \ref{fig:profiles:facc:ms:dngsmatched} and \ref{fig:massfrac:plaw:facc}, we see that the accretion fractions of the TNG100 galaxies in the profile-matched samples are very closely aligned with those of the low surface density samples.

Moving beyond the accreted mass fraction, Figure \ref{fig:profilematched:bigfig} also shows that the profile-matched samples of TNG100 galaxies are consistent with the low- surface density samples in a number of other properties. They reside in slightly lower mass halos (upper middle panel), and have lower fractions of post-infall ex-situ stellar particles (upper right panel) relative to the high surface density samples. In the outskirts, despite seeing a clean separation in the median progenitor stellar masses of TNG100 galaxies with high and low accreted fractions at fixed stellar mass (defined as Rel(\facc) above 0.5 and below -0.5, respectively; see Figure \ref{fig:mprogpercentilesoutskirtsfacc}), we are unable to differentiate the median progenitor masses (lower right panel) of the  profile-matched samples from the rest of the mass-matched galaxies. In contrast, the median values of $z_{\rm strip}$ in the outskirts of the profile-matched samples closely follow those of the low- surface density samples. This suggests that, at fixed stellar mass, the timing of mergers may be more important than the stellar masses of the disrupting galaxies when determining the shapes of the outer profiles.

We conclude from this combination of profile information and stellar particle data that -- in spite of the diversity of galaxy outskirts seen in DNGS galaxies -- the TNG100 galaxies that most closely match DNGS stellar halos are contained within a relatively small region of assembly history parameter space.

\section{Where are the ``missing'' outskirts of galaxies?}\label{sec:missing}
As detailed towards the end of Section \ref{sec:metrics}, our ability to leverage simulations to gain physical insight into the assembly histories of observed galaxies depends on how well we have characterized the outskirts of galaxies, as well as whether or not we are dealing with any biases in our observational or simulated datasets. 

In Section \ref{sec:ptcls} we made the assumption that DNGS fairly represents the observed census of stellar halos, and that TNG100 is accurate enough (over a range of observables) to justify detailed comparisons in the diffuse and relatively sparsely sampled outskirts of galaxies. Here, we assess the validity of these assumptions by exploring possible sources of bias in the observations and simulations separately.

\subsection{Observational challenges: are we ``missing'' galaxy outskirts?}\label{sec:missing:obs}
DNGS galaxies most closely resemble mass-matched TNG100 galaxies that have relatively low accreted stellar mass fractions, are hosted by low mass dark matter halos, experience relatively early accretion of satellites (and typically low mass satellites), and have higher-than-average pre-infall ex-situ fractions. However, it is important to keep in mind that observational studies of the low surface brightness outskirts of galaxies are incredibly challenging --- and this has the potential to influence any comparisons with theory in subtle (or sometimes not-so-subtle) ways.

\subsubsection{PSF and sky subtraction}
First, as mentioned in Section \ref{sec:intro}, regardless of the observational approach taken (star counts, integrated light, stacks of galaxies, etc), low surface brightness work comes with relatively large uncertainties. For integrated light, unresolved images of external galaxies, the point spread function (PSF) is typically the limiting factor \citep[e.g.][]{sandin2014}. The exact structure and extent of the PSF is a result of scattered starlight (or light from the compact centers of galaxies) from internal optics or a variable upper atmosphere. Scattered light is \textit{always} present in optical observations --- the game is simply to minimize its effects or to characterize it as robustly as possible in order to avoid biasing any measurements \citep[e.g.][among others]{slater2009,martinezlombilla2019,infantesainz2019}. 

\cite{abraham2014} demonstrated that the wide angle PSF of the Dragonfly Telephoto Array is suppressed by a factor of 10 relative to other instruments; and Figure \ref{fig:profiles:facc:sbg:dngsmatched} helps to further alleviate concerns over the PSF. We convolved the TNG100 light images with the Dragonfly PSF before measuring any surface brightness profiles, and also enforced realistic surface brightness limits by placing the convolved images into an empty region of one of the DNGS fields. We note that if the PSF were to have a significant effect, it would act to \textit{increase} the amount of light measured in the outskirts of galaxies rather than diminish it. Figure \ref{fig:profiles:facc:sbg:dngsmatched} also demonstrates that the difference in surface density profiles (Figure \ref{fig:profiles:facc:ms:dngsmatched}) is not caused by the specific conversion of light to stellar mass by \cite{merritt2016a}.

As mentioned in Section \ref{sec:data:observations}, sky subtraction presents another challenge for low surface brightness observations. Specifically, if measurements of the sky are biased high (low), then the diffuse outskirts of galaxies well be over-(under-) subtracted, and stellar halo mass estimates will be correspondingly too low (high). To deal with this, the Dragonfly data reduction pipeline implements a two-stage approach to characterizing and subtracting the sky. In the first round, we fit a $3^{\rm rd}$ degree polynomial to the background of each image after heavily masking out all sources, and create a deep (preliminary) stack of all sky-subtracted images. We then construct a much more aggressive source mask from this stack, which is then implemented in a second and more accurate round of sky modeling \citep[for more details, see][]{merritt2016a,zhang2018,danieli2019}.

Although Dragonfly is very well set up for low surface brightness imaging, it does have its limitations. The azimuthally averaged surface brightness profiles reach $30-32$ mag arcsec$^{-2}$, but these apertures cover huge areas on the sky. On smaller scales (relevant for individual stream detection) typical depths are $\sim 29$ mag arcsec$^{-2}$. This is deep enough to detect streams with luminosities of the order $10^{8} L_{\odot}$ \citep{vandokkum2019a}, but it is plausible that lower luminosity streams are missed. 

\subsubsection{Sample selection}
Unintended biases in the DNGS sample selection could also come into play. The only observable galaxy property that was explicitly used by \cite{merritt2016a} was luminosity; however, we specifically used the $B$-band luminosity, which is a relatively blue band and is therefore not an ideal tracer of the underlying stellar mass distribution and might lead to biased comparisons when selecting galaxies in TNG100 based on stellar mass alone.
A rough comparison between the stellar mass-luminosity distributions between the two samples indicates that DNGS galaxies fall anywhere between the bright and faint end of the TNG $B$-band luminosity distributions.
A related issue, as mentioned in Section \ref{smoothing:observational}, is that our TNG100 light images do not include the effects of dust, which could exacerbate the differences between the Dragonfly and TNG100 surface brightness profiles. To investigate this, we used the median $g-r$ color within $20$ kpc as a metric, and found that while there are trends between this median color and the shape of the profile in the TNG100 parent sample, there are no discernible trends within any given mass-matched sample, indicating that the more fundamental relation is between galaxy stellar mass and color. We also measured this median color for DNGS galaxies, and found that while DNGS galaxies do not always have the same colors as the median of their mass-matched samples, they do not systematically fall to bluer or redder colors. 

Similarly, no size criteria were applied when selecting DNGS galaxies, but we noted in previous sections that some of the more massive DNGS galaxies have smaller half-mass radii than their entire mass-matched sample. This complicates the interpretation of our results, as our primary metric for characterizing the stellar halos of galaxies is the mass beyond 20 kpc (see Appendix \ref{app:matching:size}). It is possible that, for TNG100 galaxies, this measurement contains a greater influence from extended (in-situ) disks than the observations, although we note that there are examples of large disks in DNGS as well (most notably M101, with spiral arms visible out to 40 kpc).

Moreover, there is a possibility of an environmental bias. We attempted to mimic the DNGS sample by including both centrals and sufficiently isolated satellite galaxies in the TNG100 parent sample, and although we could not identify any differences in the surface density profiles of the two populations, the ratio of centrals to satellites in the parent TNG100 sample is much higher than in DNGS. Furthermore, the requirement that TNG satellite galaxies in the parent sample have at most one massive ($M_{\rm stell} \geq 10^{10} M_{\odot}$) neighbor within 1 Mpc effectively minimizes any environmental effects that might be at play.

\subsubsection{Inclination angle}
DNGS galaxies span a wide range of inclination angles, and we compare each with a randomly-oriented mass-matched sample of TNG100 galaxies. In cases where DNGS galaxies are face-on (or close to face-on), we may be performing a biased comparison --- specifically, the majority of TNG100 galaxies will likely be more highly inclined than the DNGS galaxy. As noted by \cite{elias2018}, shifting a galaxy from edge-on to face-on has the effect of decreasing the surface brightness in the outskirts by up to $1.5$ magnitudes. Given the random orientations of the TNG100 galaxies, however, the majority of these corrections will be smaller than this, and the key result that DNGS galaxies tend to be most similar to low \facc TNG100 galaxies would remain unchanged. Moreover, as shown by \cite{elias2018}, decreasing the observed inclination angle of a galaxy does not create down-bending profiles such as those seen in DNGS galaxies. To verify our intuition, we measured the average axis ratio (as measured by \texttt{ellipse}) for both the DNGS and TNG100 mass-matched samples. Axis ratios are more difficult to robustly measure in the outskirts, so we restricted this analysis to the inner regions of the density profile where the $g$-band surface brightness is brighter than $25$ mag arcsec$^{-2}$. Axis ratios in DNGS span from $0.5 \leq b/a \leq 1.0$, with a mean value of 0.68;  the distribution in the TNG100 parent sample has a mean of 0.63 with 98 percent of the sample falling between 0.31-0.88. Within each mass-matched sample, we compared the profiles of galaxies with high inclination ($0.45 < b/a < 0.55$) to those with low inclination ($0.75 < b/a < 0.85$), and did not find a significant difference in the outer density slopes. Furthermore, we did not identify trends between axis ratio and profile shape within the mass-matched samples, and when we compared the observed profiles to only the profiles with axis ratios within $\pm 0.2$ our main results hold. We therefore concluded that while axis ratios are important considerations when interpreting the outskirts of galaxies, they do not prevent us from learning about the assembly histories of galaxies from their outskirts.

\subsubsection{Sample sizes}
The most obvious reason to suspect a biased comparison, however, is sample size. As it currently stands, DNGS has only eight galaxies, and is clearly dwarfed by the numbers in TNG100. As a result, it is difficult to robustly determine whether the DNGS galaxies (particularly those with very low inferred \facc) are simply very rare.

We note, however, that other observational studies have arrived at qualitatively similar results to DNGS. In particular, \cite{harmsen2017} compared the stellar halo properties of six massive edge-on disk galaxies from the GHOSTS survey \citep{radburnsmith2011,monachesi2016} with predictions from a variety of theoretical models, and found a similarly large scatter and low average accretion fraction relative to expectations from simulations. It is beyond the scope of this work to produce mock GHOSTS data from TNG100; however, we point out that \cite{harmsen2017} converted the reported estimates of stellar halo mass fractions from \cite{merritt2016a} to their own metric, and demonstrated that the two observational datasets are consistent with one another, suggesting that their stellar halo mass fractions are also below the median of TNG100 at fixed mass. More recently, \cite{rich2019} measured surface brightness profiles from deep imaging of 65 spiral galaxies in the HERON survey. Although they did not quote estimates for stellar halo masses, they showed that spiral galaxies span a large range in total diameter (outer ``envelope'' included), even at fixed stellar mass. If we consider this envelope diameter to be a proxy for the prominence of the stellar halo, then this result is consistent with the diverse galaxy outskirts reported by \cite{merritt2016a} and \cite{harmsen2017}. However, \cite{rich2019} deferred detailed comparisons with simulations to future work, and therefore we cannot yet rule out the possibility that both the average stellar halo mass fraction and degree of galaxy-to-galaxy scatter when measured from a significantly larger observational dataset might fall more in line with simulations. 
Relatedly, it is also possible that the volume of TNG100 is not large enough to capture the full diversity of merger histories and resulting galaxy outskirts.

\subsection{Difficulties of matching samples}
\label{sec:discussion:matching}
When comparing observations with simulations, sample selection is a key consideration, as is the manner in which matching is done. In this work, we matched galaxies based on their stellar mass -- specifically, the ``observer-friendly'' stellar mass obtained by integrating the surface density profiles down to $10^{4}M_{\odot}$ kpc$^{2}$. An important caveat to keep in mind is that the fact that the DNGS and TNG100 sample sizes are so different prevents us from matching the stellar mass functions between the two, meaning that correlations between stellar mass and other properties will manifest more visibly in the TNG100 dataset than in DNGS. As a workaround, and to avoid biasing our comparisons towards lower stellar masses in TNG100 as much as possible, we imposed a Gaussian distribution of stellar masses in each of our TNG100 mass-matched samples (see Section \ref{sec:data:simulations} for details). This is an improvement over simply selecting all TNG100 galaxies that fall within the DNGS stellar mass errorbars, but is still not ideal. For example, a close examination of Figure \ref{fig:profilematched:bigfig} reveals that -- even within a given mass-matched sample -- the median masses of the low surface density, profile-matched, and high surface density samples separate slightly from one another.

Another choice that we made when performing the matching was to use an integrated property (stellar mass) rather than the surface density profiles themselves. Given the correlations between galaxy stellar mass and outer profile shape, another possibility is that the DNGS galaxies which appear to lack clear profile analogs in TNG100 might be better described by lower mass TNG100 galaxies.

Moreover, we have adopted a definition of stellar mass that uses a surface density floor rather than a fixed aperture, meaning that we could be comparing galaxies of very different sizes. And indeed, as mentioned in Section \ref{sec:metrics}, not all DNGS galaxies are well matched in sizes (half-mass radii) within their mass-matched samples.
We investigated the extent to which the relatively small sizes of DNGS galaxies affects our results in Appendix \ref{app:matching:size}, and found that the 2D half-mass radii of TNG100 galaxies with high Rel(\facc) are systematically larger than the sizes of galaxies with low Rel(\facc). Furthermore, size offsets make distinguishing differences in stellar halos from differences in galaxy extents difficult, and can in some cases act to worsen discrepancies between the two samples, particularly for stellar halo metrics involving a physical boundary. Nevertheless, we emphasize that this effect cannot entirely account for our observation that DNGS galaxies most closely match TNG100 galaxies with low relative accretion fractions and appear to be missing stellar mass in their outskirts relative to the simulated galaxies.

Along similar lines, we also re-measured galaxy stellar masses by integrating the surface density profiles out to a fixed aperture of 30 kpc, and confirmed that the central results of this paper are robust to our choice of stellar mass definition.

\subsection{The role of resolution effects and simulated galaxy property convergence in TNG}\label{sec:discussion:resolution}
Another relevant issue to examine is the extent to which various physical properties of TNG100 galaxies change with resolution. All TNG simulations have also been carried out in a series of lower resolution versions, which allows us to assess this quantitatively; and, fortunately, we can refer to a number of results from the existing TNG literature for insight.

\cite{pillepich2018a} compared the fiducial TNG100 run with two lower resolution runs (TNG100-2 and TNG100-3, at factors of 8 and 64 lower mass resolution, respectively) and showed that as resolution improves, galaxy stellar masses \textit{increase} at fixed dark matter halo mass (corresponding to an overall higher normalization of the galaxy stellar mass function, particularly at low masses and low redshift). Galaxy stellar masses in TNG100-2, for example, are approximately $40$ percent lower than in the fiducial TNG100-1. On the other hand, galaxy stellar sizes in TNG \textit{decrease} towards higher resolution, and this holds true for both the in-situ and ex-situ stellar components. Later, \cite{pillepich2019} extended these convergence checks to TNG50 and showed that resolution affects the sizes of low mass galaxies ($M_{\rm stell,true} \leq 10^{9}-10^{10}M_{\odot}$) more strongly than those of high mass galaxies, and particularly at lower redshifts. They showed that galaxy sizes in TNG100 are converged to better than $20-40$ percent after $z=2$, but that the fraction of disk galaxies increases with resolution. 

Despite the resolution dependence of the stellar sizes and masses of galaxies, however, \cite{pillepich2018b} demonstrated that the relative contributions of in-situ and ex-situ stellar mass fractions as a function of either total stellar or dark matter halo mass are fully consistent between TNG100 and TNG300 (the latter has a factor of 8 lower mass resolution and a factor of 2 lower spatial resolution). This applies for dark matter halo masses down to $\sim 10^{12}M_{\odot}$, a limit which encompasses the halo masses for every one of our TNG100 mass-matched samples except for the one matched with NGC 1042. Even in this case, however, the median \facc between TNG100 and TNG300 differs by approximately 0.05.

Given all of this, we consider it entirely possible that resolution effects could be relevant to our finding that the outskirts of TNG100 galaxies appear to be, on average, more massive and extended than their mass-matched DNGS galaxies. However, there are a number of competing effects at play. We might expect, for instance, that at higher resolution the outskirts of galaxies might be more massive because they were built from more massive satellite galaxies; however, those satellite galaxies would also have smaller sizes and might disrupt less easily. The fact that the ex-situ to in-situ fraction appears to be converged also hints that resolution may have a stronger impact on the distribution, rather than the amount of accreted material, although this has not been demonstrated explicitly. Finally, comparing samples between two different resolution simulations is complicated by changes to the galaxy stellar mass functions as well as to the fraction of disk galaxies.

It is beyond the scope of this particular paper to carry out a detailed study; however, we consider this one of many promising directions to explore in the future, and the resolution-dependence of stellar and dark matter particle stripping from subhalos is already being investigated in detail by Lovell et al. (in preparation).

\subsection{Limitations of simulations: are we building galaxy outskirts too efficiently?}\label{sec:discussion:toymodels}

An important question to consider, as pointed out towards the end of Section \ref{sec:metrics}, is whether we should expect the outskirts of TNG100 galaxies to agree with observations at all. Cosmological hydrodynamic simulations of today are generally able to reproduce bulk properties of the galaxy population; however, models using different methodologies or implementing different physical models can lead to a range of predictions for observable quantities \citep[see review by][]{somervilledave2015}, and this uncertainty becomes a limiting factor when defining theoretical expectations for the properties of stellar halos.

A number of sources of tension with observations remain even for state-of-the-art simulations, and it is possible that any mismatch in the stellar halos of galaxies is simply a manifestation of these rather than a separate tension on its own. For example, a known issue with the original Illustris simulation was that the simulated galaxy disks were approximately twice as large as their observational counterparts \citep{snyder2015}. In TNG100, galaxy sizes are larger than observations by at most 0.1 dex for stellar masses of $10^{10.5}-10^{11}M_{\odot}$  \citep[within the $1\sigma$ scatter;][]{genel2018,rodriguezgomez2019}. It is nevertheless possible that extended galaxy disks make a significant contribution to estimates of stellar halos, possibly augmented by radial migration or disk heating \citep[e.g.][]{roskar2013,veraciro2014,grand2016}. Along these lines, \cite{harmsen2017} compared the stellar halo masses, metallicities, power law slopes and color gradients measured for 6 massive disk galaxies observed as part of the GHOSTS \citep{radburnsmith2011} survey with accretion-only N-body models and hydrodynamic models separately, and came to the intriguing conclusion that the accretion-only models were a better match to the data. \cite{monachesi2019} also showed that accreted-only stellar halo profiles in the Auriga simulations are a better match to GHOSTS observations of stellar halos when measured along the minor axis. This suggests that hydrodynamic simulations may overestimate the role of processes responsible for forming in-situ stellar halos. 

Additionally, \cite{genel2014} showed that the galaxy stellar mass functions in Illustris were over-abundant relative to observations at $z \geq 1$, particularly at the low mass end ($10^{8}-10^{9}M_{\odot}$). The situation in TNG100 is improved with respect to Illustris by a factor of a few, but simulated galaxies are still more abundant than observed for stellar masses of $10^{9}$ at $z \gtrsim 1$ by up to a factor of 2 \citep[depending on the observational sample at hand;][]{pillepich2018b}. 
At $z\sim  2$, observational constraints for galaxies with stellar masses around $10^{9}M_{\odot}$ are only consistent within approximately a factor of 10, limiting our ability to place strong constraints on simulations.

\begin{figure*}
    \centering
    \includegraphics[width=\linewidth]{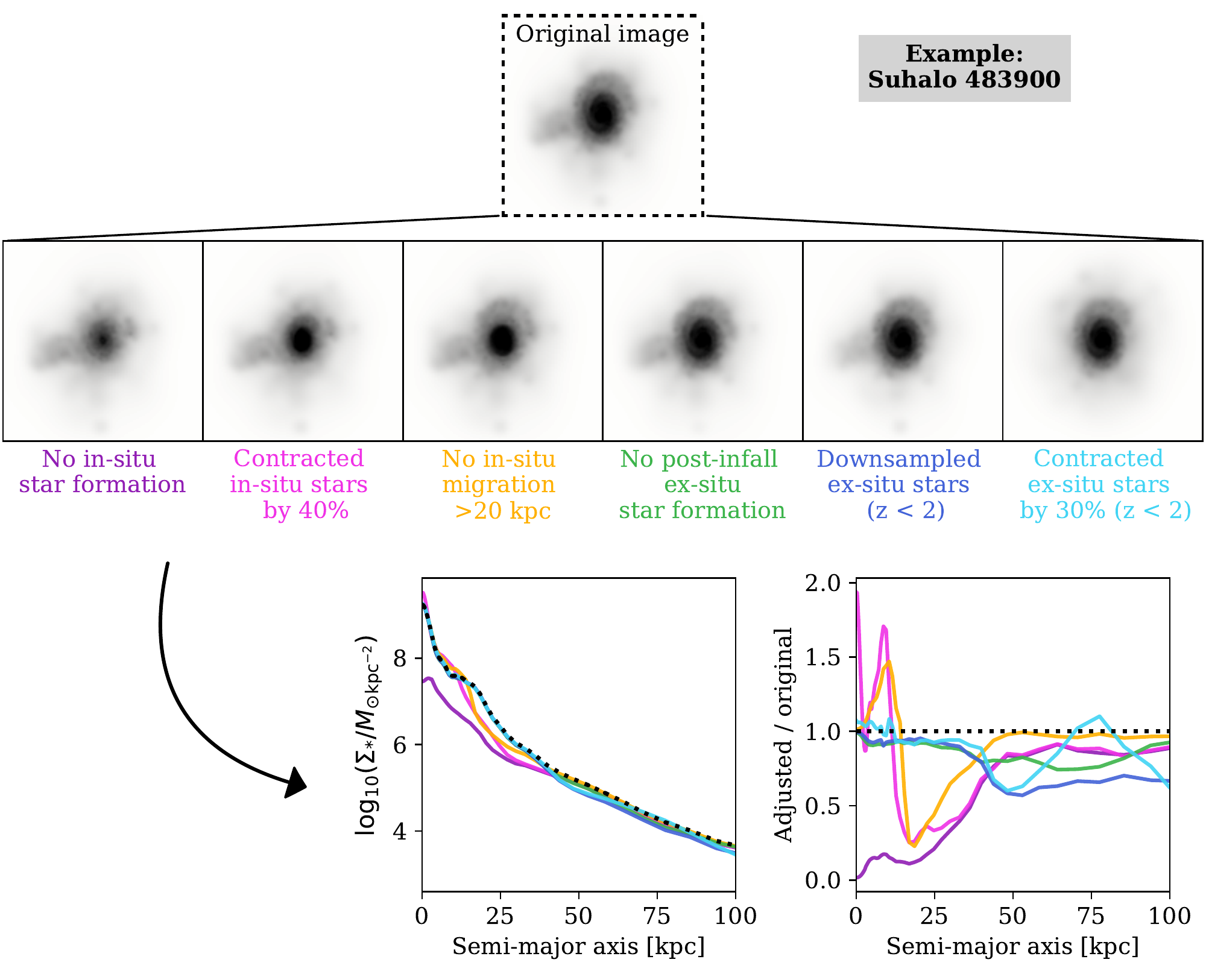}
    \caption{\textbf{An overview of our ``toy models.''} At the top, we show the original, unadjusted stellar mass map for Subhalo 483900 of TNG100 as an illustrative example. Just below are six different ``adjusted'' stellar mass maps, after applying the toy models described in Section \ref{sec:discussion:toymodels}. The adjusted surface density profiles (and the surface density ratios relative to the original) resulting directly from these five panels are shown in the bottom row. The unadjusted profile is indicated by the black dotted line in both cases. We note that the sharp features within $\sim 15$ kpc in the surface density ratios associated with the models contracting in-situ stars (pink) and preventing in-situ migration (yellow) are caused by a shift in position of the material in the spiral arms of the galaxy in these two scenarios. See text for specific details of each model.}
    \label{fig:adjustedframes}
\end{figure*}

Uncertainties in the stellar-to-halo mass relation, which is largely unconstrained below stellar masses of $10^{10}M_{\odot}$ at high redshifts ($z \geq 2$), could also manifest in the stellar halos of galaxies. If, for example, the low mass slope of this relation were slightly too shallow in TNG100, then central halos/galaxies in TNG100 would accrete subhalos/satellites with stellar masses that are too high, thereby acquiring an excess of ex-situ material per merger. We note that here, we are making the assumption that the cosmological merger rates in TNG100 are accurate, and that the \textit{number} of low mass progenitor galaxies that built up the central galaxy and its stellar halo is correct.
\cite{pillepich2018b} show the stellar-to-halo mass relation for TNG100 (with the stellar mass measured within a few different apertures; see their Figure 11) and trace its evolution from $z=1$ to $z=0$. They show that there is broad agreement between TNG100, TNG300, and Eagle, as well as a number of semi-empirical models and abundance matching results; however, some differences between the models are apparent (and see also \citealt{engler2020} for a closer look at the differences between satellite and central galaxy stellar-to-halo mass relations).

Alternatively, even if the satellite galaxies have the correct amount of stellar mass, they might disrupt too easily. \cite{vandenbosch2018} used a suite of idealized simulations to study N-body subhalo disruption, and showed that the artificial disruption of low mass halos could be prevalent in simulations, even occasionally in halos above the resolution limit. They argue that this ``overmerging'' problem holds true in cosmological simulations as well, albeit with the caveat that their work does not include baryonic effects. The authors also focus on circular orbits, which may not be applicable for the majority of infalling satellites in TNG100 \citep[see also][who demonstrate that these effects do not pose a significant problem for cosmological zoom-in simulations]{bahe2019}. \cite{errani2019} further demonstrated that low mass subhalos (near the resolution limit of a simulation) can form artificial cores, which are more easily disrupted than cusps. If some fraction of satellite galaxies in TNG100 are disrupting too early, it is possible that they deposit too much stellar mass in the outskirts of the central galaxy rather than the interior (or simply surviving as intact satellites). This mechanism therefore primarily affects the \textit{distribution} rather than the amount of stellar material.

A related possibility is that TNG100 satellite galaxies do not quench efficiently enough after crossing the virial radii of their hosts, leading to a significant contribution from ex-situ, post-infall star formation (but see Donnari et al. 2020 in prep. for a closer look at quenched fractions in TNG100).

\begin{figure*}
    \centering
    \includegraphics[width=\linewidth]{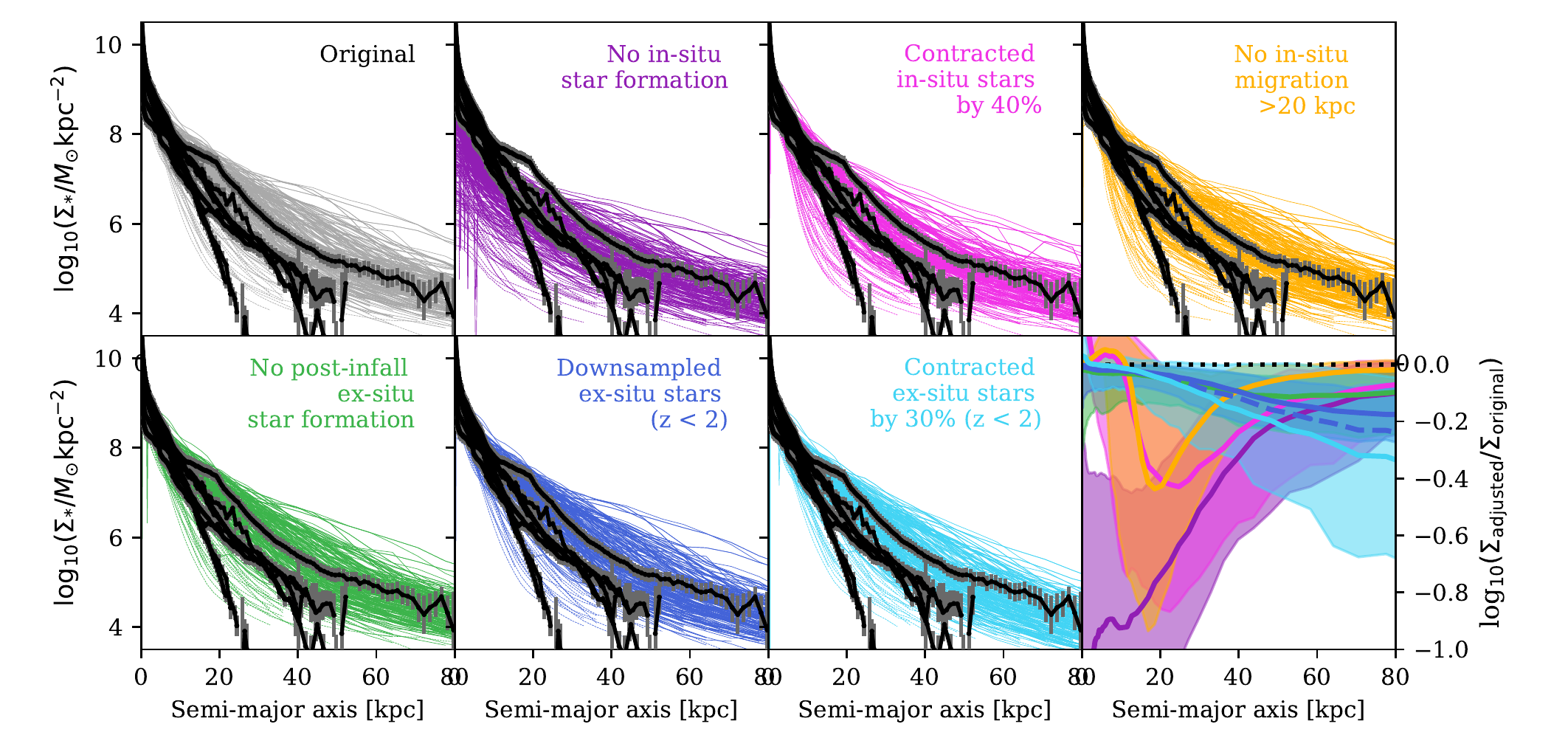}
    \caption{\textbf{DNGS vs TNG100: adjusted stellar mass surface density profiles.} Here, we provide a more detailed look at the effects of various toy models (discussed in Section \ref{sec:discussion:toymodels}) on the surface density profiles of the TNG100 galaxies. Each panel shows all eight DNGS profiles in black points, overlaid on their respective mass-matched TNG100 samples in solid lines. The top left panel shows the unadjusted TNG100 profiles, and the following panels show the adjusted profiles resulting from each of the six toy models (see text for details). Since NGC 1042 is significantly less massive than the rest of the DNGS galaxies, we distinguish its mass-matched TNG100 sample with dashed lines to visually account for the trend between profile shape and stellar mass. Finally, in the lower right panel we show the median surface density ratios of each toy model in solid lines, with the $5^{\rm th}-95^{\rm th}$ percentiles of the distribution as shaded regions. In the inner regions of the stellar halo ($\sim 20-50$ kpc), the models affecting the in-situ stellar particles are most effective at modifying the shapes of the density profiles, while beyond 50 kpc the contraction of ex-situ stars becomes most impactful. For the "downsampled ex-situ stars" model, the solid line and shaded region correspond to our default approach -- we defined a maximum progenitor stellar mass for each galaxy, and randomly removed 50 percent of the ex-situ stellar particles contributed by satellites with less massive than this threshold since $z=2$. For comparison, the dashed line shows a more extreme version of this where we exclude \textit{all} ex-situ stellar particles with $M_{\rm stell,prog} \leq 10^{9}M_{\odot}$.}
    \label{fig:adjustedprofiles}
\end{figure*}

To quantify the effect that these scenarios might have on the TNG100 galaxy population at $z=0$ --- and in particular, to explicitly explore any observable consequences --- we used the information from the stellar particle tracker to implement six different toy models, and created ``adjusted'' stellar mass maps for the central galaxies in our mass-matched samples (i.e., the ones for which we were able to track stellar particles). Figures \ref{fig:adjustedprofiles} and \ref{fig:adjustedmassmetrics} summarize the results of our findings, and we step through the details of each model in the following subsections.

\subsubsection{Excluding in-situ stellar particles}
\label{sec:toymodel:exsitu}
A hypothetical scenario in which galaxies are comprised of only accreted material is not particularly realistic, but it does delineate the maximum contribution that an overly massive (or extended) in-situ disk could have on the surface density profiles and stellar halo mass fractions in the TNG100 galaxies. To implement this, we flagged every in-situ stellar particle using the results of our stellar particle tracking code (i.e., requiring that $z_{\rm form} = z_{\rm cross} = z_{\rm strip}$). We then generated new images excluding all in-situ stellar particles.

The top row of Figure \ref{fig:adjustedframes} shows the adjusted stellar mass map for an individual galaxy (our favorite Subhalo 483900), and compares it directly to the original, unadjusted mass map. Removing the in-situ stellar particles primarily affects the inner regions of the galaxy -- this was expected, since we saw from Figure \ref{fig:fex} that the ex-situ components of galaxies typically only become dominant out at 30-50 kpc. The next row of Figure \ref{fig:adjustedframes} shows the effect of excluding in-situ stellar particles on the stellar mass surface density profiles more explicitly (purple curves). We show both the adjusted surface density profile and the ratio of the adjusted to the original, and point out that although the inner regions are pushed to substantially lower surface densities, the outskirts ($\gtrsim 40-50$ kpc) are relatively unaffected. 

In Figure \ref{fig:adjustedprofiles}, we show the surface density profiles for all eight DNGS galaxies (black points) overlaid on the TNG100 mass-matched samples (colored lines). In the top left panel, we show the unadjusted TNG100 profiles as a reference -- as we saw in Figure \ref{fig:profiles:facc:ms:dngsmatched}, the DNGS profiles lie at or below the median of the mass-matched TNG100 profiles at all radii. We show the mass-matched sample of NGC 1042 in a dashed rather than solid line to prevent comparisons between those profiles and the rest of DNGS (NGC 1042 has a low stellar mass relative to the rest of the DNGS sample; and recall the trends between stellar mass and profile shape discussed in Section \ref{sec:measured:profiles}). Removing the in-situ stellar particles results in a starkly different picture: internal to $\sim 20$ kpc, the DNGS galaxies now have higher surface densities than the TNG100 galaxies, and out in the stellar halo they appear to span the full distribution of simulated profile shapes. 

To summarize the impact of the model while minimizing the effects of any trends with stellar mass, we show the median (purple line) and scatter ($5^{\rm th}-95^{\rm th}$ percentiles; purple shaded region) of the ratio of the adjusted (excluding in-situ) to unadjusted density profiles as a function of radius for all of the mass-matched TNG100 samples in the bottom right panel of Figure \ref{fig:adjustedprofiles}. Even for this much larger sample, the effect of this toy model closely follows what we saw for our example subhalo in Figure \ref{fig:adjustedframes}. Surface densities in the inner regions of the galaxy are lower by up to an order of magnitude, and by $\sim 0.5$ dex even out to 40 kpc.

\begin{figure*}
    \centering
    \includegraphics[width=\linewidth]{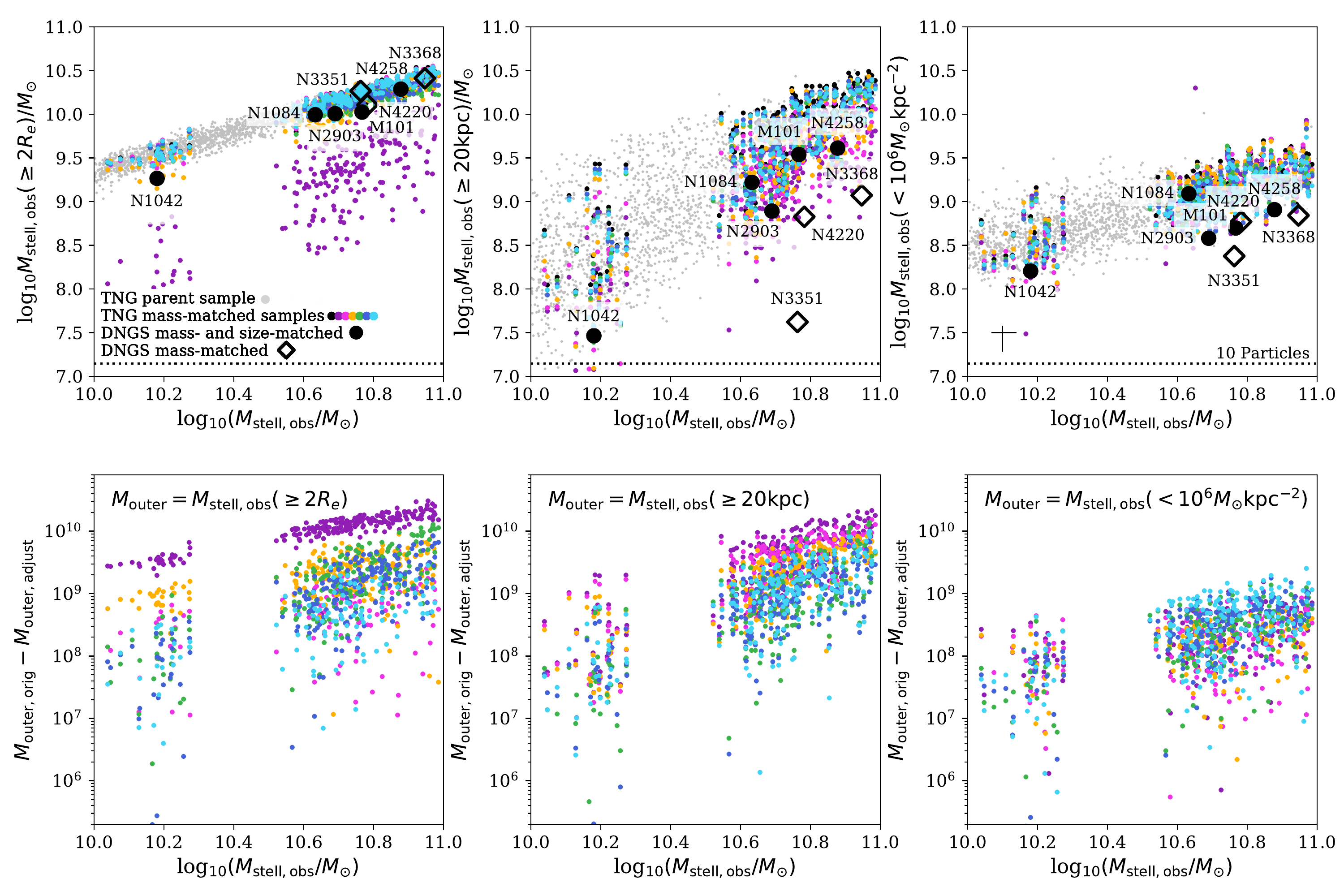}
    \caption{\textbf{DNGS vs TNG100: adjustments to the stellar halo mass estimates. Top:} We investigate how each toy model affects the three methods of estimating the mass in the stellar halo presented in Figure \ref{fig:masses:compare}. In each case, our extreme model of preventing in-situ star formation results in the most drastic change; however, the relative impacts of the more physical models vary between stellar mass definitions. For ease of comparison, the stellar mass on the x-axis is always measured from the unadjusted density profiles. The unadjusted parent sample is shown in grey points for reference, and as with Figure \ref{fig:masses:compare}, the error bar in the right hand panel illustrates the typical error in the observed galaxy and stellar halo mass measurements. \textbf{Bottom:} Whether by removing stellar particles in the stellar halo entirely or shifting them in towards the center of the galaxy, each toy model results in a reduction of stellar halo mass relative to the original images. Each panel shows the \textit{deficits} of stellar mass for each stellar halo definition. Again, for ease of comparison, the ``observed'' stellar mass on the x-axis is always measured from the unadjusted density profiles.}
    \label{fig:adjustedmassmetrics}
\end{figure*}

In Figure \ref{fig:adjustedmassmetrics}, we ask whether (or to what extent) the drop in surface densities corresponds to an improved match in stellar halo mass fractions between DNGS and TNG100. We show the stellar masses computed for the galaxies without in-situ stars (purple points), and compare them directly to the observed DNGS measurements (black symbols). Back in Section \ref{sec:metrics}, we saw how strongly the precise definition of the stellar halo mass can affect our conclusions, so we performed this comparison for all three previously explored metrics: the stellar mass beyond two half-mass radii, beyond 20 kpc, and below $10^{6}M_{\odot}$ kpc$^{-2}$. This toy model, despite not being particularly realistic, lowers outer stellar mass fractions by a median value of $0.51^{+0.28}_{-0.37}$ dex when measured beyond $20$ kpc (the errorbars encompass the $5^{\rm th}-95^{\rm th}$ percentiles), and brings the models into closer agreement with the DNGS galaxies (with NGC 3351 being the only exception). When measuring the mass below our surface density threshold the results are similar, but if we measure beyond two half-mass radii, the drop in stellar halo mass is more extreme, but has the opposite effect of \textit{increasing} the differences between observations and simulations (driven by the larger half-mass radii in this model relative to the original images). 

Additionally, in Figure \ref{fig:adjustedmassmetrics} we show the physical amount of stellar mass removed from the stellar halo by excluding in-situ stellar particles as a function of (original) galaxy stellar mass, checking all three observer-friendly definitions of the stellar halo. The amount of mass removed from the outskirts increases for higher galaxy masses, but this model is able to remove up to $\sim 10^{9}-10^{10}M_{\sun}$.

\subsubsection{Restricting the spatial extent of in-situ stellar particles}
\label{sec:toymodel:insitucontract}
Rather than remove in-situ stars entirely, our second model explores the possibility that the in-situ disks are simply too extended. We identified in-situ stellar particles in the same way as the previous section, and shifted them in towards the center of the galaxy such that their new adjusted galacto-centric distances were smaller by 40\%. This choice could potentially lead to a mismatch of up to 0.2 dex in galaxy sizes, since -- as mentioned above -- TNG100 sizes are currently compatible with observational results to better than 0.1 dex over $10^{10}-10^{11}M_{\odot}$. However, as with the previous model, we consider this to be an upper limit of sorts on the impact that a reduction in the sizes of insitu stellar populations could have on stellar halos.

Figure \ref{fig:adjustedframes} reveals a noticeably smaller disk for our example Subhalo 483900. The surface density profile and density ratio profile for this model (pink solid curves) show a drop in surface density relative to the original profile between $\sim 20-50$ kpc by up to a factor of 2. We note that the sharp feature in the ratio profile is the result of the galaxy's spiral arms shifting inwards.

Moving to Figure \ref{fig:adjustedprofiles}, we see that the profiles of the TNG100 galaxies with spatially contracted in-situ stellar particles are in better agreement with the DNGS outskirts beyond 20 kpc. As shown by the lower right panel of this figure, surface densities beyond this point can be up to 0.8 dex lower than in than the unadjusted case (although the median value is 0.4 dex).

Stellar halo masses are lower by $0.33^{+0.20}_{-0.29}$ dex when measured beyond 20 kpc, and we can see from Figure \ref{fig:adjustedmassmetrics} that, unlike the previous model, shrinking the in-situ stellar particle distribution results in a closer match to observations in all three metrics. The impact is minimized when measuring beyond 2 half-mass radii, but in general Figure \ref{fig:adjustedmassmetrics} shows that this model decreases the mass in the stellar halo by $10^{8}-10^{10}M_{\odot}$.

\subsubsection{Restricted migration of in-situ stellar particles in the outskirts}
\label{sec:toymodel:demigrate}
Our next model simply imposes that the only in-situ stellar particles that exist in the stellar halo were born there. We therefore allow the presence of extended in-situ stellar populations, but attempt to ``reverse'' any effects that might have kicked, heated, or migrated stellar particles out to radii beyond 20 kpc. Specifically, we identified in-situ stellar particles as described in Section \ref{sec:toymodel:exsitu}, and then flagged those that are currently located in the stellar halo but were born at galacto-centric distances $d_{\rm form} < 20$ kpc. We then randomly re-sampled the coordinates of each stellar particle, such that the adjusted distance is within 10\% of $d_{\rm form}$. We did not attempt to preserve the overall morphology of the migrated population of stellar particles, as in this simple model we did not distinguish between the various physical effects that could cause stellar particles to move outwards over the course of time. We then generated stellar mass maps for each using the newly adjusted coordinates.

Returning to the middle panel of Figure \ref{fig:adjustedframes}, we see that, similarly to the previous models, the largest effect on the example Subhalo 483900 seems to be in the inner regions, and we can discern that the main disk of the galaxy has shrunk in size. This is confirmed by the surface density and ratio profiles (yellow curves), where we can clearly see that material that was once located between $\sim 20-50$ kpc has been moved inwards, typically to distances around $\sim 10$ kpc (we can also once again see the signature of material in the spiral arms moving inwards in the density ratio profile). 

Figure \ref{fig:adjustedprofiles} demonstrates that the effect of this model on the mass-matched surface density profiles resembles that of the ``contracted in-situ disk'' model, although its impact drops off more rapidly at larger galacto-centric distances. This can be seen most clearly by comparing the yellow and pink solid curves in the lower right panel. The two models are similar -- in both cases we move in-situ stars towards the center of the galaxy, but here we only work with a subset of the stellar particles and shift them inwards in a non-uniform way. Figure \ref{fig:adjustedmassmetrics} shows a median change relative to the unadjusted stellar halo masses of $0.20^{+0.17}_{-0.28}$ dex, corresponding to a decrease of $10^{9}-10^{10}M_{\odot}$ (yellow points).

\subsubsection{Excluding ex-situ, post-infall stellar particles}
\label{sec:toymodel:postinfall}
In this model, we enforced a more efficient quenching of satellite galaxies, such that star formation ceases once they enter the host's dark matter halo. To do this, we flagged stellar particles that formed in a satellite galaxy \textit{after} crossing the virial radius of the halo of the central galaxy, i.e. with $z_{\rm cross} \geq z_{\rm form} > z_{\rm strip}$. We then created new stellar mass maps with these stellar particles excluded.

In Figure \ref{fig:adjustedframes}, we can see from the adjusted surface density profile and density ratio (green curves) that the effect of this model is relatively small. Visually, it is difficult to identify structural changes to Subhalo 483900, but we see from the profiles that it primarily impacts the galaxy at radii $\sim 20-80$ kpc, leading to a decrease in the stellar surface density of approximately 20\%. 
This is consistent with what we saw in Figure \ref{fig:fex}; although post-infall star formation contributes significantly to the \textit{ex-situ} stellar mass budget in the inner regions of the stellar halo $(\sim 20-50)$ kpc, it is generally $\lesssim 40$ percent of the total stellar mass.

Figure \ref{fig:adjustedprofiles} confirms the ineffectiveness of this model -- the adjusted profiles (green line and shaded regions) after preventing all post-infall star formation are entirely consistent with the original, unadjusted profiles. In the lower right panel, we see that the median density ratio is approximately 0.1 dex over all radii. Finally, we see in Figure \ref{fig:adjustedmassmetrics} (green points) that this model results in a drop in stellar halo masses of $0.07^{+0.05}_{-0.09}$ dex, or typically $10^{8}-10^{9}M_{\odot}$.

\subsubsection{Down-sampled late-time accreted contributions}
\label{sec:toymodel:mergersdownsamp}

Here, we consider the possibility that the stellar masses of satellite galaxies in TNG100 could be too high due to uncertainties in the low mass end of the stellar to halo mass relation at $z\sim 1-2$. If this were the case, it would mean that -- for a given assembly history -- a central galaxy acquires too much stellar mass per merger event \citep[by up to a factor of 2;][]{pillepich2018b}. Based on this idea, we used the data shown in Figure \ref{fig:mprogzstrippercentilesoutskirts} to define a maximum progenitor mass for each individual galaxy such that 90 percent of the ex-situ stellar particles in the outskirts ($>20$ kpc) were contributed by satellites at or below this stellar mass. We then flagged all ex-situ stellar particles that were acquired from satellite galaxies with masses less than this maximum progenitor mass since $z=2$, and randomly down-sampled these by a factor of two. At $z=1$, a factor of two fewer galaxies with stellar masses $\sim 10^{9}$ would bring the TNG100 galaxy stellar mass function into excellent agreement with available observational constraints.

In the case of our example Subhalo 483900, we can see from Figure \ref{fig:adjustedframes} that this model primarily effects the outer regions of the galaxy. This is also true for the full mass-matched sample in Figure \ref{fig:adjustedprofiles} (dark blue solid curves and shaded region), where we see that the outer density profiles can become steeper by up to $0.2$ dex. For comparison, we also tested a second way of reducing the ex-situ stellar mass in the outskirts: we simply removed all ex-situ stellar particles contributed by satellite galaxies with $M_{\rm stell,prog} \leq 10^{9}M_{\odot}$. This variation represents a hypothetical scenario in which \textit{all} subhalo disruption at these masses is due to numerical resolution leading to under-dense low mass halos that are too susceptible to disruption. The median density ratio is shown in the lower left panel of Figure \ref{fig:adjustedprofiles} (dark blue dashed curve). Applying a uniform upper limit to progenitor masses has a much stronger effect on the lower mass galaxies (this can also be inferred directly from the percentiles shown in Figure \ref{fig:mprogzstrippercentilesoutskirts}); however, outside of $\sim 50$ kpc, the median profile is slightly steeper (by $\sim 0.05$ dex) than in our default down-sampling model.

Figure \ref{fig:adjustedmassmetrics} (dark blue points) reveals that the amount of mass removed from the stellar halo by this model is relatively small, with typical values between $10^{7}-10^{9}M_{\odot}$; this corresponds to a median decrease in the stellar halo masses of $0.07^{+0.05}_{-0.10}$ dex.

\subsubsection{Artificially delayed satellite mass loss.}
\label{sec:toymodel:mergersdelay}
To explore the scenario in which accreted material is simply more spatially extended relative to observations due to premature stripping of stellar particles from their progenitor satellite galaxies (rather than being contributed by overly massive satellites), we identified ex-situ stellar particles accreted after $z=2$, and redistributed them such that their new adjusted distance was reduced to 70\% of its original value. We generated new mass maps with the adjusted coordinates, and found that this model results in half-mass radii that are on average 10\% smaller.

In contrast to the previous model, contracting satellite debris leads to a decrease in the stellar mass surface density of Subhalo 483900 by up to 50\% at large radii. Looking towards the mass-matched samples in the lower left panel of Figure \ref{fig:adjustedframes}, we see that this model becomes increasingly effective out to $\sim 100$ kpc (cyan line and shaded region).  

Similarly, Figure \ref{fig:adjustedprofiles} shows that this model is most effective in the outermost regions of the surface density profiles, resulting in TNG100 profiles (cyan curves and shaded regions) that are slightly steeper (by a median difference in ${\rm log}_{10}\Sigma$ by up to 0.4 dex) than the unadjusted profiles beyond $40-50$ kpc.

Despite this drop in surface density in the outer regions, the majority of the mass in the stellar halo exists at smaller galacto-centric distances. As a result the overall change to the stellar halo masses is only $0.09^{+0.09}_{-0.11}$ dex, with losses of $10^{7}-10^{9}M_{\odot}$ (Figure \ref{fig:adjustedmassmetrics}, cyan points). 
However, we note that this model leads to the most significant mass reduction when measuring below surface densities of $10^{6}M_{\odot}$ kpc$^{-2}$.

\subsubsection{Lessons from toy models}
The toy models described in Sections \ref{sec:toymodel:exsitu}-\ref{sec:toymodel:mergersdelay} are overly simplistic and, at times, even unphysical --- we do not claim, for instance, that all of these galaxies could have reached $z=0$ with no contribution from in-situ star formation. Likewise, prematurely ``turning off'' star formation in satellites would impact the colors, star formation histories, and gas contents of TNG100 satellite galaxies found within their host virial radii at $z=0$.  By implementing these models, we effectively assumed that it is possible to change \textit{individual} aspects of TNG100 in isolation, while in reality there are a number of non-linear effects at play, and any small change could lead to a ripple effect. We therefore consider our findings to be guidance towards potentially useful future directions rather than conclusive, quantitative results.

As we saw in Figures \ref{fig:adjustedframes}-\ref{fig:adjustedmassmetrics}, none of these models can single-handedly erase the differences between the DNGS and mass-matched TNG100 stellar halo masses or outer surface density profiles. Models that removed or resulted in a more compact distribution of in-situ stellar particles were the most effective at steepening the inner regions of the stellar halo profiles ($\sim 20-50$ kpc). These models also had the greatest impact on the stellar halo masses, removing between $10^{9}-10^{10}M_{\odot}$ of material and bringing the TNG100 stellar masses into better agreement with observations. To put this into perspective, these numbers are comparable with the typical (maximum) masses of satellite galaxies that contribute ex-situ stars to their centrals \citep[with satellite masses measured at their maximum value;][]{pillepich2018b}. 

Outside of $50$ kpc, contracting the debris of recently ($z<2$) infalling satellites had the most significant impact on the outer density slopes, followed by down-sampling ex-situ stars. Furthermore, testing an extreme scenario in which all low mass satellite disruption is artificial by removing all ex-situ stellar particles contributed by satellites with $M_{\rm stell,prog} \leq 10^{9}$ resulted in slightly steeper slopes than our default downsampling model. Taken together, this suggests that the timing of stellar particle stripping (i.e., exactly where particles end up within the host) may be more important than uncertainties in the stellar to halo mass relation when building up the outskirts of galaxies beyond $\sim 50$ kpc. These ex-situ models were less effective at changing the mass in the stellar halo than those focused on in-situ stellar particles, removing between $10^{7}-10^{9}M_{\odot}$. However, as we saw in Figure \ref{fig:mprogpercentilesoutskirtsfacc}, this amount of mass is comparable to the median progenitor stellar masses of ex-situ stellar particles in the outskirts of TNG100 galaxies, which range from $10^{7}-10^{8}M_{\odot}$ and $10^{8}-10^{9}M_{\odot}$ for the low and high Rel(\facc) populations, respectively.

Our finding that none of the toy models completely reconcile the observations with simulations is puzzling, but we speculate that the \textit{combination} of delayed stellar particle stripping with a less spatially extended in-situ stellar body might be more effective. Additionally, it is important to keep measurement uncertainties in mind, particularly when interpreting the stellar halo mass estimates. As noted previously, the error bar in Figure \ref{fig:adjustedmassmetrics} span the $5^{\rm th}-95^{\rm th}$ percentiles of the distribution, which are $\sim 0.2$ dex and comparable to the effect that some of our toy models have on the stellar halo mass fractions. We stress, however, that since the stellar halo masses were measured for the simulations and observations in an identical way, mass measurement errors should not systematically affect the average differences between the two. The crudeness of the toy models should also not be taken lightly, and more sophisticated, self-consistent models may be more informative.

\section{Summary and outlook}\label{sec:conclusions}
In this paper, we presented a comparison between $\sim 1800$ simulated galactic stellar halos in TNG100 and eight observed stellar halos in the Dragonfly Nearby Galaxies Survey. We selected all disk galaxies from TNG100 in the stellar mass range $10^{10}-10^{11}M_{\odot}$ that were either centrals or satellites residing in loose groups, and produced 2D stellar mass maps as well as 2D mock $g$ and $r$ band light images for each at the spatial resolution of Dragonfly imaging. For the mock images, we also convolved the ``raw'' TNG100 light images with the Dragonfly PSF and placed them in a relatively empty region of one of the DNGS fields in order to impose realistic surface brightness limits. We measured azimuthally averaged stellar surface density (or surface brightness) profiles for every galaxy in this parent sample, and then 
defined smaller stellar mass-matched samples of 50 TNG100 galaxies for each DNGS galaxy, matching on the total stellar mass integrated down to $10^{4}M_{\odot}$ kpc$^{-2}$. Finally, we tracked individual stellar particles across the entire simulation output from $z=20$ to $z=0$ in order to tie the information encoded in the outer density profile to the ``true'' assembly history for the TNG100 galaxies.

The central result of this work is that the outskirts of galaxies contain information about the assembly histories of galaxies, and that, specifically, DNGS galaxies appear to be ``missing'' stellar mass (or light) beyond 20 kpc relative to their mass-matched analogs in TNG100 
(or, conversely, TNG100 galaxies that are mass-matched to those in DNGS typically feature more massive or more extended stellar populations outside of $\sim20$ kpc).
However, comparisons between observations and simulations are sensitive to the choice of methodology -- for example, whether to use the entire surface density profile or any estimate of the stellar halo mass or slope --  as well as details of sample matching. We emphasize that combining multiple individual metrics or summary statistics is necessary in order to obtain a complete picture of stellar halos.

We showed that DNGS galaxies have surface density profiles most similar to TNG100 galaxies with low accreted stellar mass fractions for their stellar mass. 
Furthermore, when we paired information about individual stellar particles in the outskirt of TNG100 galaxies with their profile shape information, we found that the subset of TNG100 galaxies that most closely match DNGS galaxies in both stellar mass \textit{and} profile shape reside in low mass dark matter halos, experience early accretion of massive ($\geq 10^{9}M_{\odot}$) satellites, accrete fewer massive satellites overall, and have a higher fraction of ex-situ stellar particles that formed before crossing the virial radius.

In an effort to understand the underlying cause of the differences in observed and simulated stellar halos as well as the apparent ``clumping'' in assembly history parameter space, we applied six simplistic toy models to the TNG100 galaxies and investigated the effect that they had on the profile shapes and overall stellar halo masses: we excluded all in-situ stars; contracted the spatial extent of in-situ stars by 40\%; shut off the migration of in-situ stars beyond 20 kpc; removed any ex-situ stellar particles that formed within satellites after crossing the virial radius of the host; down-sampled ex-situ stellar particles by a factor of two for $z \leq 2$; and contracted the spatial extent of ex-situ stellar particles acquired over $z \leq 2$ by 30\%. We showed that each model resulted in somewhat lower stellar halo masses; however, no single scenario was able to fully explain the differences in the outer profiles \textit{and} the stellar halo masses between observations and simulations. Instead, we speculate that a combination of still possibly overly-extended in-situ stellar populations and premature satellite disruption could be the culprit, and advocate that future studies take a step beyond our toy models towards a more robust, self-consistent investigation. 

Looking ahead, a more complete understanding will be achieved through a combination of ever-larger observational samples, which will be critical for more robustly characterizing the diversity in the outskirts of galaxies. The Dragonfly Edge-On Survey (DEGS; C. Gilhuly et al., submitted) is underway and will provide a detailed look at over a dozen edge-on disk galaxies, and the Dragonfly Ultra Wide Survey will cover 10,000 deg$^{-2}$ on the sky down to $\sim 28.8$ mag arcsec$^{-2}$ on $1'$ scales, facilitating the study of thousands of stellar halos. WFIRST will provide an exciting an complementary dataset, as it will allow star counts surveys to be carried out efficiently out to $\sim 20$ Mpc, and the HERON survey \citep{rich2019, mosenkov2020} has already taken a promising step forward.

Furthermore, it is clear that stellar halos are extremely sensitive testbeds of several aspects of galaxy formation models. In order to carry out fair and meaningful comparisons with observed stellar halos, state-of-the-art cosmological simulations need to match observed (or derived) properties such as galaxy stellar sizes and stellar masses (particularly of lower mass satellites at $z \sim 1-2$) to better than a factor of 1.5-2 (0.2-0.3 dex). Moreover, they require a high enough resolution to adequately sample the sparse outskirts of galaxies in stellar particles and to prevent the early disruption of satellite galaxies, as well as volumes large enough to properly match the galaxy populations found in observational surveys and identify true analogs. Finally, even with future larger samples of observed stellar halos, we will need to construct suitable and sophisticated methods to identify analogs between observed and simulated samples in order to avoid biased comparisons.

\section*{Acknowledgements}
We thank the anonymous referee for their thoughtful comments
which improved the quality of this work.
We also extend a huge thank-you to the entire staff at
New Mexico Skies for their incredible support and assistance over the years. 
A. Merritt thanks Morgan Fouesneau for a number of helpful discussions along the way (esp. regarding \texttt{emcee}), and Shany Danieli for coining the phrase ``missing outskirts.'' 
F. Marinacci acknowledges support through the Progrm "Rita Levi Montalcini" of the Italian MIUR.
All authors acknowledge and are
grateful for the cmocean \citep{cmocean} and astropy packages \citep{astropy:2013,astropy:2018}, and also acknowledge the usage of the HyperLeda database (http://leda.univ-lyon1.fr).


\bibliographystyle{mnras}
\bibliography{tng}

\appendix 

\section{Effects of image generation choices on density profiles}\label{app:images}

\begin{figure*}
    \centering
    \includegraphics[width=\linewidth]{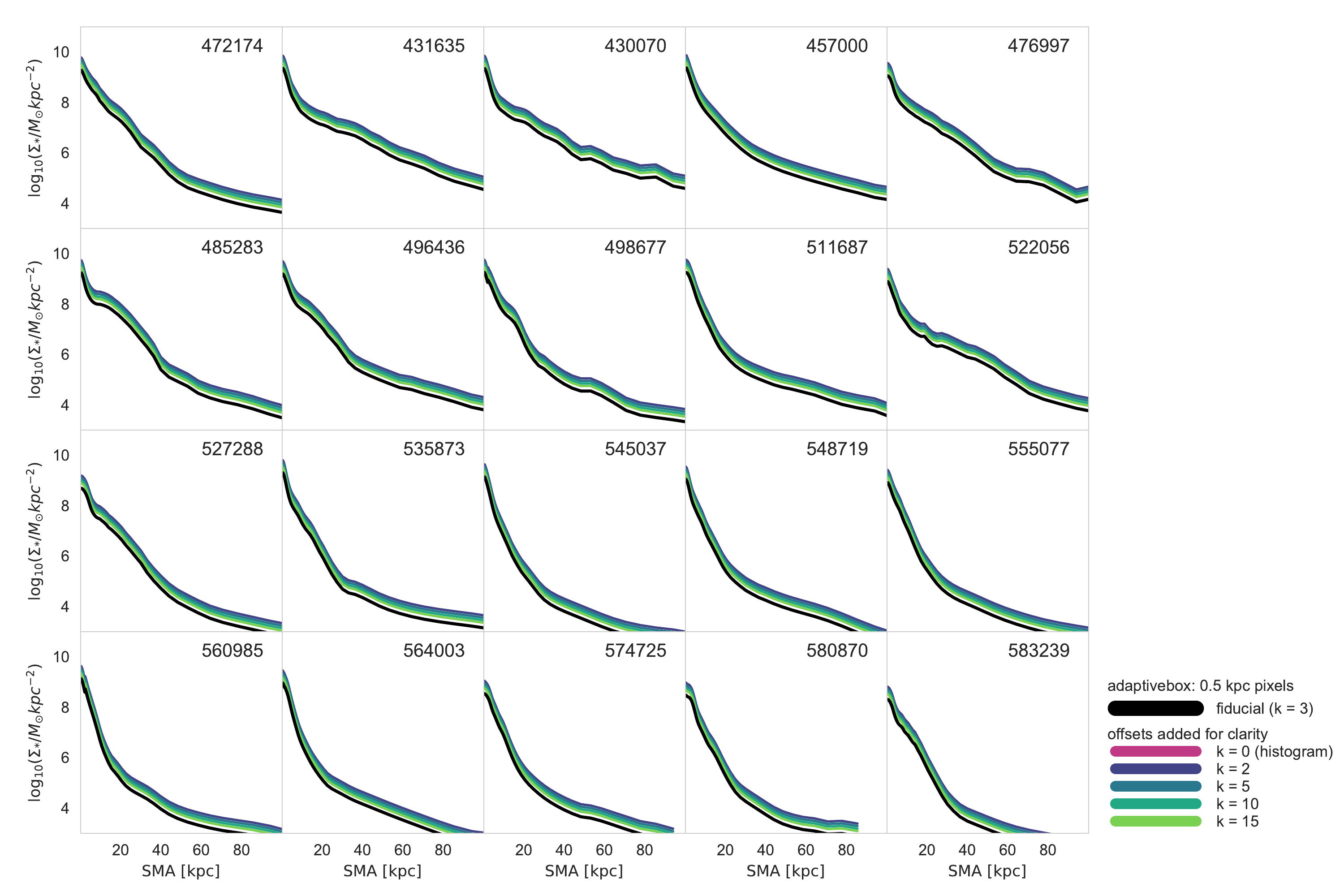}
    \caption{\textbf{Changing kernel sizes.} The fiducial profiles (setting $k=3$ in \texttt{adaptivebox}) are shown in black and compared to the results from images generated using different values of $k$ (colored lines, with small offsets applied). The galaxies are the same set shown in Figures \ref{fig:library1} and \ref{fig:library2}, and are in order of decreasing mass (left to right, top to bottom). Even in the outer reaches of the profile, the choice of $k$ does not have a significant effect.}
    \label{fig:convergence:kernelsize}
\end{figure*}

\begin{figure*}
    \centering
    \includegraphics[width=\linewidth]{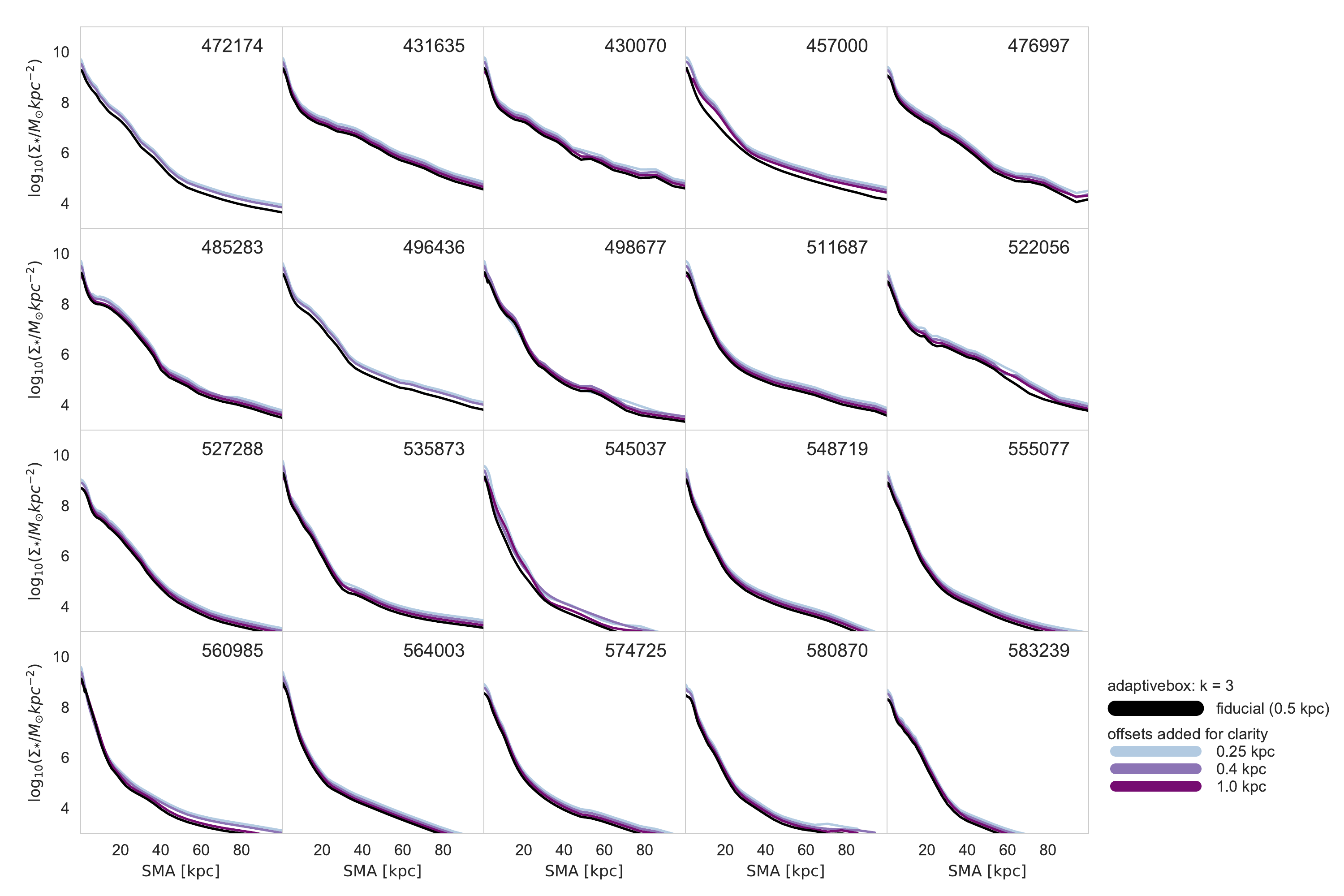}
    \caption{\textbf{Changing 3D pixel sizes.} Black lines show the fiducial profiles (setting $\Delta X = 0.5$ kpc in \texttt{adaptivebox}); colored lines show the effect of choosing larger or smaller pixel sizes at this stage. Some small-scale variation is discernible; however the overall profile shapes are independent of the choice of pixel sizes.}
    \label{fig:convergence:pixelsize}
\end{figure*}

When running \texttt{adaptivebox}, we were forced to make a set of choices about the way that the mass (or light) in a given stellar particle contributes to its nearby pixels. In particular, we choose the physical size of the 3D grid as well as the width of the Gaussian kernel based on the distance to the $k^{\rm th}$ nearest neighbor (these have fiducial values of $\Delta X = 0.5$ kpc and $k = 3$, respectively).

To explore the extent to which these choices influenced the resulting stellar mass surface density profiles, we generated several different versions of each galaxy shown in Figures \ref{fig:library1} and \ref{fig:library2}, stepping through a range of values for both pixel size ($\Delta X = 0.25, 0.4, 1.0$) and kernel size ($k = 2, 5, 10, 15$). As mentioned in Section \ref{sec:measured:images}, these 20 galaxies are equally spaced over the full stellar mass range covered by our parent sample; this is important in case any of these effects are mass-dependent.

Figure \ref{fig:convergence:kernelsize} shows that changing the value of $k$ has essentially no effect on the surface density profiles. The fiducial profiles for each galaxy are shown in black, and we overplotted the profiles generated from images with alternate values of $k$ in blue and green, with small offsets applied to ease comparisons (the lines would otherwise fall on top of one another).

As a more conservative check, we also measured the surface density profiles for the $k=0$ case; i.e., directly from 2D histograms of stellar mass such that the only ``smoothing'' that has been applied is placing the stellar particles into their nearest pixel bin (we used the fiducial value of $0.5$ kpc here). Figure \ref{fig:convergence:kernelsize} shows these profiles in magenta curves. Although the histogram-based profiles are significantly noisier (particularly in the outskirts), they are in good agreement with the profiles measured from the smoothed images, and we are therefore confident that we have not biased our results in any significant way by choosing to smooth the images.

The situation is similar for choices of physical pixel sizes --- Figure \ref{fig:convergence:pixelsize} once again displays the fiducial profiles in black, alongside the profiles associated with different values of $\Delta X$ in color with small offsets applied. Some small scale differences are noticeable (in the ``bumps and wiggles'') between the surface density profiles, but overall we can see that changing the pixel size has a negligible effect on the shape and overall structure of the profiles.

Figures \ref{fig:convergence:kernelsize} and Figure \ref{fig:convergence:pixelsize} also demonstrate, encouragingly, that the robustness of our image generation scripts is not a function of stellar mass. It is important to keep in mind, however, that the question of whether there are enough stellar particles to reliably characterize the relatively sparse outskirts of these simulated galaxies is distinct from (and much harder to quantify than) the question of whether or to what extent the details of our methodology are influencing our results.

\section{On the subtleties of matching samples}\label{app:matching}

\subsection{Size-matching at fixed stellar mass}\label{app:matching:size}
\begin{figure*}
    \centering
    \includegraphics[width=\linewidth]{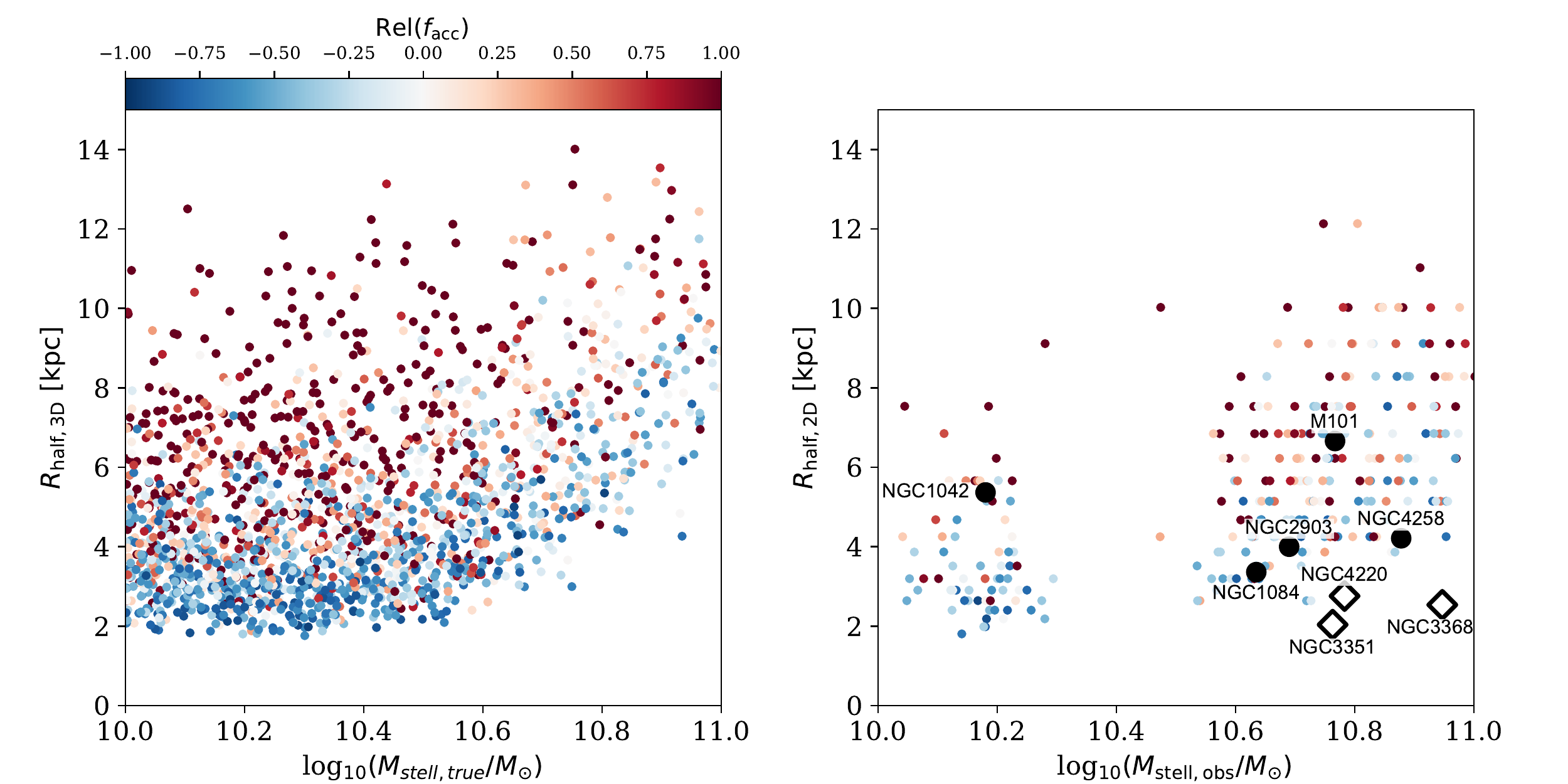}
    \caption{\textbf{The dependence of the size-mass relation on assembly history. Left Panel:} The size-mass distribution of the TNG100 parent galaxy sample, where both size and stellar mass are their ``true'' values -- i.e., stellar masses are the sum of all stellar particles bound to the galaxy, and sizes are measured in 3D. The color-coding shows the relative accretion fraction, such that red/blue points indicate galaxies with high/low accretion fractions for their stellar mass. \textbf{Right panel:} Here, we focus only on the mass-matched TNG100 galaxy samples, and plot observer-friendly values. Stellar masses are integrated from the surface density profiles down to a threshold of $10^{4} M_{\odot}$ kpc$^{2}$, and half-mass radii are measured in 2D; the color-coding is once again Rel(\facc). DNGS galaxies are shown in black points, and we indicate those with and without size-matches in the TNG100 mass-matched samples with filled circles and open diamonds, respectively. The apparent discreteness of the sizes is due to the placement of the elliptical annuli.}
    \label{fig:matching:sizemass}
\end{figure*}

The sizes of galaxies can add another layer of complexity to the task of identifying analogs between two samples.
\cite{genel2018} and \cite{rodriguezgomez2019} demonstrated that the sizes of TNG100 galaxies are in good agreement with large samples of observed galaxies, when using observer-friendly size definitions measured on mock images. Between stellar masses of $10^{10.5}-10^{11}M_{\odot}$, the simulated and observed sizes are in very good agreement, with systematic offsets of at most 0.1 dex (within the 1 $\sigma$ scatter).
On the other hand, as noted previously in Sections \ref{sec:metrics:masses} and \ref{sec:discussion:matching}
, some of the DNGS galaxies (NGC 3351, NGC 3368, and NGC 4220) are have smaller sizes (2D half-mass radii) than every one of their stellar mass-matched TNG100 samples.
The fact that some DNGS galaxies fall outside of this scatter is not necessarily cause for concern -- with only 8 galaxies, we cannot make any robust statements about offsets between these two specific samples, and can only say that it is possible that DNGS coincidentally happens to contain small galaxies.
Regardless, this raises the question of whether -- or to what extent --the differences we find between the DNGS and TNG100 stellar halos are due to mismatches in galaxy sizes.

In the left panel of Figure \ref{fig:matching:sizemass}, we show the distribution of ``true'' stellar masses and sizes for the full parent sample of TNG100 galaxies. Stellar masses are the sum of all stellar particles bound to the galaxy, and sizes are half-mass radii measured in 3D. The points are color-coded by the relative accretion fraction, and we can see clearly that at fixed $M_{\rm stell,true}$, galaxies with higher/lower accretion fractions have larger/smaller sizes. This is expected, as a more active assembly history leads to more material at larger radii, thus increasing the radius that contains half of the stellar mass of the galaxy.

In the right hand panel of Figure \ref{fig:matching:sizemass}, we shift into observer space. We show only the TNG100 galaxies in the stellar mass-matched samples; here, stellar masses are measured by integrating the surface brightness profiles down to $10^{4}M_{\odot}$ kpc$^{-2}$ and sizes are 2D half-mass radii. The colors once again reflect Rel(\facc), and we see the same correlation with size. We also overlay the DNGS galaxies, and, following the scheme of Figures \ref{fig:masses:compare} and \ref{fig:adjustedmassmetrics}, indicate those with and without size matches with filled and open points, respectively. 

A comparison between Figure \ref{fig:matching:sizemass} and Figure \ref{fig:profiles:facc:ms:dngsmatched} shows that the three DNGS galaxies that are smaller than their entire mass-matched samples also have surface densities that are at or below their analog TNG100 surface density profiles. We note, however, that these are not the only galaxies to have lower surface densities in the outskirts -- NGC 1084, NGC 2903 and M101 fall into this category as well despite having size-matched galaxies, and NGC 1042 reverses the trend by exhibiting a surface density profile that lies near the median of its mass-matched TNG100 profiles and a size that is towards the upper end of the distribution.

\begin{figure*}
    \centering
    \includegraphics[width=\linewidth]{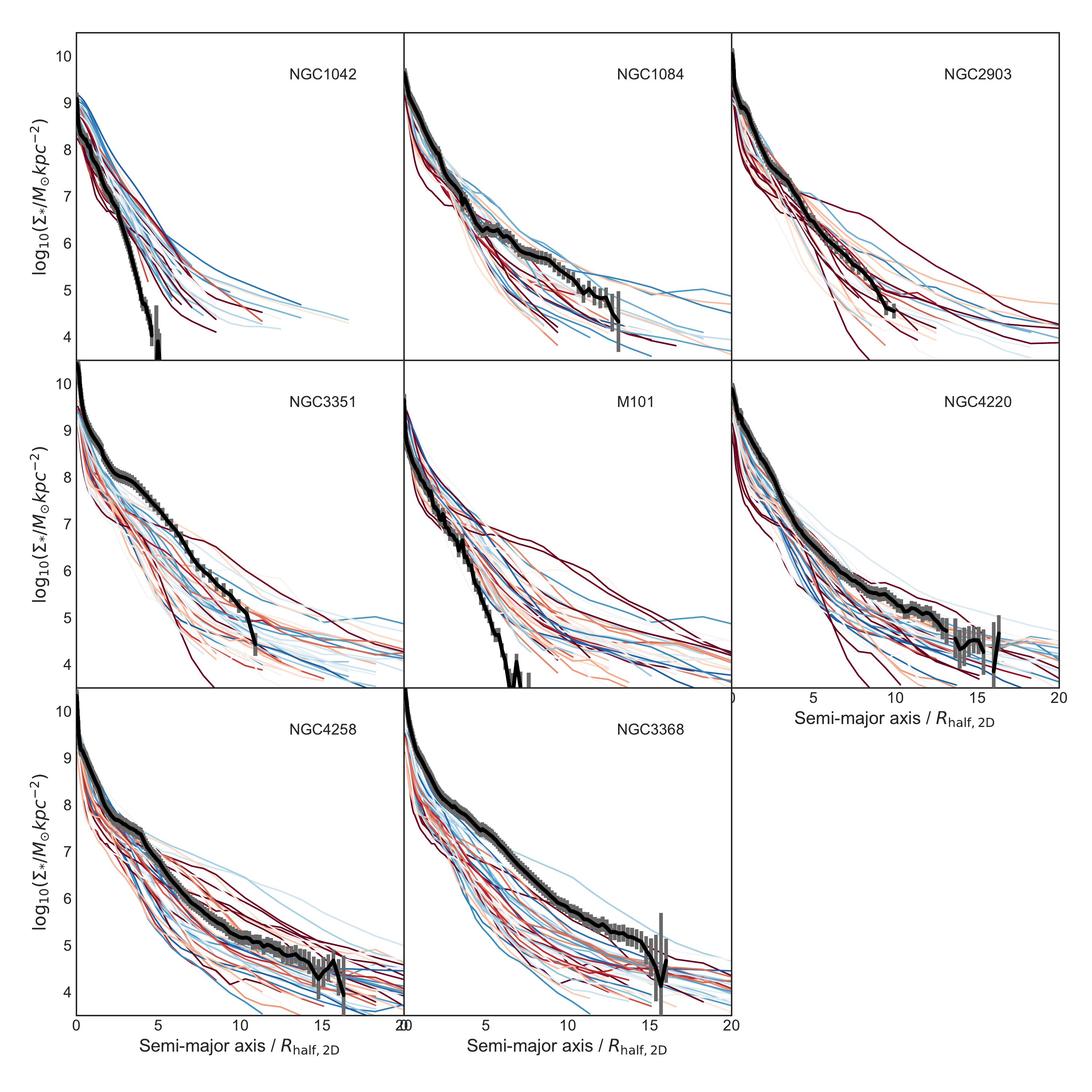}
    \caption{\textbf{DNGS vs TNG100 re-scaled stellar mass surface density profiles.} Each panel here shows a comparison between the surface density profile of a DNGS galaxy and the surface density profiles of stellar mass-matched TNG100 galaxies, where we have re-scaled the x-axis by the 2D half-mass radii of each galaxy. Similar to Figure \ref{fig:profiles:facc:ms:dngsmatched}, we only show the profiles out as far as the distance to the $50^{\rm th}$ outermost stellar particle. The color coding describes the relative distance of each TNG100 galaxy in \facc from the median \facc at fixed $M_{\rm stell,true}$ (see Figures \ref{fig:profiles:facc:ms} or \ref{fig:matching:sizemass}). DNGS galaxies lie above their TNG100 analogs in some cases, due to their relatively small sizes. We note that when we scale the surface density profiles in this way, we can no longer identify any correlation between the outer profile shape and Rel(\facc).}
    \label{fig:matching:scaledprofiles}
\end{figure*}

Figure \ref{fig:matching:scaledprofiles} shows the surface density profiles of each DNGS galaxy as a function of galacto-centric distance scaled by the 2D half-mass radius. We overlay the TNG100 mass-matched samples, color-coded by Rel(\facc). Unlike what we saw in Figure \ref{fig:profiles:facc:ms:dngsmatched}, some of the re-scaled DNGS surface density profiles lie \textit{above} their TNG100 counterparts as a result of their relatively small sizes. We emphasize that when we scale the profiles in this way, we erase the correlation between the outer profile shape and relative accretion fraction.

Turning to stellar halo masses, we note that in Figure \ref{fig:masses:compare}, the three DNGS galaxies without size matches are the only ones to fall outside of the entire distribution of stellar halo masses when measured beyond 20 kpc. However, when measured outside of 2 half-mass radii or below surface densities of $10^{6}M_{\odot}$ kpc$^{-2}$, they cannot be easily separated from the rest of the DNGS sample. Additionally, we once again find that in spite of having a \textit{larger} size than most of its mass-matched sample, NGC 1042 is at the low end of the stellar halo mass distribution irrespective of the choice of measurement. 

Taken together, we conclude that the small sizes of NGC 3351, NGC 3368, and NGC 4220 may act to increase any existing differences relative to their mass-matched samples, particularly when measuring stellar halos outside of 20 kpc. However, given the behavior of the rest of the sample it seems unlikely that this can fully reconcile the stellar halos of the two samples of galaxies. These findings are also consistent with the results of Section \ref{sec:discussion:toymodels}, where we saw that reducing the spatial extent of either the ex-situ or in-situ stellar components lead to an improved agreement between DNGS and TNG100 stellar halos.

\bsp	
\label{lastpage}
\end{document}